\documentclass{aa}  
\newcommand{\HII}{H\,{\scriptsize II}}

\newcommand{\CII}{[C\,{\scriptsize II}]}

\newcommand{\NII}{[N\,{\scriptsize II}]}

\newcommand{\OI}{[O\,{\scriptsize I}]}

\newcommand{\kms}{km s$^{-1}$}

\newcommand{\msol}{M$_{\odot}$}   
\usepackage[version=3]{mhchem}
\usepackage{graphicx}
\usepackage{txfonts}
\begin{document} 

\title{Anatomy of the massive star-forming region S106}
\subtitle{The [O\,{\small I}] 63 $\mu$m line observed with GREAT/SOFIA as a versatile diagnostic tool for
the evolution of massive stars}

\author{N. Schneider\inst{1,2} 
\and
M. R\"ollig\inst{1}
\and
R. Simon\inst{1}
\and 
H. Wiesemeyer\inst{3}
\and 
A. Gusdorf\inst{4}
\and 
J. Stutzki\inst{1}
\and  
R. G\"usten\inst{3}
\and
S. Bontemps\inst{2} 
\and
F. Comer\'on\inst{5}
\and
T. Csengeri\inst{3}
\and
J.D. Adams\inst{6}
\and
H. Richter\inst{7}
}

\institute{I. Physikalisches Institut, Universit\"at zu K\"oln,
Z\"ulpicher Str. 77, 50937 K\"oln, Germany\\
\email{nschneid@ph1.uni-koeln.de}
\and 
OASU/LAB, Universit\'e de Bordeaux, 33615 Pessac, France
\and 
Max-Planck Institut f\"ur Radioastronomie, Auf dem H\"ugel 69, 53121 Bonn, Germany
\and 
LERMA, Obs. de Paris, ENS, PSL Research University, CNRS, Sorbonne Universit\'es, UPMC Univ.Paris 06, Paris, France
\and 
European Southern Observatory, Alonso de C\'ordova 3107, Vitacura, Santiago, Chile
\and 
SOFIA, USRA, NASA/Armstrong Flight Research Center, 2825 East Avenue P, Palmdale, CA 93550, USA
\and 
DLR, Rutherfordstraße 2, 12489 Berlin-Adlershof, Germany
%
 }

\date{draft of \today}

\abstract {The central area (40$'' \times$40$''$) of the bipolar
  nebula S106 was mapped in the \OI\, line at 63.2 $\mu$m (4.74 THz)
  with high angular (6$''$) and spectral (0.24 MHz) resolution, using
  the GREAT heterodyne receiver on board SOFIA. The spatial and
  spectral emission distribution of \OI\ is compared to emission in
  the CO 16$\to$15, \CII\ 158 $\mu$m, and CO 11$\to$10 lines,
  mm-molecular lines, and continuum.  The \OI\ emission is composed of
  several velocity components in the range from --30 km s$^{-1}$ to 25
  km s$^{-1}$.  The high-velocity blue- and redshifted emission
  (v=--30 to --9 km s$^{-1}$ and 8 to 25 km s$^{-1}$) can be explained
  as arising from accelerated photodissociated gas associated with a
  dark lane close to the massive binary system S106 IR, and from
  shocks caused by the stellar wind and/or a disk--envelope
  interaction. At velocities from --9 to --4 km s$^{-1}$ and from 0.5
  to 8 km s$^{-1}$ line wings are observed in most of the lines that
  we attribute to cooling in photodissociation regions (PDRs) created
  by the ionizing radiation impinging on the cavity walls. The
  velocity range from --4 to 0.5 km s$^{-1}$ is dominated by emission
  from the clumpy molecular cloud, and the \OI, \CII, and high-J CO
  lines are excited in PDRs on clump surfaces that are illuminated by
  the central stars. Modelling the line emission in the different
  velocity ranges with the KOSMA-$\tau$ code constrains a radiation
  field $\chi$ of a few times 10$^4$ and densities n of a few times
  10$^4$ cm$^{-3}$.  Considering self-absorption of the \OI\ line
  results in higher densities (up to 10$^6$ cm$^{-3}$) only for the
  gas component seen at high blue- and red velocities. We thus confirm
  the scenario found in other studies that the emission of these lines
  can be explained by a two-phase PDR, but attribute the high-density
  gas to the high-velocity component only.  The dark lane has a mass
  of $\sim$275 M$_\sun$ and shows a velocity difference of $\sim$1.4
  km s$^{-1}$ along its projected length of $\sim$1 pc, determined
  from H$^{13}$CO$^+$ 1$\to$0 mapping. Its nature depends on the
  geometry and can be interpreted as a massive accretion flow (infall
  rate of $\sim$2.5 10$^{-4}$ M$_\odot$/yr), or the remains of it,
  linked to S106 IR/FIR. The most likely explanation is that the
  binary system is at a stage of its evolution where gas accretion is
  counteracted by the stellar winds and radiation, leading to the very
  complex observed spatial and kinematic emission distribution of the
  various tracers.}
 

\maketitle


\section{Introduction}

\begin{figure}
\centering
\includegraphics[width=9cm, angle=0]{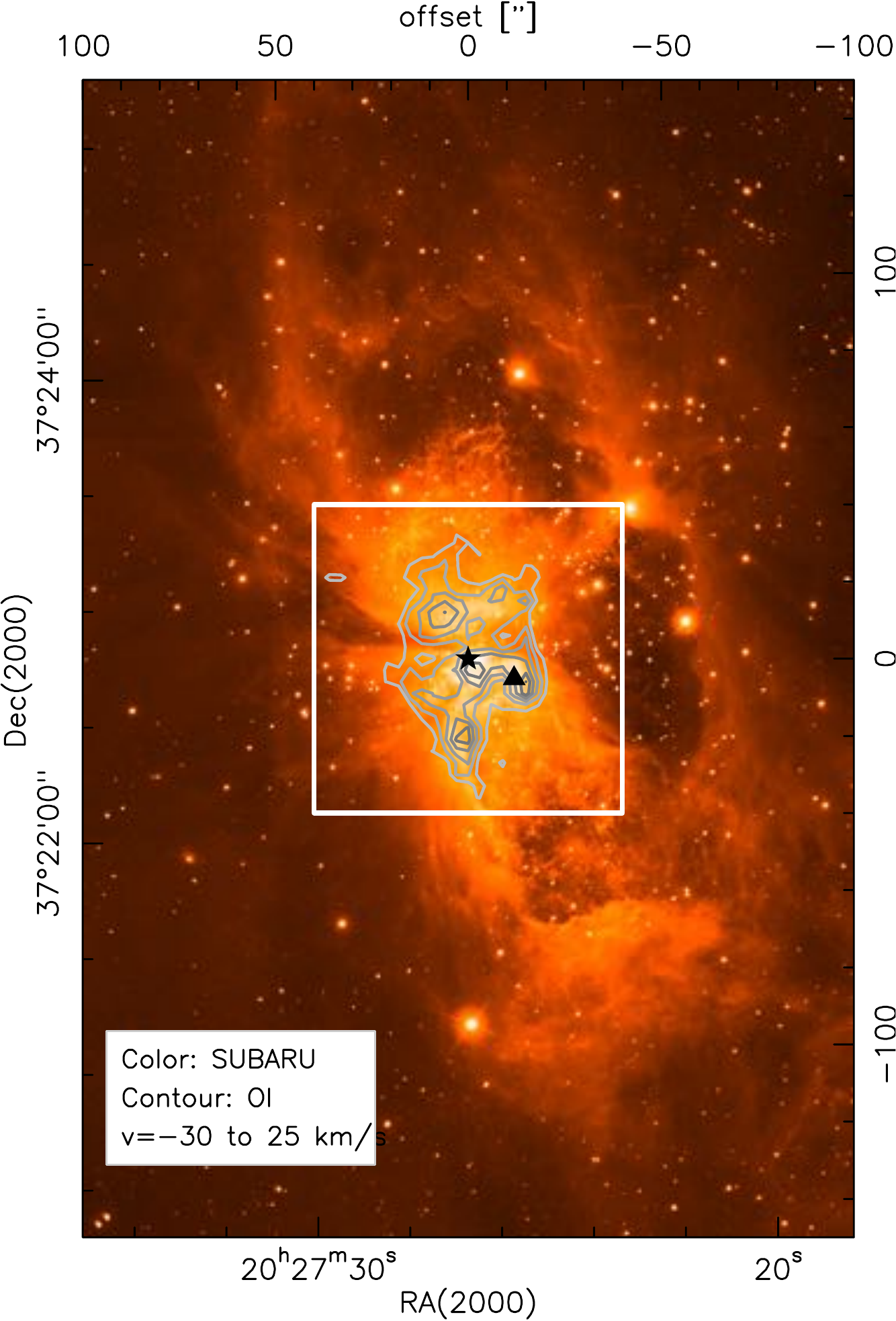}
\caption{Near-IR image (filters at J (1.25 $\mu$m), H (1.65 $\mu$m),
  and K$'$ (2.15 $\mu$m)) of S106 taken with Subaru (Oasa et
  al. 2006), outlining the bipolar emission nebula, with contours of
  velocity integrated (--30 to 25 km s$^{-1}$) \OI\ emission (136 to
  456 K km s$^{-1}$ in steps of 64 K km s$^{-1}$). The area mapped
  with SOFIA in \OI\ and CO 16$\to$15 emission is indicated with a
  white polygon. The star indicates the position of the S106 IR binary
  system and the triangle the position of the young stellar object
  (YSO) S106 FIR.}
\label{overview}
\end{figure}

%
\noindent {\bf Massive star formation} \\
\noindent Massive stars form by accretion of mass in one of three
ways: the core accretion model similar to low-mass stars via a
prominent disk \citep{McKee2002}, the competitive accretion scenario
in a clustered environment with a very small accretion disk
\citep{Bonnell2007,Bate2012}, or the fragmentation-induced starvation
view \citep{Peters2011} by fragmentation from gravitational
instability in dense accretion flows.  Only the second and third
scenarios straightforwardly explain why massive stars often form in
multiple systems and are found preferentially at the centre of stellar
clusters \citep{Bontemps2010}. Regardless of the scenario, what is
known so far about how the different stages can be observationally
traced, in particular with regard to the far-infrared (FIR), can be
summarized as follows \citep[see also][]{Tan2014}:
\vspace{0.2cm}

\noindent {\sl Pre-stellar phase:} Massive dense clumps (masses from a
few tens to a few thousands of M$_\odot$, size $\sim$1 pc, density
n$\ge$10$^6$ cm$^{-3}$, temperature T$\sim$15 K) are the locations
where high-mass stars form
\citep[e.g.][]{Beuther2002,Tan2013,Motte2007}.  They often show
subfragmentation into several smaller cores, i.e. pre- and
protostellar objects, of size scale $\ll$0.1 pc
\citep[e.g.][]{Bontemps2010}, but there are also examples for isolated
massive cores \cite[e.g.][]{Duarte-Cabral2013,Csengeri2017}.
Observationally, these cold dense clumps and cores are best traced by
dust emission and molecular lines (such as N$_2$H$^+$) in the mm and
sub-mm wavelength range.  In contrast, molecular line emission in the
FIR is not observed in the cold dense gas phase because the energy
levels of rotational transitions are too high. Only light hydrides
such as CH$^+$, CH, NH, OH, etc., can be seen in more diffuse
gas. Molecular lines with lower excitation than the background
continuum temperature can be observed in absorption.  A good example
is NH$_3$ with its transitions from non-metastable to metastable
levels, which have low excitation temperatures.  For example,
\citet{Wyrowski2012} successfully observed the 1.8 THz NH$_3$ line in
absorption in high-mass star-forming clumps.
\vspace{0.2cm} 

\noindent {\sl Protostellar/stellar phase:} Massive protostars
contract so quickly that there is virtually no pre-main sequence
stage.  Hydrogen burning starts almost instantaneously, and they are
main sequence objects while still accreting. However, no well-defined
Keplerian disk has been detected yet, only disk-like structures on
size scales smaller than 2000 AU
\citep[e.g.][]{Beltran2011,Sanchez-Monge2013}.  The envelope--disk
interaction leads to an atomic jet of at least partly ionized gas
orthogonal to the disk, which can then drive a molecular outflow
\citep[e.g.][]{Kuiper2011}. For low-mass protostars and massive
protostars in the core accretion model the bipolar radio jet and
molecular outflow are well collimated with a time-dependent activity
and opening angle \citep{Bontemps1996}.  For massive protostars in the
competitive accretion scenario, the jet/outflow is often poorly
collimated.  Surrounding lower-mass protostars can cause multiple
overlapping, randomly aligned outflows so that the overall outflow
pattern can be rather complex \citep{Hunter2008}. A gas distribution
with a highly fragmented and filamentary structure is predicted in the
fragmentation-induced starvation scenario where the dense accretion
flows should be directly observable. Radiative heating leads to a
higher Jeans mass so that fewer but more massive stars form
\citep{Peters2010a}.

Generally, outflows from massive protostars are more massive and
energetic \citep{Beuther2002,Duarte-Cabral2013} than those from
low-mass protostars. Extreme UV photons (EUV, energy $>$13.6 eV) lead
to ionization of the protostellar outflows and create an
outflow-confined \HII\ region. One class of such \HII\ regions are
those with a morphology of an hourglass shaped parsec-scale bipolar
nebula such as S106.  Another example is associated with the evolved
star MWC349A \citep{Menten2012} which in addition shows a subparsec
(sub-pc) nebula that is explained as being due to a biconical outflow
of ionized gas pinched at the waist by a small disk seen edge-on
\citep{Cohen1985}.  Typical of many bipolar nebulae is also a belt of
dense cold gas in the equatorial plane of the nebula
\citep{Gvaramadze2010}.  Though it is commonly accepted that a disk
oriented perpendicular to the symmetry axis of the \HII\ region plays
a crucial role in shaping the small-scale nebula/\HII\ region and the
large-scale bipolar nebula, it is not clear how this process works in
detail. In addition, the nature of the dense gas structure in the
nebula waist, where the existing star or stars are embedded, is not
clear.  It might be the remains of an accretion flow, as was shown in
numerical simulations \citep{Peters2010a}, or the result of a shock
that leads to an expansion and ultimately the dispersal of the belt
\citep{Menten2012}.

The earliest stages of hyper-compact,
ultra-compact, and compact \HII\ regions are sometimes difficult to
identify because the massive protostar is deeply embedded in dense
gas, and even cm-emission can become optically thick. Ionized
collimated jets have been seen in radio continuum for some massive
protostars \citep[e.g.][]{Gibb2003,Guzman2014}.  At the stage when a
larger cavity is blown out, entrained gas can also arise from the ablation
of gas at the cavity borders to the surrounding molecular cloud,
caused by the stellar wind. Generally, as soon as the massive star
reaches the main sequence, all forms of radiative feedback (thermal
heating, ionization, radiation pressure on dust) and mechanical
feedback (stellar winds from the stellar surface, protostellar
winds/outflows from the magneto-centrifugally driven flows powered by
accretion) become important. Observationally, it is challenging to
disentangle the different processes and to establish an evolutionary
sequence for the protostellar phase because it is short (a few
10$^3$ yr, \citealt{Duarte-Cabral2013}) and because of the small number
of objects to study.  An additional complication is that most massive
stars form as binary or multiple systems, so that the observed jet and
outflow patterns are even more complex and depend on the properties of
each system.

It is the objective of this study and follow-up studies on the
star-forming region S106 to identify and potentially better
characterize these detailed phases by combining the spatial, velocity,
and intensity information of atomic and molecular line observations
and continuum from the mm to the FIR. In this first paper we focus on
interpreting the \OI\ 63 $\mu$m emission observed in the immediate
environment of the exciting source S106 IR (see below) in the
framework of a photodissociation region (PDR).  A possible shock
origin of the \OI\ emission will be discussed and modelled in a
subsequent paper, followed by a study of the large-scale structure of
the bipolar nebula using \CII\ 158 $\mu$m SOFIA data. \\

\begin{figure}
\centering
\includegraphics[width=7cm, angle=0]{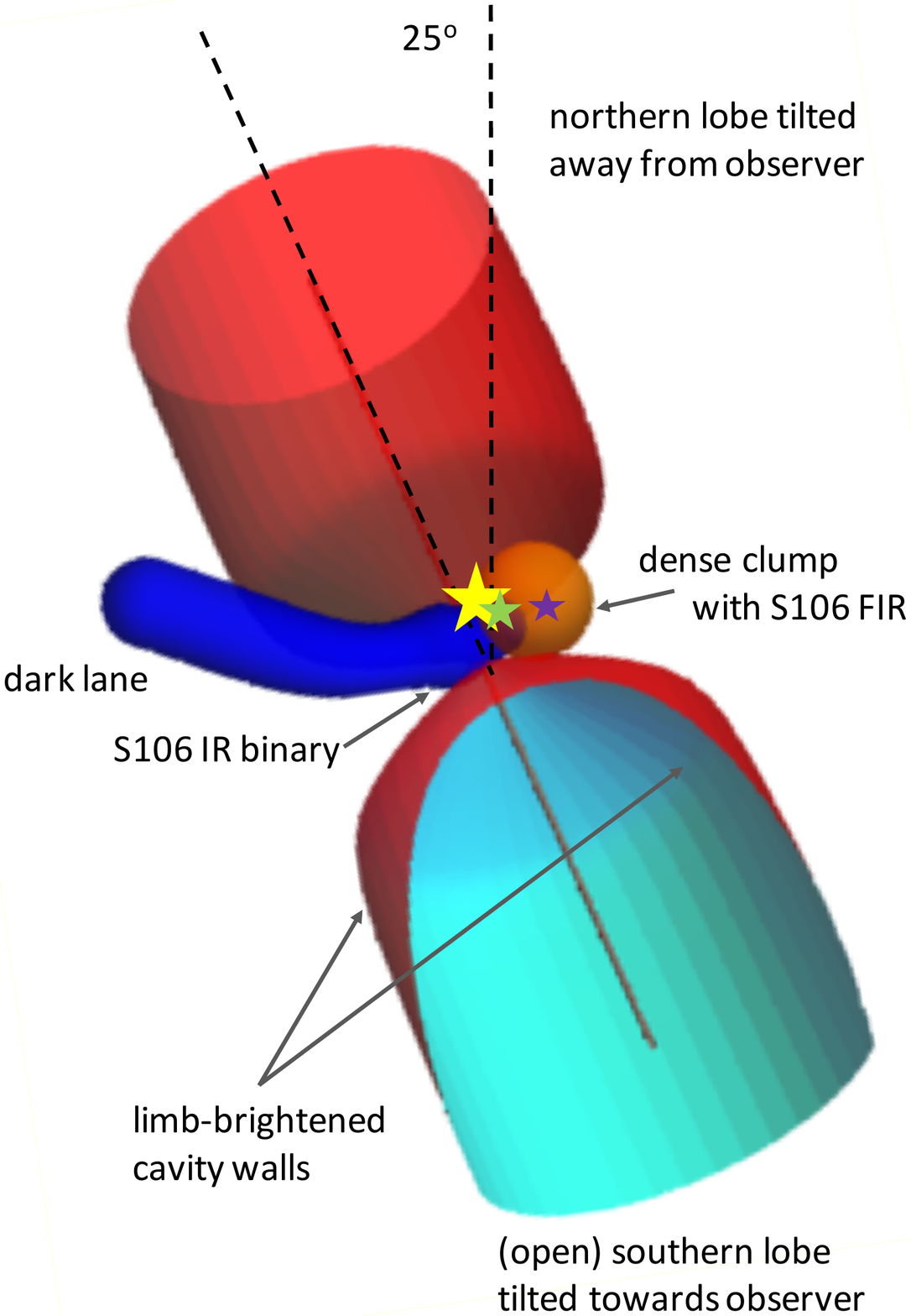}
\vspace{-1cm}
\caption{Schematic view of the S106 region as seen on the sky.
  The nebula is tilted $\approx$25$^\circ$ to the east in the plane of
  the sky; the northern lobe, which is more obscured by foreground gas
  and dust, is inclined ($\approx$15$^\circ$) away from the observer.
  The southern lobe is inclined towards the observer and lacks the
  front side. The bipolar nebula and the \HII\ region are embedded in
  an extended molecular cloud.}
\label{cartoon}
\end{figure}
  
\noindent{\bf S106} \\
\noindent 
The bipolar \HII\ region S106 in Cygnus X is an enigmatic object that
has received considerable attention owing to its eye-catching
appearance. Its distance was estimated to be 1.7 kpc
\citep{Schneider2007} and then determined with maser parallax
measurements to be 1.3$\pm$0.1 kpc \citep{Xu2013}. In the recent Gaia
DR2 catalogue, the distance is derived from the trigonometric parallax
to be 1.671 kpc (+738 pc, -392 pc). In this paper, we use a value of
1.3 kp.  S106 IR was always thought to be a single late O to early B
star (see \citealt{Hodapp2008} for a summary).  However, recent
observations \citep{Comeron2018} show that it is a close (separation
$<$0.2 AU), massive binary system, most likely consisting of a late O
and a late B star, being responsible for the bipolar emission nebula
(Fig.~\ref{overview}).  Hereafter, we refer to this binary system as
S106 IR.  A small (only slightly larger than the beam of $\sim$30
mas), edge-on disk-like feature around the two stars was discovered by
cm-interferometry \citep{Hoare1996, Gibb2007}, but its exact
evolutionary status is not clear \citep{Adams2015}.  The system
already shows signatures of main sequence stars, emitting copiously in
the UV (where the O star dominates) and driving an ionized wind with a
velocity of 100--200 \kms\ \citep{Bally1983,Simon1982} into the
bipolar cavity.  Associated with this system is an IR cluster with
more than 160 members \citep{Hodapp1991}. The centroid of the cluster
lies about 30$''$ west and 15$''$ north of S106 IR and is
symmetrically distributed about this location.

Figure~\ref{cartoon} shows a cartoon of S106 with its known features 
and those observed to date. The two lobes of the \HII\ region are inclined
with respect to a vertical north--south axis. In addition, the northern
(southern) lobe is tilted away from (towards) the observer
\citep{Solf1982}.  A large portion of the front part of the southern
cavity facing the observer has been eroded so that the backside of the
cavity is visible in the optical and IR.  The northern lobe is
obscured by foreground gas and dust. The strongest nebular emission in
the optical and near-IR arises from the shell-like structures that
represent the ionization front and from the limb-brightened edges of
the cavity walls.  High-velocity emission of the \CII\ 158 $\mu$m fine
structure line and high-J CO lines \citep{Simon2012} was interpreted
as arising from swept-up material at the front and back sides of the
expanding wind-driven cavity.

Another prominent feature related to S106 is a dark lane dividing the
two lobes (indicated in dark blue in the cartoon) and lying apparently
in front of S106 IR.  It was initially interpreted as the shadow of a
large-scale disk \citep{Bally1983,Bieging1984}, but dust continuum
observations \citep{Vallee2005,Motte2007,Simon2012,Adams2015} revealed
it to be an elongated high column-density feature associated with the
extended molecular cloud \citep{Schneider2007}.  Dust observations
also show strong emission $\approx$15$''$ west of S106 IR, coinciding
with two clusters of H$_2$O masers \citep{Stutzki1982,Furuya1999}.
The dust peak was named S106 FIR by \citet{Richer1993}, and
interpreted as a Class 0 young stellar object (YSO).  The molecular
gas close to S106 IR is highly clumped \citep{Schneider2002} and a
number of authors interpreted their observations of ionic and
fine-structure lines \citep{vandenAncker2000, Schneider2003,
  Simon2012, Stock2015} as arising from photodissociation regions
(PDRs) on the surfaces of these individual clumps around S106
IR. However, most of the studies of the FIR fine-structure cooling
lines of S106 were hampered because only line integrated fluxes were
observed and used for PDR modelling. Here we take a different approach
in that we benefit from the unprecedented velocity resolution of the
German REceiver for Astronomy at Terahertz frequencies (GREAT) on
board the Stratospheric Observatory for Far-Infrared Astronomy
(SOFIA), and apply PDR modelling to line ratios in different velocity
ranges so that we can determine more precisely the density and
radiation field in individual portions of the gas.

In addition to getting a better understanding of the nature of the
S106 star-forming complex itself, one of the objectives of this paper
is putting our results into a larger context of how massive stars form
and evolve.  The paper is structured as follows.  Section~\ref{obs}
gives observational details and Sect.~\ref{pdr} outlines how we
realize the PDR modelling. In Sect.~\ref{results}, we present maps of
the observed lines and give results of PDR modelling. In
Sect.~\ref{discuss}, we develop a physical model of the S106 region
and put the findings into a larger context of massive star
formation. Section~\ref{summary} summarizes the results of the paper.

\section{Observations} \label{obs} 

\begin{figure}
\centering
\includegraphics[width=8.5cm, angle=0]{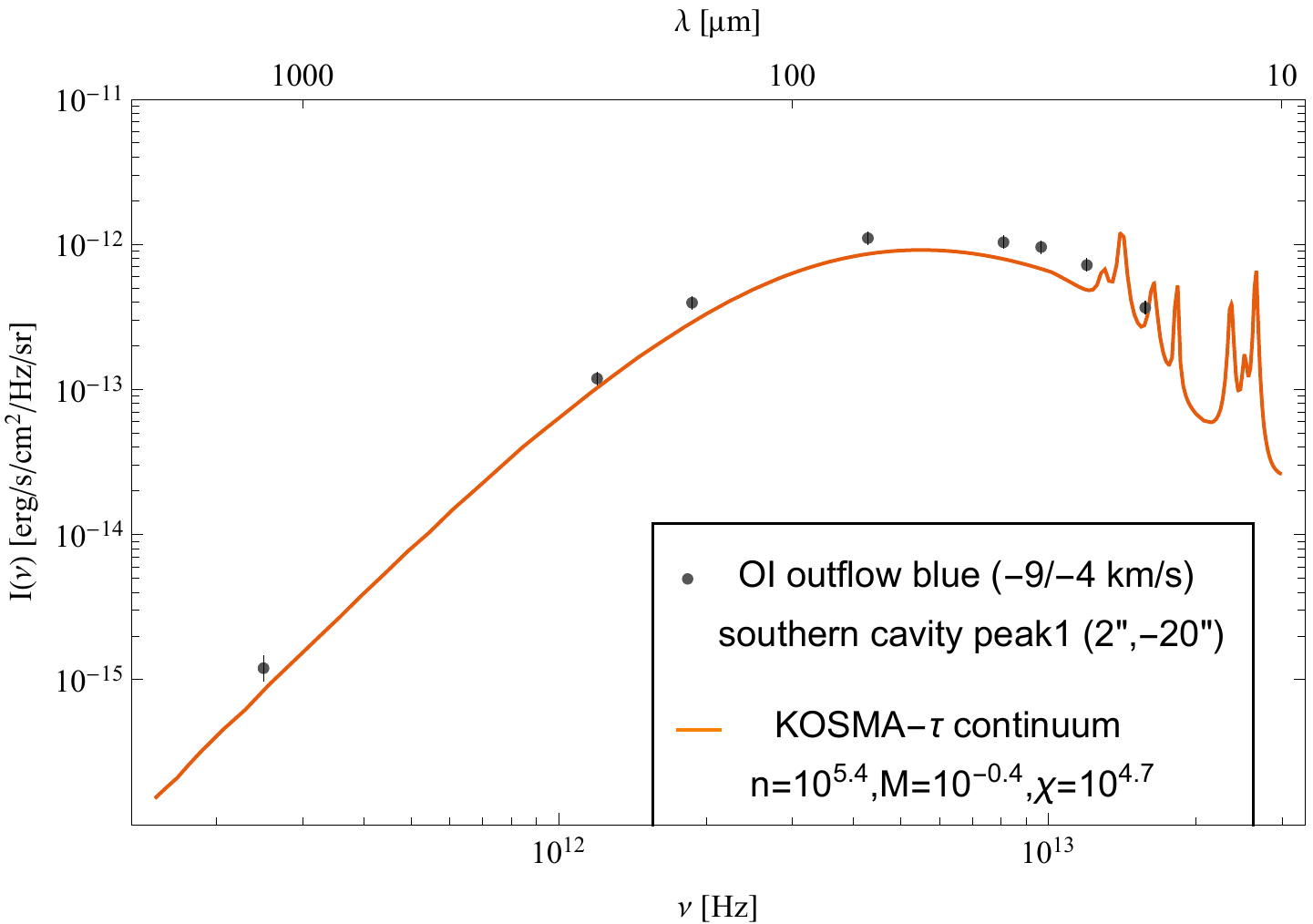}
\caption{One example (position 3a, southern cavity) for KOSMA-$\tau$
  PDR model continuum emission fitting the observed continuum
  observations (black points) of MAMBO (1.2mm), {\sl Herschel} (PACS:
  70, 160, SPIRE: 250 $\mu$m), and FORCAST (19.7, 25.3, 31.5, 37.1
  $\mu$m), assuming a unity beam filling factor. The error for the
  flux values is generally small (see Sect.~2.2). All data were
  smoothed to a common angular resolution of 15$''$. The {\sl
    Herschel}/SPIRE 250 $\mu$m data point has a resolution of 18$''$.}
\label{Fig:cont}
\end{figure}
\subsection{SOFIA} \label{sofia} 

The \OI\ $^3$P$_1 \rightarrow$ $^3$P$_2$ atomic fine structure line at
4.74477749 THz (63.2 $\mu$m) and the CO 16$\to$15 rotational line at
1.841345 THz (162.8 $\mu$m) were observed with the heterodyne receiver
GREAT\footnote{German Receiver for Astronomy at Terahertz. GREAT is a
  development by the MPI f\"ur Radioastronomie and the
  KOSMA/Universit\"at zu K\"oln, in cooperation with the MPI f\"ur
  Sonnensystemforschung and the DLR Institut f\"ur Planetenforschung.}
on board SOFIA during one flight on December 15, 2015, from
Palmdale/California. The \OI\ line was observed in the single-pixel H
channel (now upgraded to a 7-pixel array), and the CO line in the
single-pixel L2 channel. The average system temperatures over the
total bandpass were 3641 K and 5059 K for \OI\ and CO, respectively.
We employed a fast Fourier transform spectrometer (AFFTS).  The centre
 intermediate frequency (IF) was 1455 MHz for the \OI\ line and 1000
MHz for the CO line.  The frequency resolution of the original
\OI\ and CO data sets is 0.244 MHz, giving a velocity resolution of
$\sim$0.04 km s$^{-1}$ (16384 spectrometer channels).

Beam-switched on-the-fly maps with a scan length of 90$''$, a step
size of 3$''$, and a dump time of 1s were performed. The chopper throw
was 360$''$ with a chop angle of 45$^{\circ}$ anti-clockwise from
+R.A.  The map centre position refers to S106 IR
(R.A., Dec.)(J2000)$=(20^h27^m26^s.74,37^\circ22'47''.9)$. The same
area was covered three times while scanning in R.A., and two times in
Dec. Procedures to determine the instrument alignment and telescope
efficiencies, antenna temperature and atmospheric calibration, as well
as the spectrometers used, are described in \citet{Heyminck2012} and
\citet{Guan2012}. Tracking was done on the H-channel and the
co-alignment between the H- and L2-channel is determined to be
$\sim$2$''$.  All line intensities are reported as main beam
temperatures scaled with main-beam efficiencies of 0.69 and 0.68 for
\OI\ and CO, respectively, and a forward efficiency of 0.97.  The main
beam sizes are 15.3$''$ for the L2 channel and 6.1$''$ for the H
channel.

The telluric \OI\ line, originating from the mesosphere, contributes
as a narrow feature at the velocity of the bulk emission of the cloud
(approximately +3 km s$^{-1}$). We corrected for the absorption
following the procedure described in Appendix A in
\citet{Leurini2015}, assuming that the profile can be characterized by
a Gaussian. In summary, for the opacity correction the absorption
strength was adjusted so as to achieve an adequate interpolation
between adjacent unaffected spectral channels.

The calibrated \OI\ and CO spectra were further reduced and analysed
with the GILDAS software, developed and maintained by IRAM. From the
spectra, a third-order baseline was removed and spectra were then
averaged with 1/$\sigma^2$ weighting (baseline noise).  The mean
r.m.s. noise temperatures per 0.6 km s$^{-1}$ velocity bin are 1.9 K
for \OI\ (6$''$ beam) and 1.6 K for CO (15$''$ beam).  We estimate
that the absolute calibration uncertainty is  $\sim$20\%.

For some overlays and PDR modelling, we use \CII\ 158 $\mu$m and
$^{12}$CO 11$\to$10 data observed with SOFIA from 2012
\citep{Simon2012} and new observations from May 2015 and December 2016
(Simon et al., in prep.).

\begin{figure*}
\centering
\includegraphics[width=13cm, angle=0]{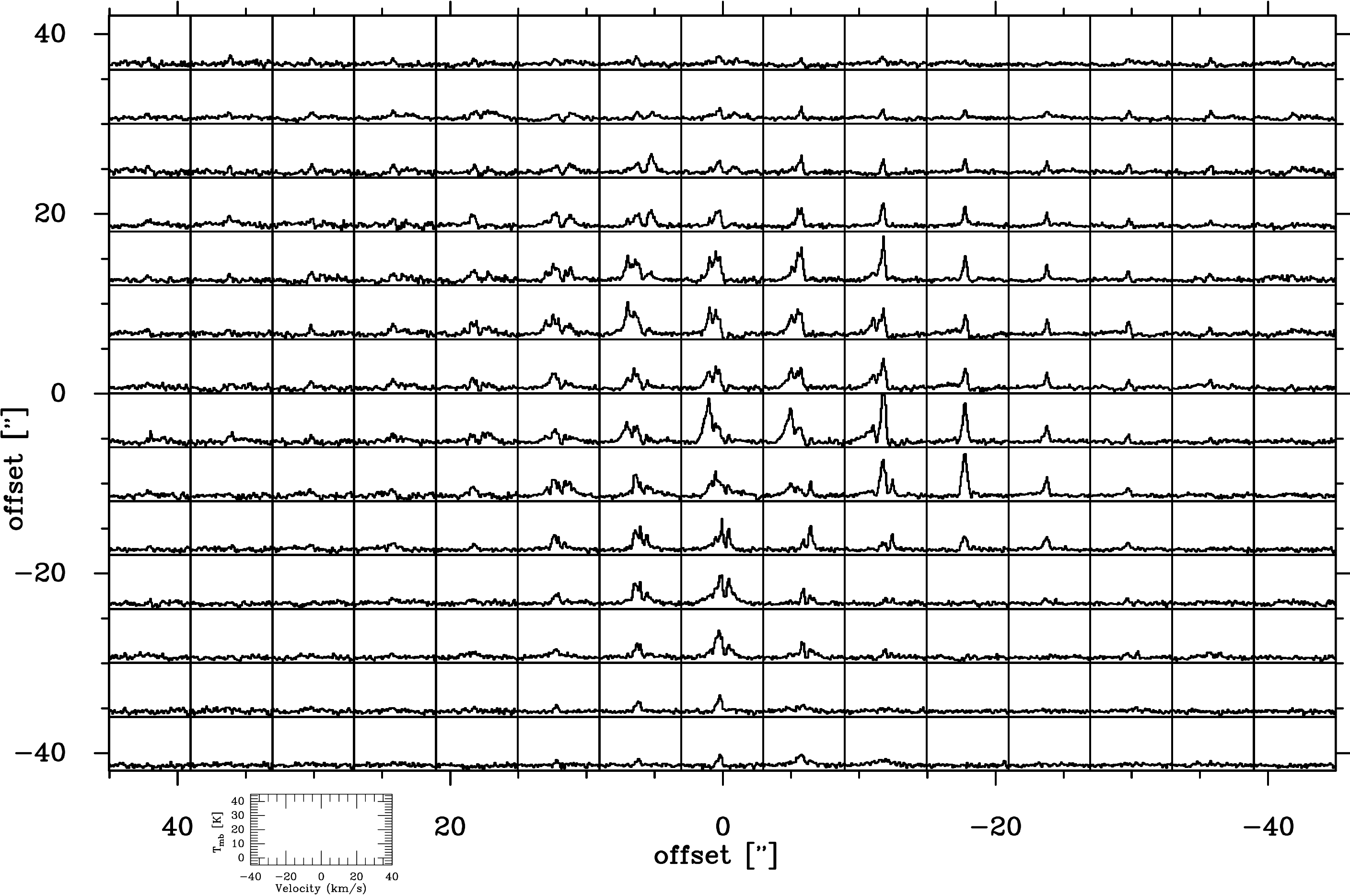}
\caption{Spectral map of \OI\ emission with a velocity resolution of 1
  km s$^{-1}$ on a 6$''$ beam-sampled grid in the velocity range from -40
  to 40 km s$^{-1}$ and main beam brightness temperature range from -4 to
  40 K.}
\label{spectra-map}
\end{figure*}

\subsection{Complementary data} \label{other-data} 

\noindent {\bf IRAM 30m} \\
We use molecular line data from the IRAM\footnote{IRAM is supported by
  INSU/CNRS (France), MPG (Germany), and IGN (Spain)} 30m telescope,
obtained in January 2016 (PI: N. Schneider). The observations were
performed with two settings, one with the high-resolution spectrometer
FTS50 attached to the EMIR E0 receiver (at 3mm wavelength), and one
with the lower resolution FTS200 in combination with EMIR E1 (at 1mm
wavelength). A large number of molecular lines ($\sim$20) was
observed, covering a range from 85.3 to 106.1 GHz and from 89.8 to
110.5 GHz, respectively. We performed on-the-fly maps that consist of
one horizontally and one vertically scanned map of size 300$''$ at 3mm
and 168$''$ at 1mm (larger than is shown in Figure~1).  The quality of
the data is very good, thus only baselines of first order were
removed. For the H$^{13}$CO$^+$ 1$\to$0 data we employ here, the
average system temperature during the observations was 88 K and the
average baseline rms for spectra smoothed to 15$''$ angular resolution
and with a velocity resolution of 0.17 km s$^{-1}$ is 0.16 K. \\

\noindent {\bf Other instruments} \\
For overlays and PDR modelling, we use published or archived continuum
data from {\sl Herschel} (70, 160, 250, 350, 500 $\mu$m) and FORCAST
(19.7, 25.3, 31.5, 37.1 $\mu$m) \citep{Adams2015}, {\sl Spitzer} (3.6,
4.5, 5.8, 8.0 $\mu$m), MAMBO (1.2mm) \citep{Motte2007}, the VLA
\citep{Bally1983}, and Subaru \citep{Oasa2006}. The flux uncertainties
for the continuum data are $\sim$10\% for PACS, SPIRE, and FORCAST,
and $\sim$20\% for MAMBO (see references above for these values).

\begin{table}[htb]
\begin{center}
\caption{Overview of the most important model parameters. All abundances
are given with respect to the total H abundance. }\label{parameter}
\begin{tabular}{lll}
\hline\hline
\multicolumn{3}{c}{\rule[-3mm]{0mm}{8mm}\bf Model Input Parameters}\\ \hline
\rule[2mm]{0mm}{2mm}He/H&0.0851&(1)\\
O/H            & 4.47 10$^{-4}$  & (2)\\
C/H            & 2.34 10$^{-4}$  & (2)\\
$^{13}$C/H      & 3.52 10$^{6}$   & (3)\tablefootmark{a}\\
$^{18}$O/H      & 8.93 10$^{-7}$  & (4)\tablefootmark{b}\\
N/H            & 8.32 10$^{-5}$  & (2)\\
S/H            & 7.41 10$^{-6}$  &(2)\\
F/H            & 6.68 10$^{-9}$  &(2)\\
$Z$            & 1              &solar metallicity\\
$\zeta_{CR}$    & 2 10$^{-16}$ s$^{-1}$ & CR ionization rate (5)\\
$R_\mathrm{V}$  &3.1             & visual extinction/reddening (7,8) \\
$\sigma_\mathrm{D}$& 1.75 $^{-21}$~cm$^2$& UV dust cross section per H (8)\\
$\langle A(\lambda)/A_\mathrm{V} \rangle$&$3.339$&mean FUV extinction\\
$\tau_\mathrm{UV}$&$3.074 A_V$&FUV dust attenuation\\
$v_b$ & 1~km~s$^{-1}$&Doppler width\\
$n_0$&$10^{2,\ldots,7}$~cm$^{-3}$&total surface gas density\\
$M$&$10^2$~\msol&cloud mass\\
$\chi$ & $10^{0\ldots,6}$&FUV intensity w.r.t. (6)\tablefootmark{c}\\
$\alpha$&1.5&density power law index\\
$R_\mathrm{core}$&$0.2 R_\mathrm{tot}$&size of constant density core\\
$N_\mathrm{tot}/A_\mathrm{V}$& 1.62 10$^{21}$~cm$^{-2}$&(8)\\
\hline\\
\end{tabular}
\end{center}
\vspace{-1cm}
\tablebib{
(1) \citet{Asplund2005}; (2) \citet{Simon-Diaz2011}; (3) \citet{Langer1990}; (4) \citet{Polehampton2005}; (5) \citet{Hollenbach2012}; 
(6) \citet{Draine1978}; (7) \citet{Roellig2013}; (8) \citet{Weingartner2001a}.}
\tablefoot{
\tablefoottext{a}{based on a $^{12}$C/$^{13}$C ratio of 67}
\tablefoottext{b}{based on a $^{16}$O/$^{18}$O ratio of 500}
\tablefoottext{c}{$\chi =1.71~G_0$ where G$_0$ is the mean ISRF from \citep{Draine1978}.}
}
\end{table}

\section{Photodissociation region modelling} \label{pdr} 

We employ the KOSMA-$\tau$ PDR model \citep{Stoerzer1996,Roellig2006}
to derive local physical conditions from the observed line intensities
and continuum fluxes. In the following, we explain the model and our
approach in more detail because this methodology will also be used for
subsequent studies.

\subsection{Model description} \label{model} 

The KOSMA-$\tau$ PDR model numerically computes the energetic and
chemical balance in a spherical cloud that is externally irradiated.
We here use a non-clumpy model approach, i.e.  a single, spherical PDR
with a density gradient similar to a Bonnor--Ebert sphere, and a cloud
mass of $M=100$~M$_\odot$ (the results depend only weakly on the model
clump mass and we ran models with a parameter range between 10$^{-3}$
M$_\odot$ and 10$^3$ M$_\odot$).  The full numerical computation
scheme of KOSMA-$\tau$ involves three steps. First, the continuum
radiative transfer code MCDRT \citep{Szczerba1997} is used to compute
the thermal balance of all dust components (see below) as well as the
far-ultraviolet (FUV) radiative transfer within the model cloud, and
the emergent continuum radiation \citep{Roellig2013}. We include a
variety of different dust models, for example the MRN model
\citep{mrn} and the dust models by \citet{Weingartner2001a}. By
assuming that the dust temperature is independent of the gas
temperature, we then use the MCDRT output as input for the second step
where the KOSMA-$\tau$ code computes the chemical and physical state
of the gas. In Fig.~\ref{Fig:cont} we show as an example a comparison
between the observed mid-IR and FIR data at the position 3b (offsets
5$''$,--25$''$) and the best fitting dust continuum model. We
determined in this way the dust continuum at all positions (nine in
total) where we performed PDR modelling.

We assume a local chemical steady state, i.e. a balance between
chemical formation and destruction processes of all involved species 
and compute their local densities and thermal equilibrium, i.e. a
balance between all heating and cooling processes, resulting in local
gas and dust temperatures.  This is done for all radial grid positions
in an iterative way to reach the numerical convergence criteria which
is that the column density deviations between iterations are less
than 1\%.  The result is the radial density and temperature profile of
the model clump. In a last step the profile is used to perform the
radiative transfer computations, giving the spectral line emission of
the model cloud for comparison with observations
\citep{Gierens1992}.\footnote{The KOSMA-$\tau$ code was part of the
  PDR comparison benchmark study \citep{Roellig2007}.}

We use the most recent CO self-shielding functions by
\citet{Visser2009} and assume a Doppler line width according to the
linewidth--size relation as given by \citet{Larson1981}. The
photo-electric heating is computed according to
\citet{Weingartner2001b} (model 4 from their Table 2). The formation
of H$_2$ on grain surfaces follows
\citet{Cazaux2002,Cazaux2004,Cazaux2010erratum} and formation on polyaromatic hydrocarbon (PAH)
surfaces is suppressed. For more details see \citet{Roellig2013}.  In
addition, we use the UMIST Database for Astrochemistry chemical
network \citep{McElroy2013} including all isotopic reaction variants
including $^{13}$C and/or $^{18}$O isotopes \citep{Roellig2013b}. In
total 3766 reactions are considered and the 227 species that are
included in the chemistry are listed in
Table~\ref{species}. Numerically problematic reaction rate coefficients
have been rescaled according to \citet{Roellig2011}. The formation of
CH$^+$ and SH$^+$ is computed using state-to-state reaction rates
\citep{Agundez2010,Nagy2013}.

\subsection{Model fitting} \label{fit} 

Table~\ref{parameter} summarizes the model parameters.  To derive
local physical conditions, we compare observed line
ratios\footnote{All ratios are computed from intensities in erg
  s$^{-1}$cm$^{-2}$sr$^{-1}$. See Table~2 for conversion factors.}
with the corresponding model line ratios to find the best fitting
model parameters. We use two line ratios, ${\rm [CII]}_{158\mu m}/{\rm
  [OI]}_{63\mu m}$ and CO(16-15)/CO(11-10), and compare them against
$({\rm [CII]}_{158\mu m}+{\rm [OI]}_{63\mu m})/\Sigma_{FIR_{range}}$.
We note that we do not use here the total FIR intensity
$\Sigma_{FIR_{total}}$, which is commonly used in PDR models
\citep{Stock2015} because we work with velocity resolved data. The
parameter $\Sigma_{FIR_{range}}$ is thus only the continuum intensity
in one particular velocity range. We define five velocity intervals
for the \CII\ and \OI\ emission as explained in
Sect.~\ref{results}. Observationally, however, we can only determine
the total FIR intensity $\Sigma_{FIR_{total}}$ obtained over all
velocity ranges. For that, we derive $\Sigma_{FIR_{total}}$ by
numerically integrating the continuum data between 10 and 1000 $\mu$m,
stemming from FORCAST\footnote{To convert the FORCAST data from
  MJy/pixel to MJy/sr we use the conversion factor $7.213\times
  10^{10}$~pixel/sr.}, {\sl Herschel}, and MAMBO (see
Sec.~\ref{other-data}) using a linear
interpolation. Figure~\ref{Fig:cont} shows an example of this
procedure.  For the determination of $\Sigma_{FIR_{range}}$ we then
assume that gas and dust are well mixed and that the absolute \OI\ 63
$\mu$m line integrated intensity in all velocity ranges is higher than
that of \CII\ and CO (which is the case in each velocity range for
each observed/analysed position, see Table~\ref{tab1}). With these two
assumptions, the fraction of \OI\ line integrated intensity in one
velocity range to that in the total velocity range is the same as the
fraction of $\Sigma_{FIR_{range}}$ to $\Sigma_{FIR_{total}}$.

All maps are on the same angular resolution of 15$''$ (also the mid-
and FIR continuum data), except  the CO 11$\to$10 map, which has a
resolution of $\sim$20$''$, and the {\sl Herschel}/SPIRE 250$''$ data,
which have a resolution of 18$''$. In the following, we thus focus on
the values for density and UV field obtained from the ${\rm
  [CII]}_{158\mu m}/{\rm [OI]}_{63\mu m}$ ratio since beam filling
factors are eliminated to the first order and because only the \OI\ and
\CII\ lines show emission in all velocity ranges, in contrast to the
CO lines that show no emission (or only marginal amounts) at the highest
velocities. The model continuum intensities are provided by MCDRT; the
model line intensities are computed by KOSMA-$\tau$.

\section{Results} \label{results} 

\begin{figure}
\centering
\includegraphics[width=8.5cm, angle=0]{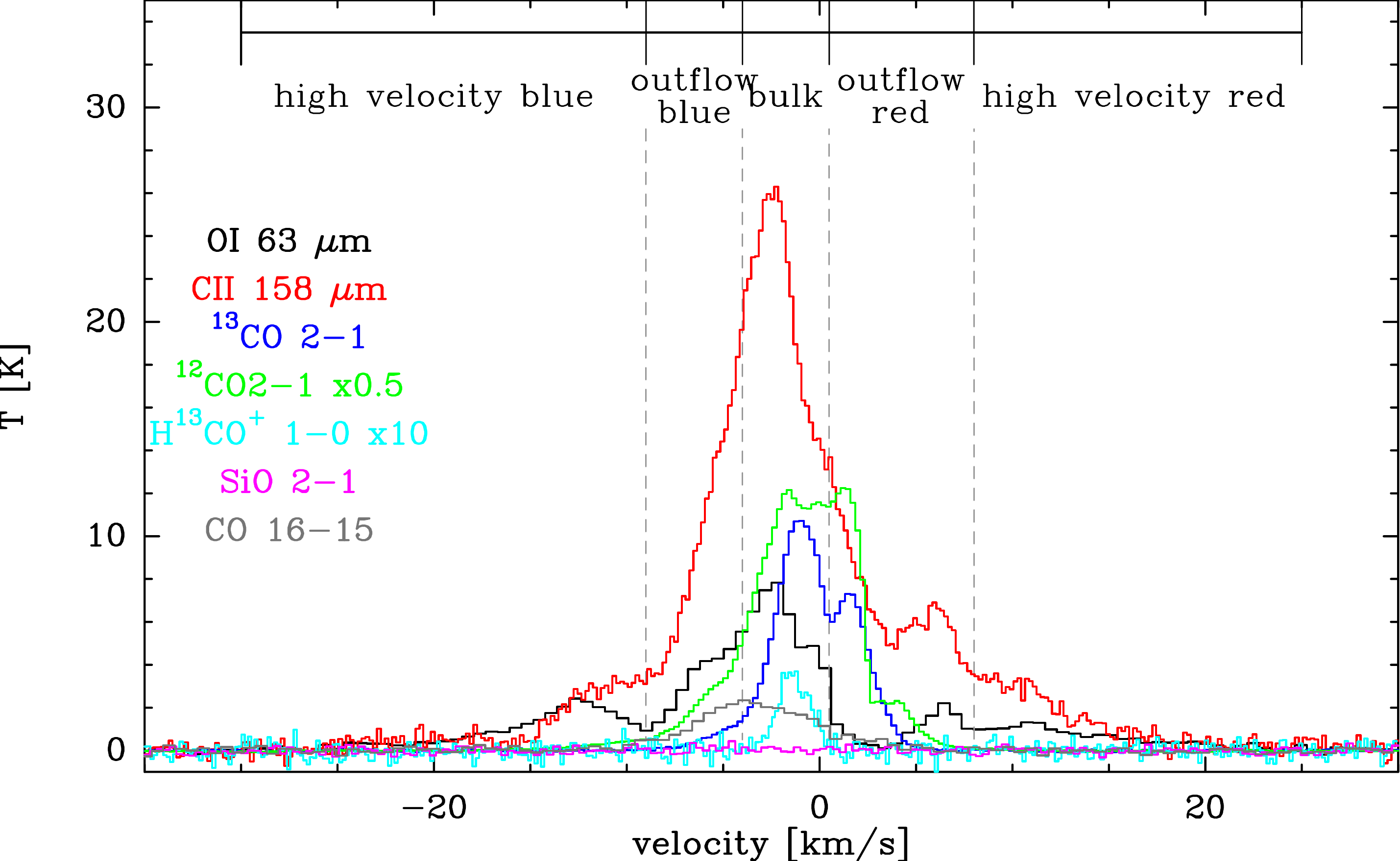}
\caption{Spatially averaged spectrum of molecular and atomic lines
  (the spectral resolution is typically 0.2--0.6 km s$^{-1}$). All
  spectra within the central area of 40$''\times$40$''$ around S106 IR
  were included. Based upon the different spectral features, we define
  five major velocity ranges. The local standard of rest (lsr)
  velocity of the S106 molecular cloud is approximately --3 km
  s$^{-1}$.}
\label{spectra}
\end{figure}

\subsection{\OI\ Spectral line map and average spectrum} \label{oi-line} 

Figure~\ref{spectra-map} shows the \OI\ spectral line map of the
region outlined in Fig.~\ref{overview} on a 6$''$ grid, representing
the original resolution of the data. The line profiles are very
complex, showing several velocity components and extended wing
emission, particularly in the blue range. Some of the spectral
features can be explained by self-absorption  (see below).
Figure~\ref{spectra} displays the average spectrum (across the
40$''\times$40$''$ area indicated in Fig.~\ref{overview}) of
\OI\ together with other tracers such as \CII\ and molecular lines.
Based upon the \OI\ line profile, the emission from other tracers, and
earlier studies of S106 \citep{Schneider2002, Schneider2003}, we
define and label five distinct velocity ranges to characterize the
velocity distribution in S106. We note  that the terms `high-velocity' and
`outflow' used here are  descriptive, comprising all gas dynamics
(molecular and atomic)  provoked by stellar feedback
(radiation, wind, shocks). In the following sections, we  single out
the possible origin of each of the different observed features.

\begin{itemize}
  \item {\bf High-velocity blue emission (HV-blue)} \newline
    v=--30 to --9 km s$^{-1}$, significant emission only observed
    in \OI\ and \CII, no emission in optically thin lines.
  \item {\bf Blue outflow emission} \newline
    v=--9 to --4 km s$^{-1}$, prominent emission in all
    optically thick lines in the form of extended wings.
  \item {\bf Bulk emission} \newline
    v= --4 to 0.5 km s$^{-1}$, commonly defined velocity range for the
    bulk emission of the associated molecular cloud, peak emission for
    all observed lines, substructure in line profiles for optically thick
    lines, strong self-absorption features in the \OI\ line.
  \item {\bf Red outflow emission} \newline
    v= 0.5 to 8 km s$^{-1}$, broad line wings for CO lines, individual components in
    \CII\ and \OI. At +3 km s$^{-1}$, there is a prominent dip in the
    \OI\ line and in all other line tracers that is due to
    absorption\footnote{As already discussed and shown in more detail
      in \citet{Schneider2003}, the dip seen in \OI, \CII,\, and
      $^{12}$CO 2$\to$1 spectra gets filled in with emission in the
      corresponding $^{13}$CO 2$\to$1 spectra.  The ratio
      $^{12}$CO/$^{13}$CO  drops to values lower than 1 for this
      velocity range over a large area on the western side of S106 IR
      (see Fig.~8 in Schneider et al. 2003).  The lack of $^{12}$CO
      (and \OI\ and \CII) emission around +3 km s$^{-1}$ is thus due
      to absorption of these lines becoming optically thick in cold
      foreground material. This cold gas originates from a molecular
      clump, associated with the extended molecular cloud.}.
  \item {\bf High-velocity red emission (HV-red)} \newline
      v=8 to 25 km s$^{-1}$, significant emission only observed in \CII\ and \OI.
\end{itemize} 

\begin{figure}
\centering
\includegraphics[width=8cm, angle=0]{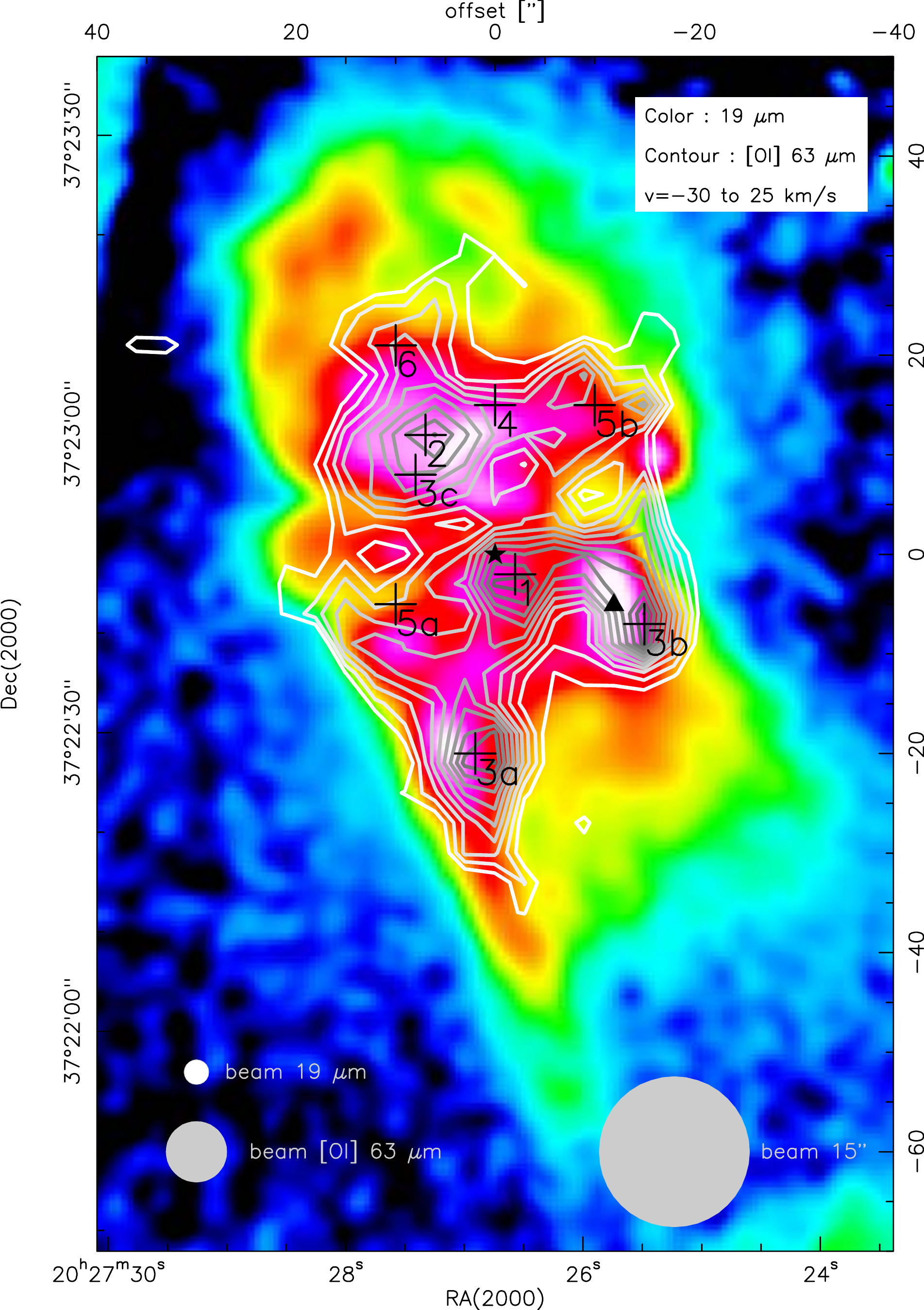}
\caption{\OI\ emission of the total velocity range  from  --30 to 25 km
  s$^{-1}$ in contours overlaid on a map of 19 $\mu$m emission
  (FORCAST, \citealt{Adams2015}). Positions for which we show
  individual spectra of various lines and report the line integrated
  intensities in Table~2 (in a 15$''$ beam) are indicated with {\bf a
    cross} and labelled 1 to 6. The angular resolution of the 19
  $\mu$m emission and the \OI\ beam are indicated in the panel. For
  comparison, the 15$''$ resolution we used for PDR modelling is also
  given. Positions 1 to 6 do not always correspond to
  peaks in \OI\ emission, but also to peak emission in high- and low-J
  CO lines (see text for details).}
\label{OI-F19}
\end{figure}

\begin{figure*}
\centering
\includegraphics[width=7.5cm, angle=0]{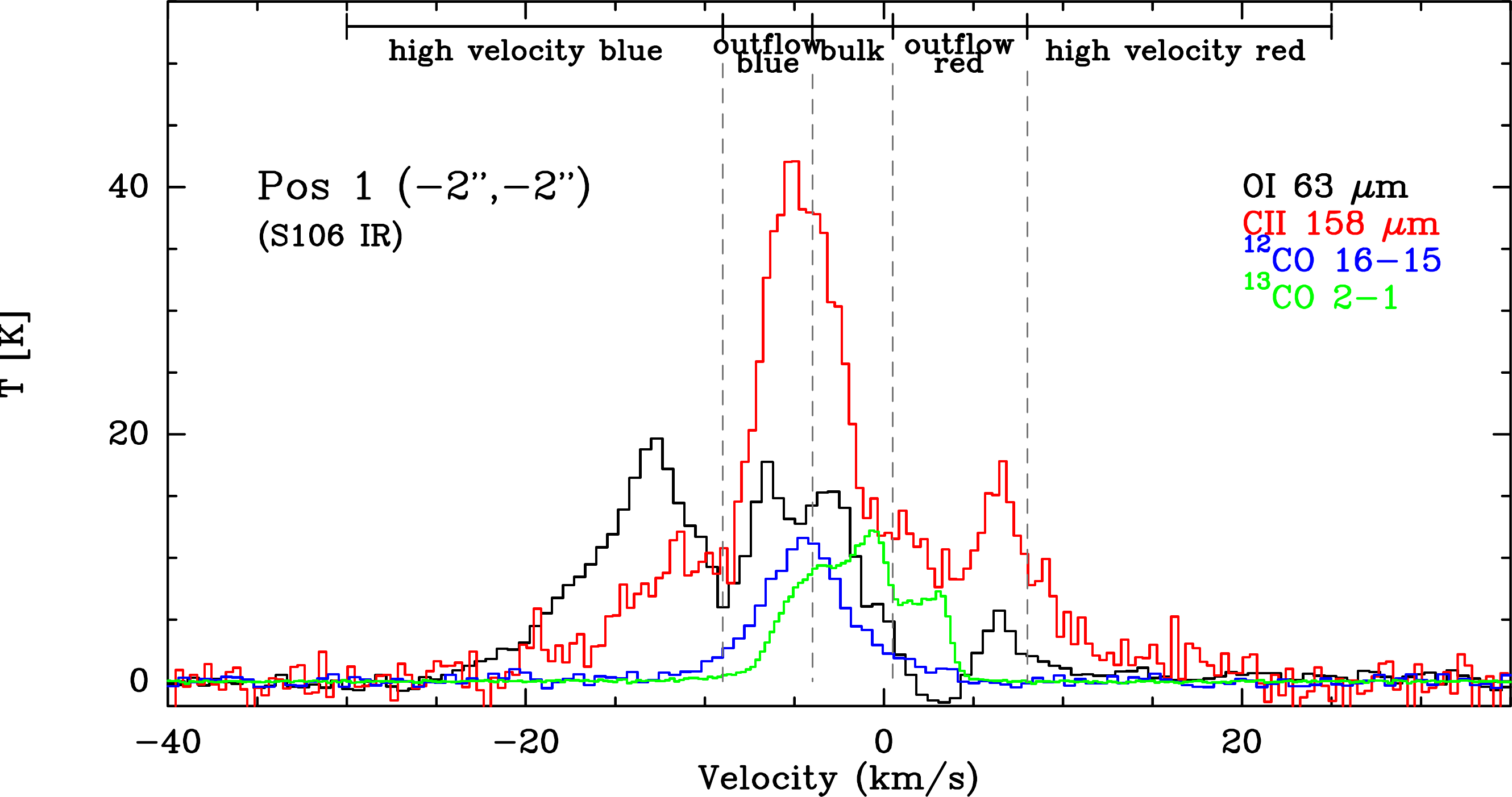}
\includegraphics[width=7.5cm, angle=0]{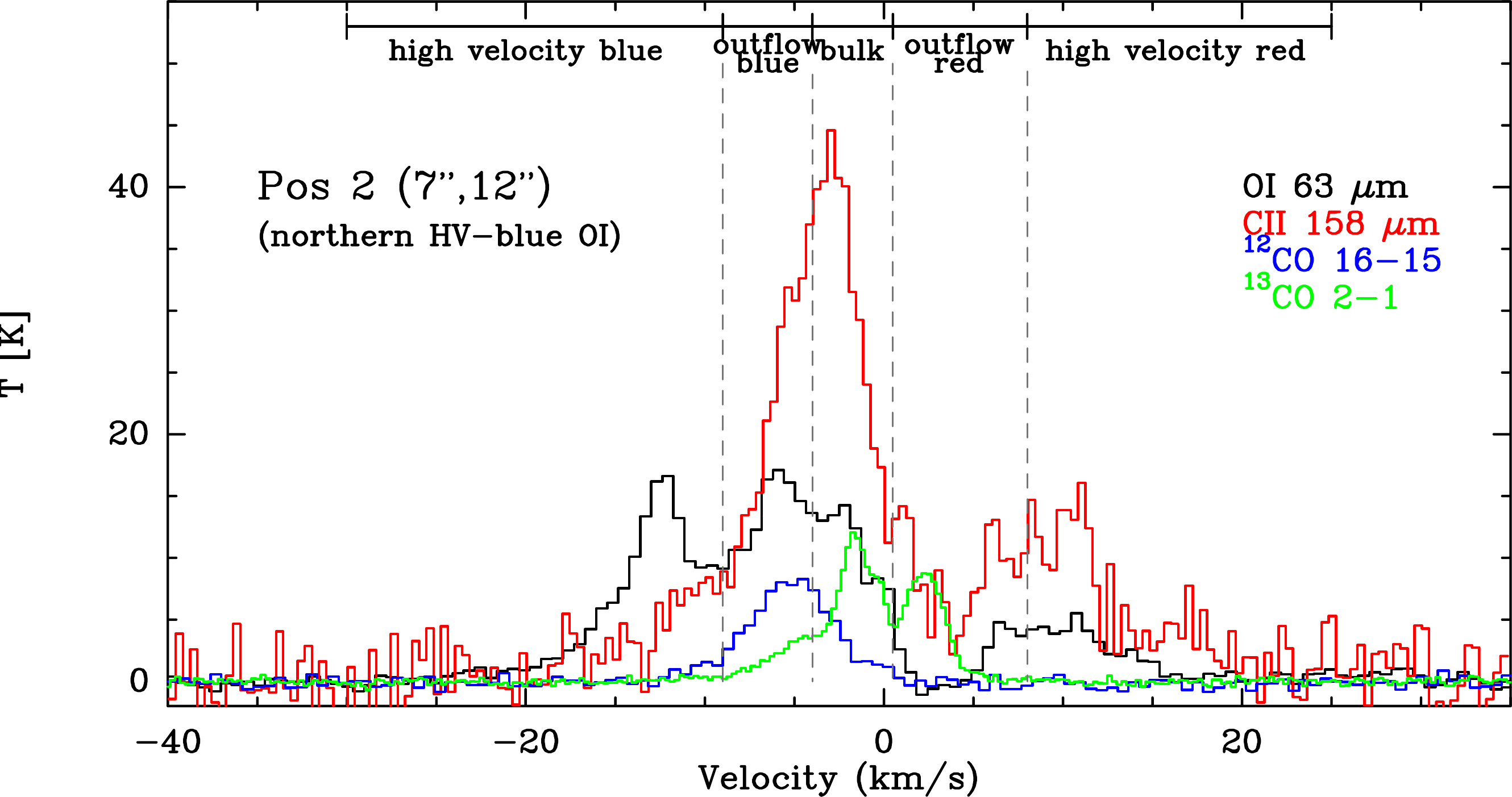}
\includegraphics[width=7.5cm, angle=0]{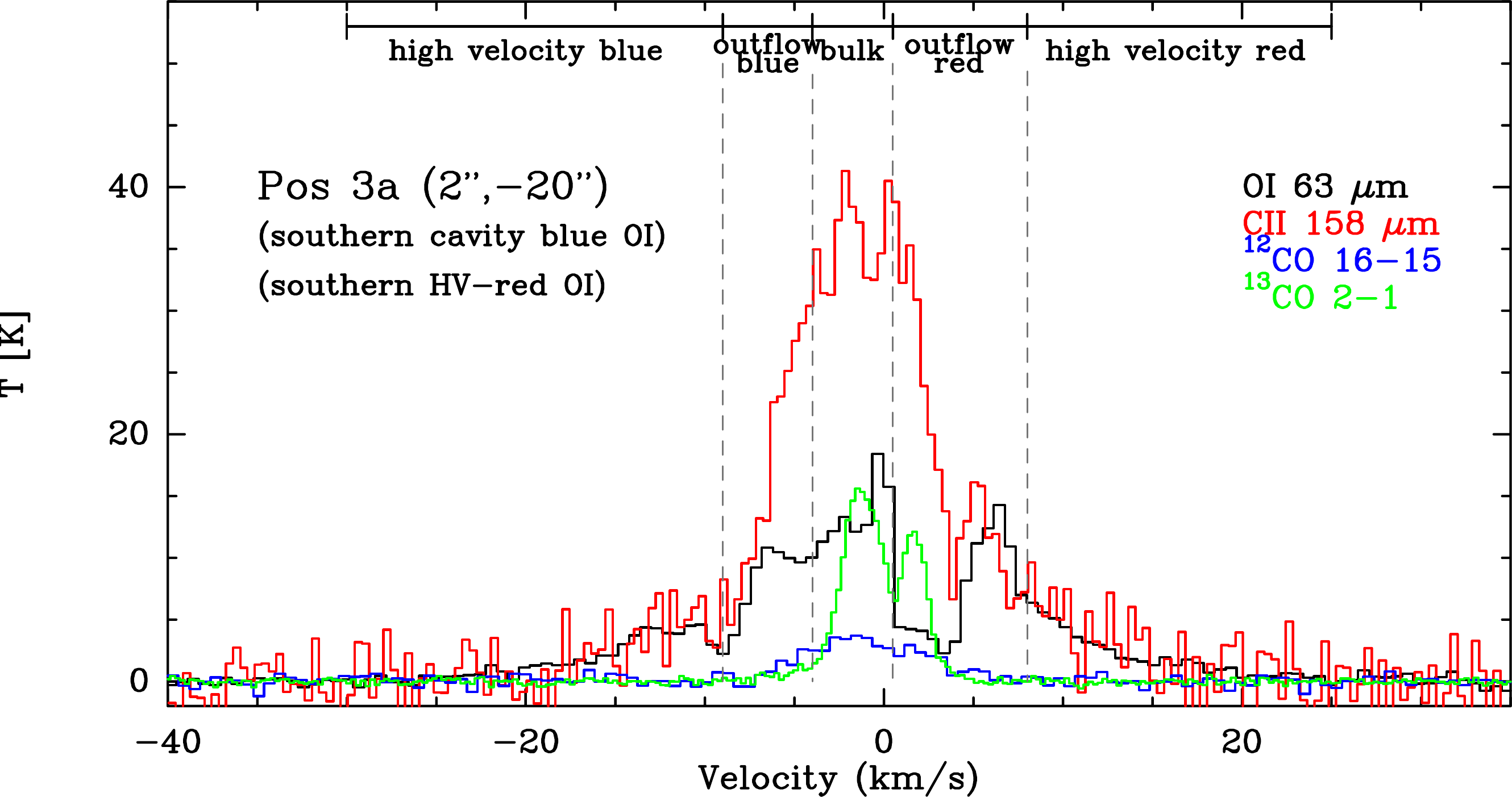}
\includegraphics[width=7.5cm, angle=0]{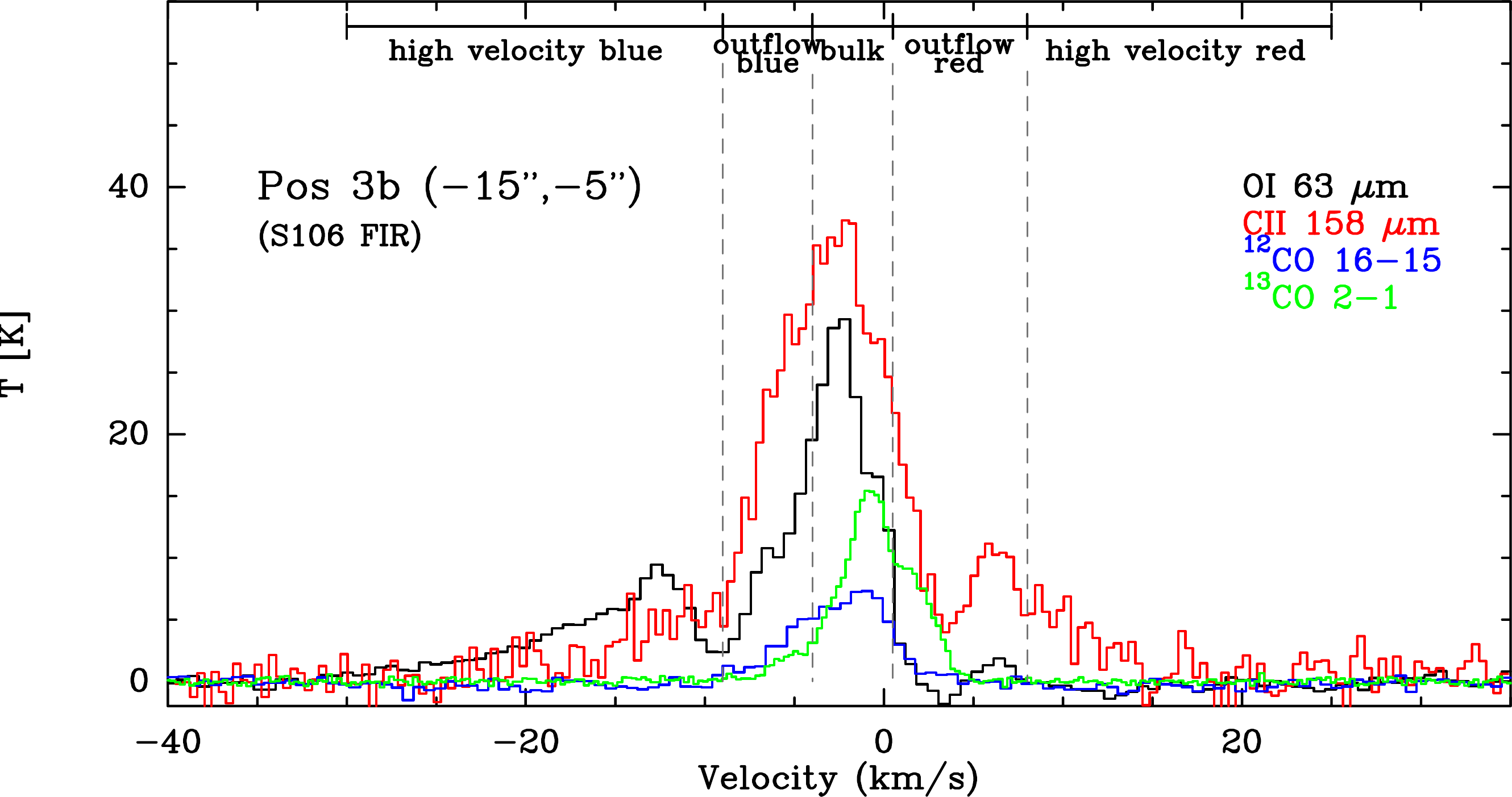}
\includegraphics[width=7.5cm, angle=0]{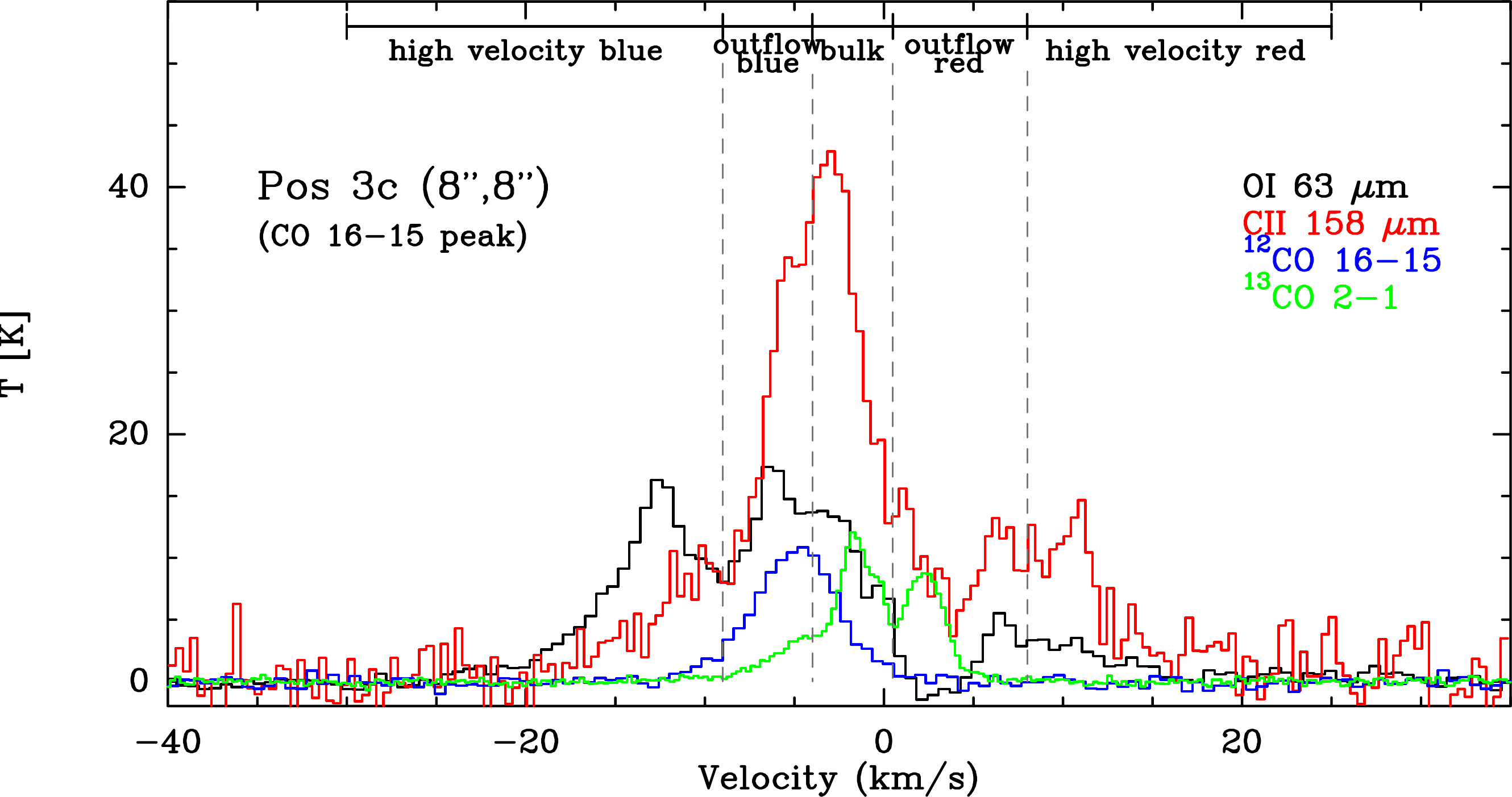}
\includegraphics[width=7.5cm, angle=0]{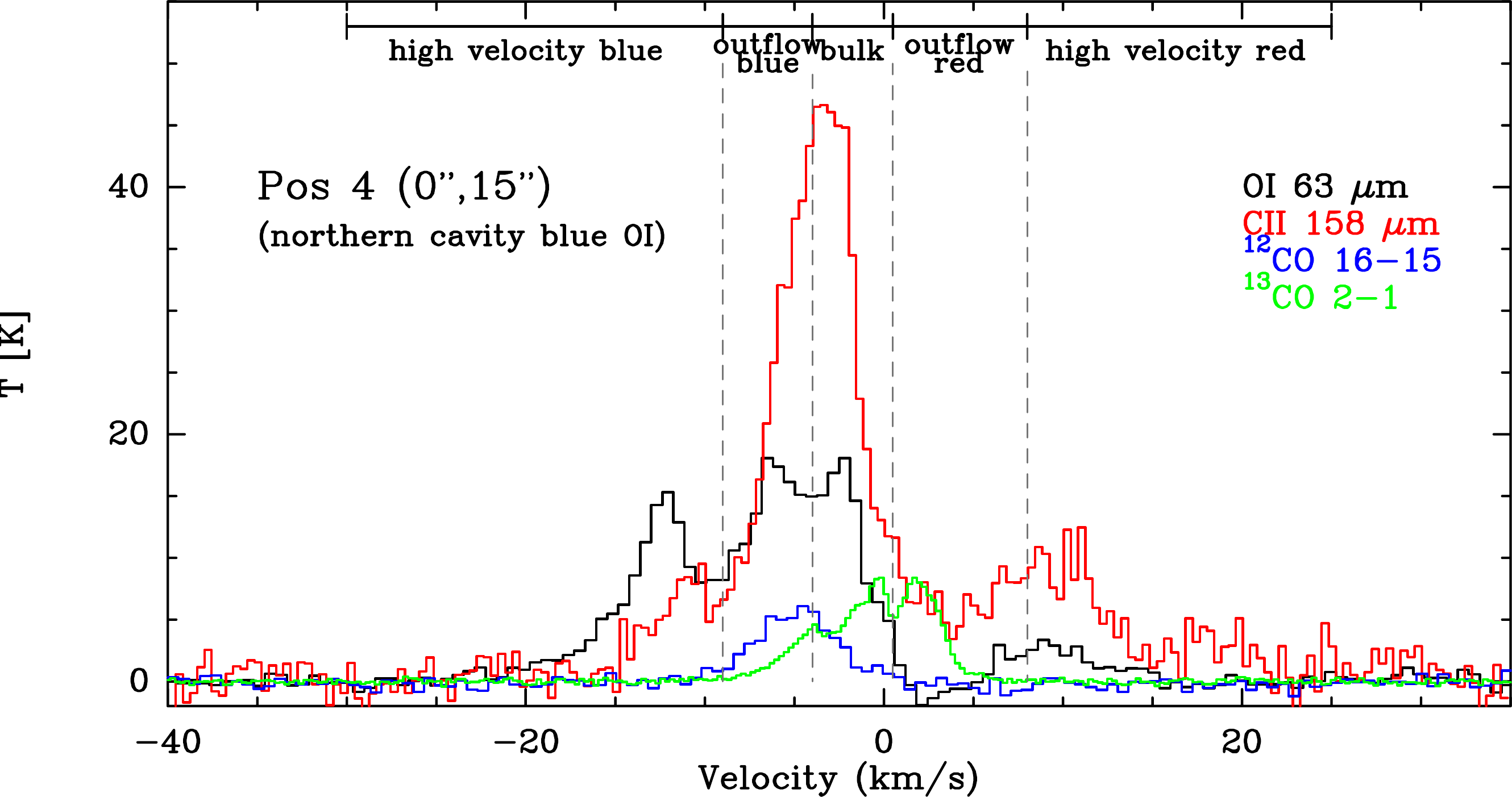}
\includegraphics[width=7.5cm, angle=0]{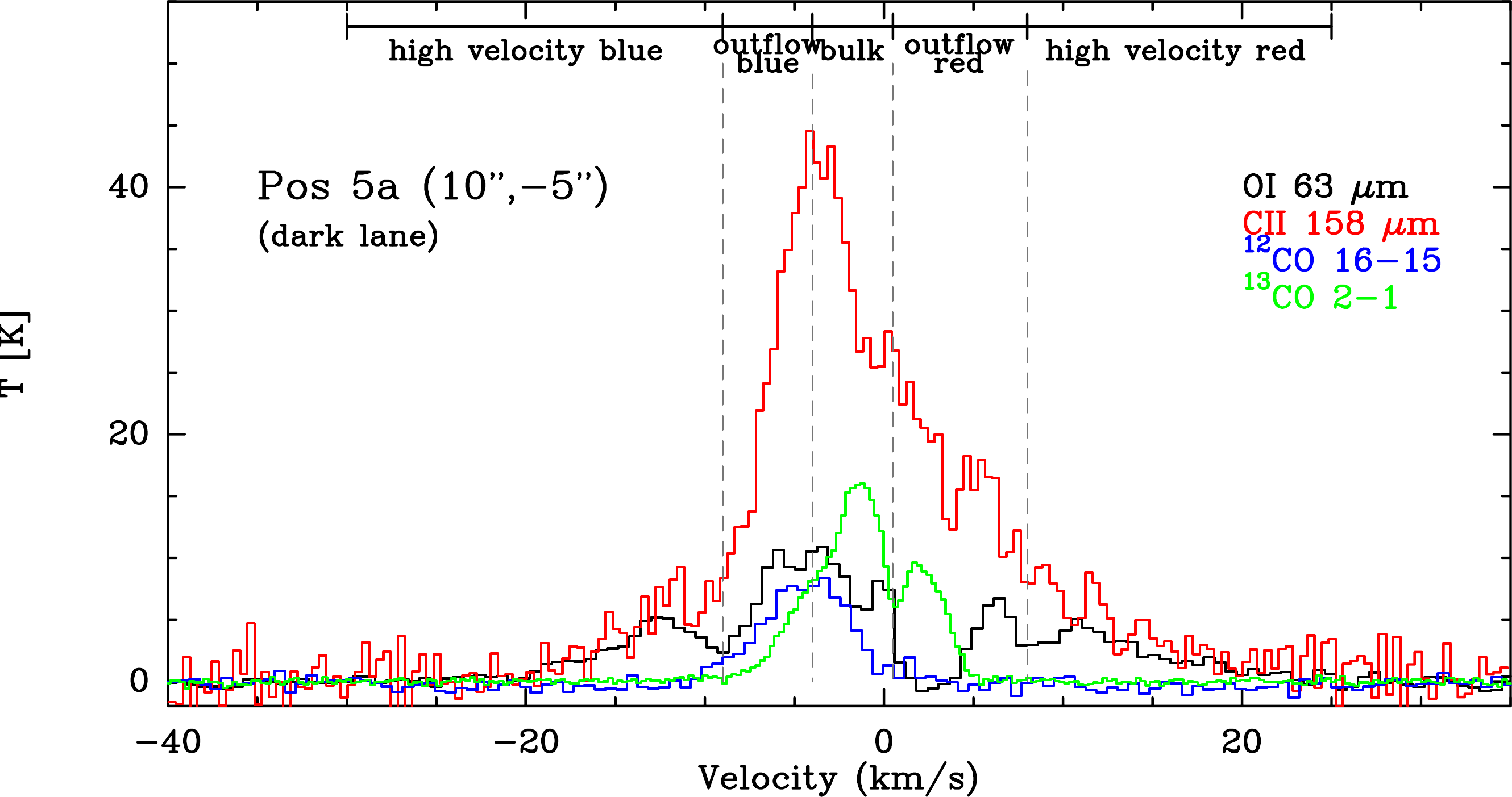}
\includegraphics[width=7.5cm, angle=0]{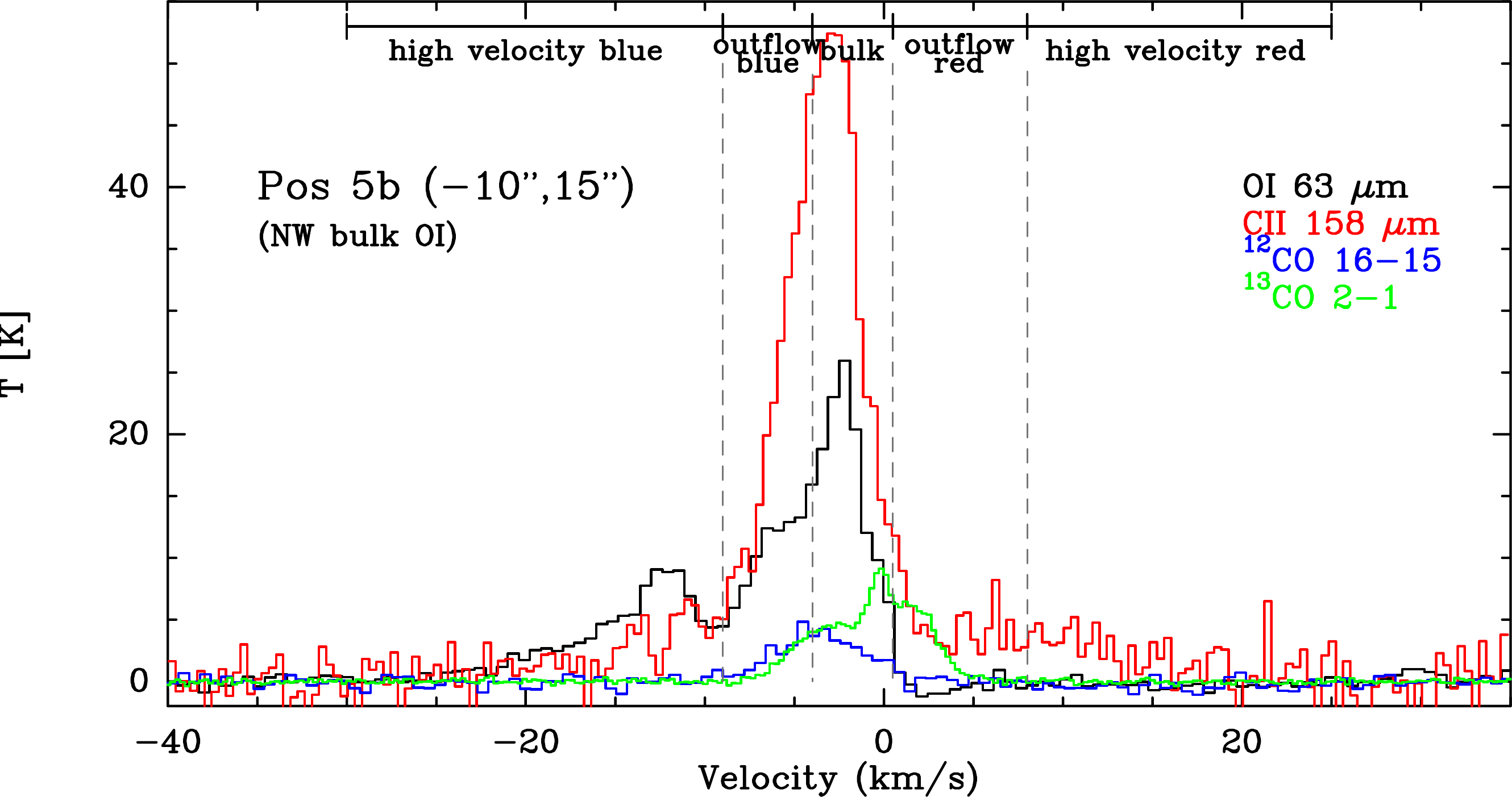}
\includegraphics[width=7.5cm, angle=0]{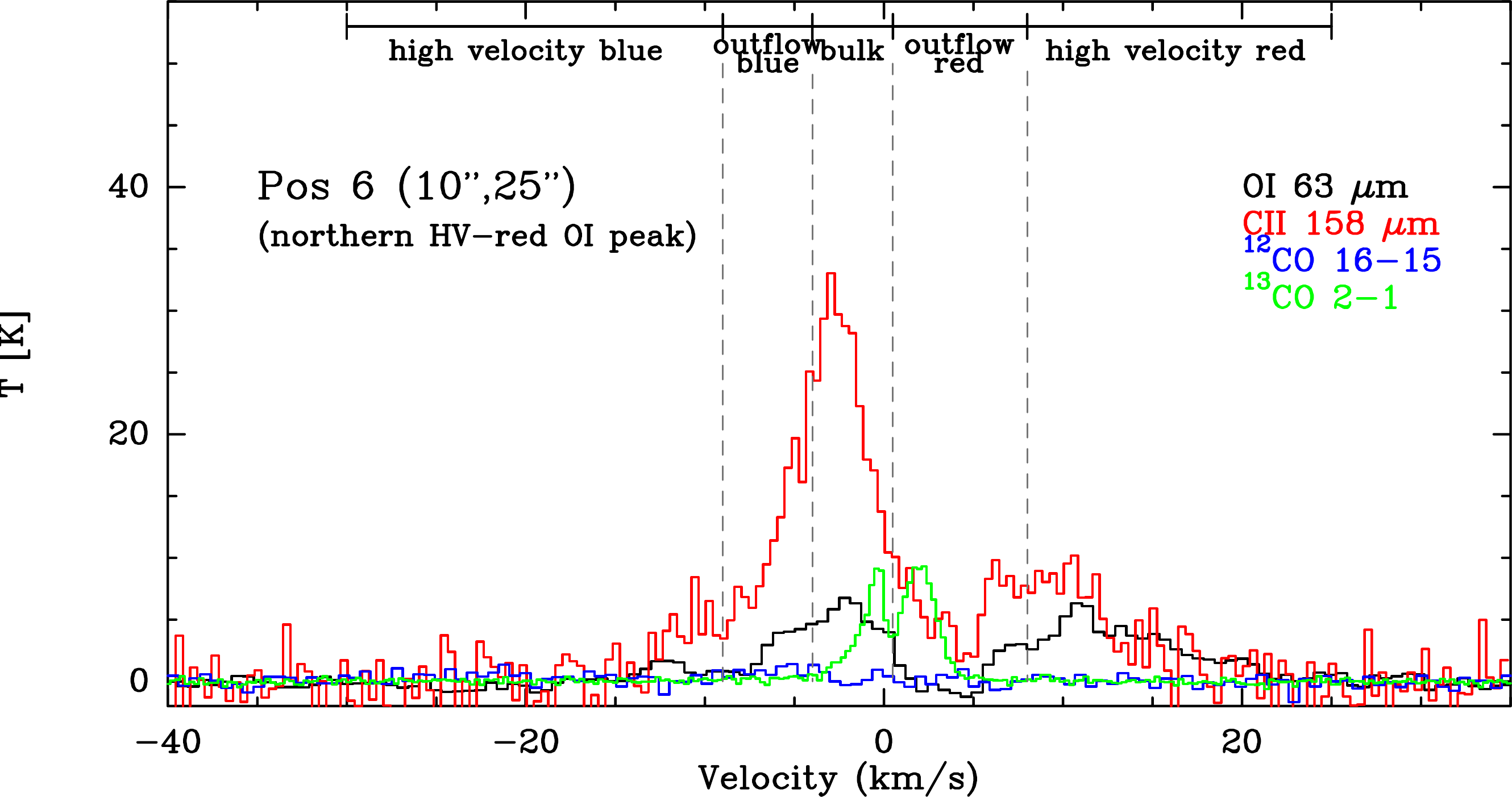}
\caption{Individual spectra with 15$''$ angular resolution at the positions indicated in Fig.~5.}
\label{spectra-single}
\end{figure*}

\begin{figure*}
\centering
\includegraphics[width=15cm, angle=0]{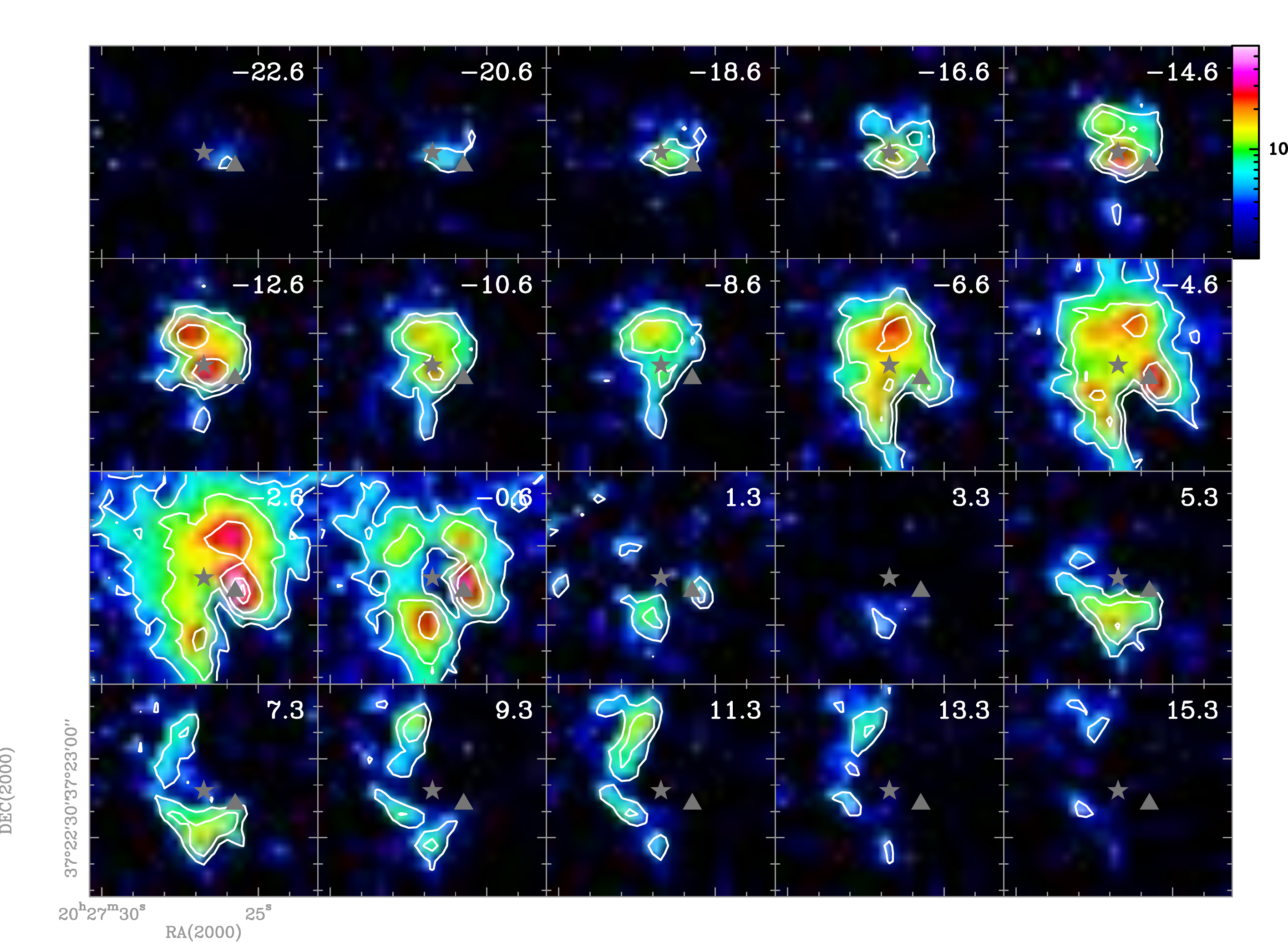}
\caption{Channel map of \OI\ emission in 2 km s$^{-1}$ velocity
  bins. The contour levels are 5, 10, 20, 45, and 60 K km
  s$^{-1}$. The logarithmic colour  scale ranges from 1.5 to 60 K km
  s$^{-1}$. The star indicates the binary system S106 IR and the
  triangle S106 FIR.}
\label{oi-chan}
\end{figure*}

\subsection{Overview of \OI, \CII, and CO emission in S106} \label{lines} 

Because the line profiles of \OI\ and other atomic and molecular line
tracers are very complex and show a strong positional variation, we
show the spectra for the positions where we performed PDR modelling in
Fig.~\ref{spectra-single}, an \OI\ channel map (Fig.~\ref{oi-chan}),
and overlays of different tracers in the main velocity ranges
(Appendix B). The spectra are extracted at the peak of emission in
different line tracers in the respective velocity range and reflect
different physical environments. They are indicated in
Fig.~\ref{OI-F19} where we plot the overall line integrated
\OI\ emission as contours on a dust map at 19 $\mu$m
\citep{Adams2015}. This overlay indicates that the total \OI\ emission
correlates very well with the warm dust, while the most prominent
peaks of \OI\ emission (positions 2, 3a, 3b) are slightly shifted with
regard to the peaks of 19 $\mu$m emission.  The line integrated
intensities and the total FIR continuum (used for PDR modelling) are
listed in Table~\ref{tab1}. We note that positions 3a and 3b appear more
than once  in the table because there are peaks of emission in
\OI\ at different velocities. \\

\begin{table*} 
  \caption{Velocity integrated intensities in a 15$''$ beam (20$''$
    for CO 11$\to$10) at the locations indicated in Fig.~4. Each
    position reflects a peak of emission in a certain velocity range
    and one or more  tracers and/or a source position: {\bf Pos 1}
    is the southern HV-blue \OI\ peak, closest to the central binary
    system S106 IR; {\bf Pos 2} is the northern HV-blue \OI\ peak;
    {\bf Pos 3a} is the southern cavity blue outflow \OI\ peak and the
    southern HV-red \OI\ peak; {\bf Pos 3b} corresponds to the YSO
    S106 FIR, associated with H$_2$O maser emission; {\bf Pos 3c} is
    the CO 16$\to$15 emission peak; {\bf Pos 4} is the northern cavity
    blue outflow \OI\ peak; {\bf Pos 5a} is the bulk CO 2$\to$1 peak
    located in the dark lane; {\bf Pos 5b} is the NW bulk \OI\ peak;
    {\bf Pos 6} is the northern HV-red \OI\ peak. The first column for
    each line gives the intensity in the respective velocity range,
    the second column the total line integrated (--30 to + 25 km
    s$^{-1}$) intensity, and the third column the percentage of how
    much of the individual velocity integrated line contributes to the
    total line flux. The last column gives the total FIR flux,
    determined as explained in Sect.~3.2.  The values for the red
    outflow velocity range (v=0.5 to 8 km s$^{-1}$) should be treated
    with care because the \OI\ and \CII\ lines suffer from absorption
    and the CO lines are weak or show no emission. The overall
    uncertainty of the values given in this table is $\sim$20\%.}
\label{tab1}
\centering  
\begin{tabular}{lc|lll|lll|lll|lll|l}     
\hline\hline
         &             &                       &                      &     &        &               &    &                    &           & & & & & \\     
Pos.     & Offset      & \OI\tablefootmark{a}  & \OI\tablefootmark{a} & \%  & \CII\tablefootmark{b}  & \CII\tablefootmark{b}   & \% & $^{12}$CO\tablefootmark{b} &
$^{12}$CO\tablefootmark{b} &   \%           &  $^{12}$CO\tablefootmark{b} & $^{12}$CO\tablefootmark{b} & \% & Total \\  
& ($''$,$''$) &                   &        &     &  &  &  & {\small 16$\to$15} & {\small 16$\to$15}  &  & {\small 11$\to$10} & {\small 11$\to$10} &
& FIR\tablefootmark{c}\\     
&  &  & \tiny{(total\tablefootmark{d})} &   &   & \tiny{(total)} &   &  & \tiny{(total)} &  &  &\tiny{(total)} &  & \\     
\hline  
\multicolumn{12}{l}{\rule[-3mm]{0mm}{8mm} {\bf High-velocity blue {\sl v=--30 to --9 km s$^{-1}$}}}\\
{\bf 1}  & (-2,-2)     & 13.4  &  28.9 & 46 &  45.0 & 308 & 15 &  3.6 & 45.7 &   8 &  4.7 & 36.3 & 13 & 24.1 \\
{\bf 2}  & (7,12)      &  8.0  &  24.2 & 33 &  30.0 & 314 & 10 &  1.0 & 25.8 &   4 &  3.9 & 30.0 & 13 & 23.4 \\
\hline
\multicolumn{12}{l}{\rule[-3mm]{0mm}{8mm} {\bf Outflow blue  {\sl v=--9 to --4 km s$^{-1}$}}}\\
{\bf 3a} & (2,-20)     &  4.3  &  21.5 & 20 &  82.8 & 292 & 28 &  4.6 & 25.4 &  18 &  6.2 & 24.8 & 25 & 15.3 \\
{\bf 3b} & (-15,-5)    &  4.9  &  22.4 & 22 &  65.0 & 245 & 27 &  7.1 & 24.7 &  29 &  8.5 & 24.3 & 35 & 17.5 \\
{\bf 3c} & (8,8)       &  6.7  &  24.7 & 27 &  69.9 & 331 & 21 & 22.3 & 34.1 &  65 & 15.4 & 34.0 & 45 & 24.8 \\
{\bf 4}  & (0,15)      &  7.2  &  23.2 & 31 &  50.5 & 273 & 18 & 13.6 & 16.1 &  84 & 12.0 & 25.5 & 47 & 19.7 \\
\hline
\multicolumn{12}{l}{\rule[-3mm]{0mm}{8mm} {\bf Cloud bulk emission {\sl v=--4 to 0.5 km s$^{-1}$}}}\\
{\bf 3b} & (-15,-5)    &  9.6  &  22.4 & 43 &  96.4 & 245 & 39 &  4.6 & 24.6 &  19 & 10.9 & 24.3 & 45 & 17.5 \\
{\bf 5a} & (10,-5)     &  4.8  &  22.1 & 21 &  98.9 & 337 & 29 &  4.6 & 43.1 &  11 & 14.8 & 37.7 & 39 & 20.4 \\
{\bf 5b} & (-10,15)    &  7.4  &  19.0 & 39 &  57.4 & 205 & 28 &  -   &  -   &  -  &  8.1 & 20.3 & 40 & 12.4 \\
\hline
\multicolumn{12}{l}{\rule[-3mm]{0mm}{8mm} {\bf Outflow red {\sl v=0.5 to 8 km s$^{-1}$}}}\\
{\bf 3a}  & (2,-20)    &  4.7  &  21.5 & 22 &  38.2 & 292 & 13 &  6.9 & 25.7 &  27  & 4.0 & 24.5 & 16 & 15.3\\ 
\hline
\multicolumn{12}{l}{\rule[-3mm]{0mm}{8mm} {\bf High-velocity red {\sl v=8 to 25 km s$^{-1}$}}}\\
{\bf 6}   & (10,25)    &  5.4  &  14.7 & 37 &  16.8 & 170 & 10 &  2.1 & 19.5 &  11 & 0.6 &  1.50 & 4 &  9.8\\
{\bf 3a}  & (2,-20)    &  3.2  &  21.5 & 15 &  46.1 & 292 & 16 &  0.5 & 25.7 &  2  & 0.3 &  24.5 & 1 & 15.3\\ 
\hline                                
\end{tabular}
\tablefoot{
\tablefoottext{a}{In units of [10$^{-3}$ erg s$^{-1}$ sr$^{-1}$ cm$^{-2}$]. Conversion I(OI) [erg s$^{-1}$ sr$^{-1}$ cm$^{-2}$] = 1.0943 10$^{-4}$ I(OI) [K km s$^{-1}$].}
\tablefoottext{b}{In units of [10$^{-5}$ erg s$^{-1}$ sr$^{-1}$ cm$^{-2}$]. Conversion I(CII) [erg s$^{-1}$ sr$^{-1}$ cm$^{-2}$] = 7.0354 10$^{-6}$ I(CII) [K km s$^{-1}$].
  Conversion I(CO16-15) [erg s$^{-1}$ sr$^{-1}$ cm$^{-2}$] = 6.3953 10$^{-6}$ I(CO16-15) [K km s$^{-1}$]. Conversion I(CO11-10) [erg s$^{-1}$ sr$^{-1}$ cm$^{-2}$] =
    2.0835 10$^{-6}$ I(CO11-10) [K km s$^{-1}$].}
\tablefoottext{c}{In units of [erg s$^{-1}$ sr$^{-2}$ cm$^{-2}$].}
\tablefoottext{d}{The total velocity range is from --30 to +25 km s$^{-1}$.}
}
\end{table*}

\noindent {\bf High-velocity blue emission} \\
\noindent
The most prominent feature of the \OI\ spectra is the {\sl
  high-velocity blue (HV-blue)} emission that appears as a single
component with an extended blue wing at temperatures up to 20 K at
positions {\bf 1, 2, and 3c} in Fig.~\ref{spectra-single}. The dip in
emission at --9 km s$^{-1}$ is most likely not caused by
self-absorption because the \CII\ line has a very similar emission
profile (though it cannot be excluded that both lines are affected by
self-absorption). The channel map (Fig.~\ref{oi-chan}) clearly shows
how the HV-blue emission starts south-west of S106 IR at v=--20.6 km
s$^{-1}$ and then gradually develops a north-eastern component. Both
peaks together are best visible at v=--12.6 km s$^{-1}$. \\

\noindent {\bf Blue outflow emission} \\
\noindent
The channel map shows how the northern and southern cavity lobes are
outlined in \OI\ emission in the {\sl blue outflow} velocity range
(panels   --8.6 km s$^{-1}$ to --4.6 km s$^{-1}$).  Position {\bf Pos
  3c} represents the peak of blue outflow emission in the CO 16$\to$15
line, and {\bf Pos 4} is the peak position of the blue outflow in
\OI\ emission in the northern cavity. \\

\noindent {\bf Bulk emission} \\
\noindent
Generally, the {\sl bulk emission} velocity range shows a very clumpy
distribution around S106 IR (panels --2.6 and --0.6 km s$^{-1}$). The
`hole' in the emission around S106 IR is partly caused by
self-absorption, but also reflects a real lack of gas since all
molecular line tracers show no emission (see Sec.~\ref{lane}). The
\OI\ spectra (Fig.~\ref{spectra-single}) are very complex for this
velocity range, showing mostly several components, which can be due to
several PDR layers on clump surfaces along the line of sight and/or
intrinsic self-absorption effects. Future observations of the more
optically thin \OI\ 145 $\mu$m line may help to decide between the
different possibilities. The \OI\ 145 $\mu$m/63 $\mu$m ratio
determined from velocity unresolved data \citep{Schneider2003,
  Stock2015} yield a value of 0.15--0.17, indicating that one or both
of the lines is optically thick.  We note that the molecular cloud
velocity is --1 km s$^{-1}$, indicated by the $^{13}$CO 2$\to$1 line
in (Fig.~\ref{spectra-single}), while the \OI\ line has a centre
velocity of typically --4 km s$^{-1}$.  A very prominent peak of
emission is found at position {\bf Pos 3b} which arises from a single
clump SW of S106 IR (position S106 FIR, location of the H$_2$O maser),
clearly visible in panel --2.6 km s$^{-1}$ in the channel map.
Position {\bf 5a} represents the `dark lane', which is strong in
$^{13}$CO emission and weaker in the atomic and high-J CO lines, and
{\bf 5b} is the NW peak of \OI\ emission.  \\

\noindent {\bf Red outflow } \\
\noindent
The velocity range from 0.5 to 8 km s$^{-1}$ is strongly affected by
absorption, basically all \OI\ spectra show a dip between 1 and 3 km
s$^{-1}$ (see footnote on page 6). Significant emission is only found
around v=8 km s$^{-1}$ in the southern cavity at {\bf Pos 3a}. \\

\noindent {\bf High-velocity red emission} \\
\noindent
The \OI\ emission at velocities higher than v=8 km s$^{-1}$ is very
different from the HV-blue emission; it is more diffuse and extends
further away from S106 IR, mostly outlining the eastern cavity
walls. Position {\bf  6} represents the \OI\ peak in the northern
cavity.

\begin{figure*}
\centering
\includegraphics[width=13cm, angle=0]{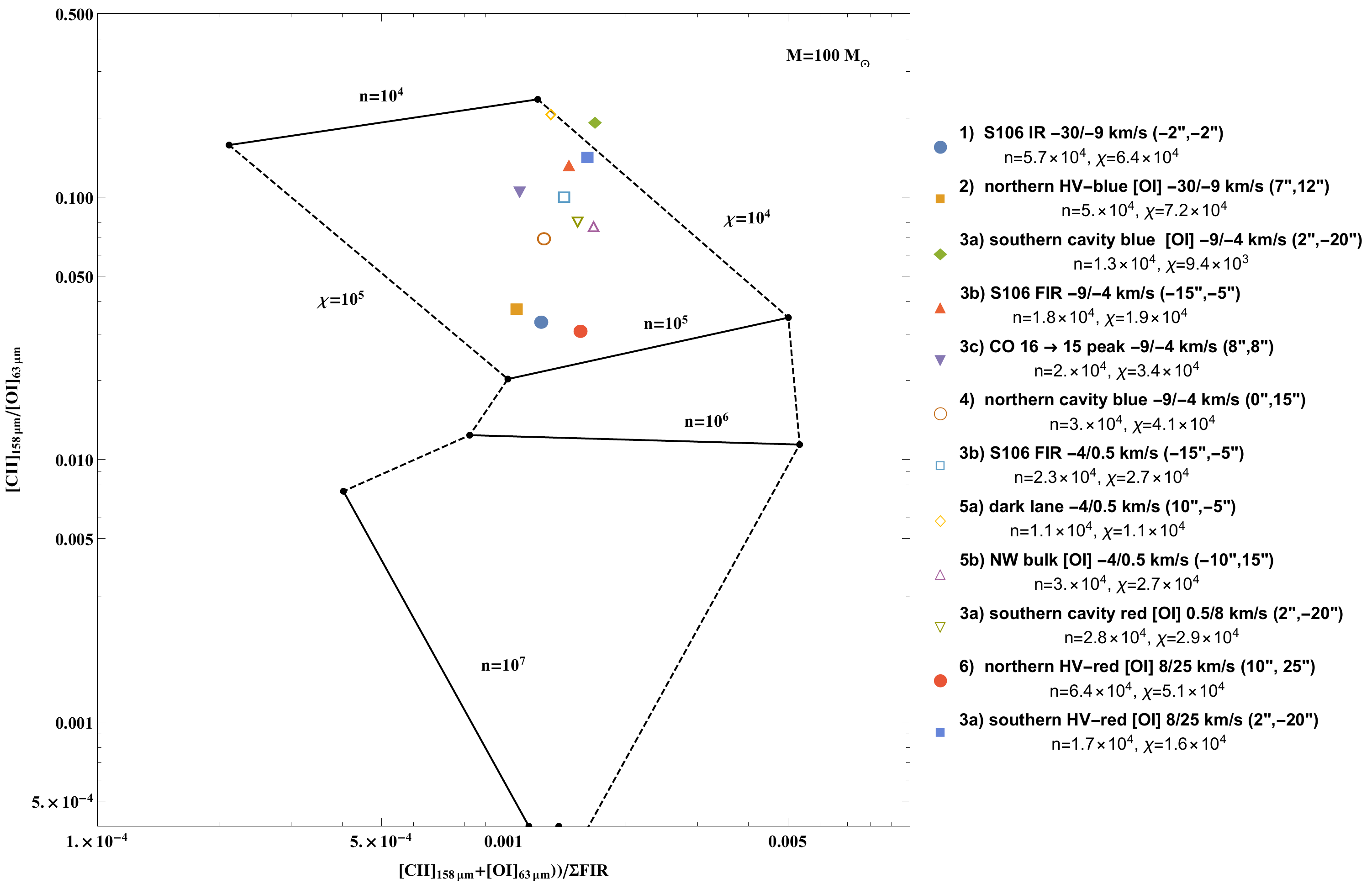}
  \caption{PDR diagnostic diagrams for observed fine-structure lines
    based on the KOSMA-$\tau$ PDR model for a clump of
    $M=100$~M$_\odot$. $[CII]_{158\mu m}/[OI]_{63\mu m}$ against
    $([CII]_{158\mu m}+[OI]_{63\mu m})/\Sigma_{FIR}$, where
    $\Sigma_{FIR}$ is the continuum intensity between 10 and 1000~$\mu
    m$ in the respective velocity range determined as described in
    Sect.~3.2. The various markers indicate the different positions
    corresponding to different velocity ranges.  The density $n$ and
    the Draine field $\chi$ are given for each position.}
  \label{Fig:ratio1}
\end{figure*}

\begin{figure*}
\centering
\includegraphics[width=13cm, angle=0]{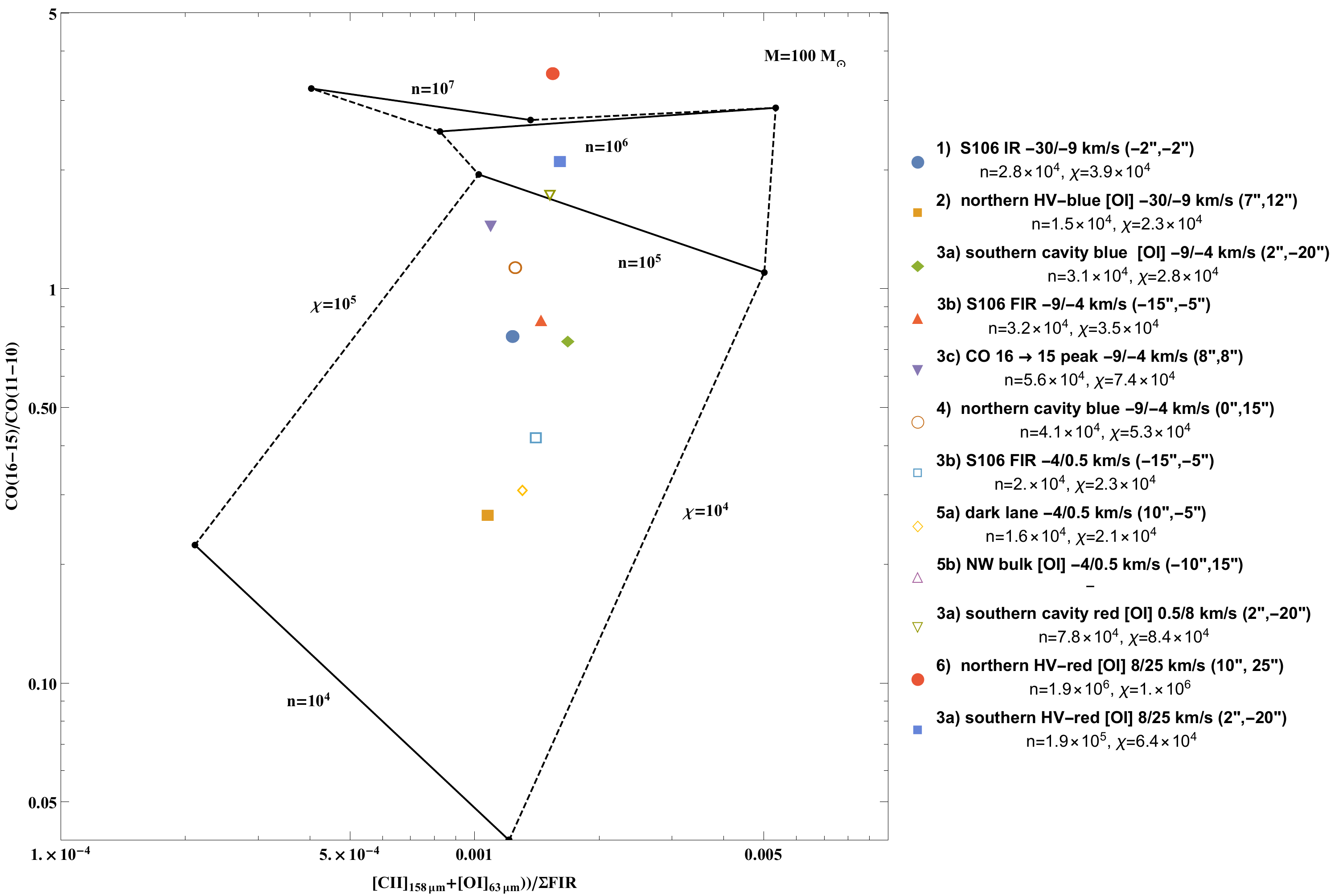}
  \caption{PDR diagnostic diagrams for observed fine-structure lines
    based on the KOSMA-$\tau$ PDR model for a clump of
    $M=100$~M$_\odot$. CO(16-15)/CO(11-10) against $([CII]_{158\mu
      m}+[OI]_{63\mu m})/\Sigma_{FIR}$, where $\Sigma_{FIR}$ is the
    continuum intensity between 10 and 1000~$\mu m$ in the respective
    velocity range determined as described in Sec.~3.2. The various
    markers indicate the different positions corresponding to
    different velocity ranges. The density $n$ and the Draine field
    $\chi$ are given for each position.}
  \label{Fig:ratio2}
\end{figure*}

\subsection{Shocks in S106} \label{shocks}

We did not detect the SiO (2$\to$1) or (5$\to$4) lines, classical
tracers of stationary C-type (e.g. \citealt{Schilke97},
\citealt{Gusdorf081}), J-type (e.g. \citealt{Guillet09}), or
non-stationary CJ-type (e.g. \citealt{Gusdorf082}) shocks. However, the
presence of shocks that can give rise to the \OI\ emission observed at
high blue and red velocities and at outflow velocities cannot be
excluded. These non-detections could indeed mean that the
shocks are either self-irradiated (see the study of shocks with
radiative precursors by \citealt{Hollenbach1989}) or externally
irradiated (e.g. \citealt{Lesaffre2013}). In the latter case, the
irradiation could come from the FUV field of the central
protostars. In shocks where irradiation plays a significant role, SiO
is expected to be converted to Si and Si$^+$. These species can be
observed through three transitions: the $^3$P$_1$--$^3$P$_0$ at
129.68~$\mu$m and the $^3$P$_2$--$^3$P$_1$ at 68.47~$\mu$m for the Si
lines, and the $^2$P$_{3/2}$--$^2$P$_{1/2}$ at 34.82~$\mu$m. The
ground-state Si transition at 129.68~$\mu$m has been marginally
detected by {\sl Herschel}-PACS, contrary to its excited counterpart
at 68.47~$\mu$m. None of the Si lines has been detected by ISO in
contrast to the  Si$^+$ transition, which was observed. The presence
of atomic and ionized silicon in the gas phase, combined with the
detection of optical and near-infrared lines reported by
\citet{Riera89}, hints at the presence of irradiated shocks in the S106
region. Though we briefly discuss possible shocks in
Sect.~\ref{hv-blue-red}, a more sophisticated attempt to characterize
their type and the role they play in the excitation of \OI\ and
\CII\, lies out of the scope of the present study, and will be
investigated in a separate publication.

\begin{table*}[htb]
\begin{center}
  \caption{Data points and model results from Figure~\ref{Fig:ratio1}
    for $\mathrm{[CII]}_{158\mu m}/\mathrm{[OI]}_{63\mu m}$ against
    $(\mathrm{[CII]}_{158\mu m}+\mathrm{[OI]}_{63\mu
      m})/\Sigma_\mathrm{FIR_{range}}$. The abbreviation `range'
    signifies that we scale the total FIR flux with the fractional
    \OI\ intensity in this velocity range (see text for further
    details). The values for density n and radiation field $\chi$
    (Cols. 7 and 8) are given for the nominal observed \OI\ intensity
    and for an \OI\ intensity increased by a factor of two (in
    parentheses).} \label{tab:ratio1}
\begin{tabular}{lclccccc}
\hline\hline \rule[-3mm]{0mm}{8mm}
Position & velocity range &               & Offset &[CII]/[OI]&$\frac{\mathrm{[CII]}+\mathrm{[OI]}}{\Sigma_\mathrm{FIR_{range}}}$&$\log n$&$\log \chi_\mathrm{Draine}$\\
         & [km s$^{-1}$] & &$({''},{''})$&$(\times 10^{-2})$&$(\times 10^{-3})$& [cm$^{-3}$]&\\ \hline
\multicolumn{8}{c}{\rule[-3mm]{0mm}{8mm} {\bf high-velocity blue}}\\ \hline 
{\bf 1}  & -30/-9       & S106 IR                   & (-2,-2)  &  3.4  & 1.24 & 4.8 (6.4) & 4.8 (4.4)\\
{\bf 2}  & -30/-9       & northern HV-blue \OI\     & (7,12)   &  3.8  & 1.08 & 4.7 (5.0) & 4.9 (4.6) \\ \hline
\multicolumn{8}{c}{\rule[-3mm]{0mm}{8mm} {\bf outflow blue}}\\ \hline
{\bf 3a} & -9/-4        & southern cavity blue \OI\ & (2,-20)  & 19.4  & 1.68 & 4.1 (4.5) & 4.0 (4.0) \\
{\bf 3b} & -9/-4        & S106 FIR                  & (-15,-5) & 13.3  & 1.45 & 4.3 (4.6) & 4.3 (4.2) \\
{\bf 3c} & -9/-4        & CO 16$\to$15 peak         & (8,8)    & 10.5  & 1.10 & 4.3 (4.6) & 4.5 (4.4) \\
{\bf 4}  &  -9/-4       & northern cavity blue \OI\ & (0,15)   &  7.0  & 1.26 & 4.5 (4.8) & 4.6 (4.4) \\ \hline
 \multicolumn{8}{c}{\rule[-3mm]{0mm}{8mm} {\bf cloud bulk emission}}\\ \hline
{\bf 3b} & -4/0.5       & S106 FIR                  & (-15,-5) & 10.1  & 1.41 & 4.4 (4.7) & 4.4 (4.3) \\
{\bf 5a}  & -4/0.5      & dark lane                 & (10,-5)  & 20.8  & 1.31 & 4.1 (4.4) & 4.1 (4.1) \\ 
{\bf 5b}  & -4/0.5      & NW bulk \OI\              & (-10,15) &  7.8  & 1.67 & 4.5 (4.8) & 4.4 (4.2) \\ \hline
\multicolumn{8}{c}{\rule[-3mm]{0mm}{8mm} {\bf outflow red}}\\ \hline
{\bf 3a} & 0.5/8       & southern cavity red \OI\  & (2,-20)  &  8.1  & 1.52 & 4.4 (4.8) & 4.5 (4.3) \\ \hline
\multicolumn{8}{c}{\rule[-3mm]{0mm}{8mm} {\bf high-velocity red}}\\ \hline 
{\bf 6}  & 8/25         & northern HV-red \OI\      & (10, 25) &  3.1  & 1.54 & 4.8 (6.3) & 4.7 (4.4) \\
{\bf 3a} & 8/25         & southern HV-red \OI\      & (2,-20)  & 14.3  & 1.61 & 4.2 (4.6) & 4.2 (4.1) \\
\hline\\
\end{tabular}
\end{center}
\end{table*}

\begin{table*}[htb]
\begin{center}
  \caption{Data points and model results from Figure~\ref{Fig:ratio2}
    for CO(16-15)/CO(11-10) against $(\mathrm{[CII]}_{158\mu
      m}+\mathrm{[OI]}_{63\mu m})/\Sigma_\mathrm{FIR_{range}}$.  The
    values for density n and radiation field $\chi$ (Cols. 7 and 8)
    are given for the nominal observed \OI\ intensity and for an
    \OI\ intensity increased by a factor of two (in
    parentheses).}  \label{tab:ratio2}
\begin{tabular}{lclccccc}
\hline\hline \rule[-3mm]{0mm}{8mm}
Position & velocity range &                & Offset &$\frac{\mathrm{CO}(16-15)}{\mathrm{CO}(11-10)}$&$\frac{\mathrm{[CII]}+\mathrm{[OI]}}{\Sigma_\mathrm{FIR_{range}}}$&$\log n$&$\log \chi_\mathrm{Draine}$\\
         & [km s$^{-1}$] &  &$({''},{''})$& &$(\times 10^{-3})$&[cm$^{-3}$] &\\ \hline
\multicolumn{8}{c}{\rule[-3mm]{0mm}{8mm} {\bf high-velocity blue}}\\ \hline
{\bf 1}  & -30/-9       & S106 IR                   & (-2,-2)  & 0.76 & 1.24 & 4.4 (4.6) & 4.6 (4.3) \\
{\bf 2}  & -30/-9       & northern HV-blue \OI\     & (7,12)   & 0.27 & 1.08 & 4.2 (4.3) & 4.4 (4.1) \\ \hline
 \multicolumn{8}{c}{\rule[-3mm]{0mm}{8mm} {\bf outflow blue}}\\ \hline
{\bf 3a} &  -9/-4       & southern cavity blue \OI\ & (2,-20)  & 0.74 & 1.68 & 4.5 (4.6) & 4.4 (4.1) \\
{\bf 3b} &  -9/-4       & S106 FIR                  & (-15,-5) & 0.84 & 1.45 & 4.5 (4.7) & 4.5 (4.2) \\
{\bf 3c} &  -9/-4       & CO 16$\to$15 peak         & (8,8)    & 1.44 & 1.10 & 4.7 (4.9) & 4.9 (4.6) \\
{\bf 4}  &  -9/-4       & northern cavity blue \OI\ & (0,15)   & 1.13 & 1.26 & 4.6 (4.8) & 4.7 (4.4) \\ \hline
\multicolumn{8}{c}{\rule[-3mm]{0mm}{8mm} {\bf cloud bulk emission}}\\ \hline
{\bf 3b} & -4/0.5       & S106 FIR                  & (-15,-5) & 0.42 & 1.41 & 4.3 (4.4) & 4.4 (4.1) \\
{\bf 5a} &  -4/0.5      & dark lane                 & (10,-5)  & 0.31 & 1.31 & 4.2 (4.3) & 4.3 (4.0) \\
{\bf 5b} &  -4/0.5      & NW bulk \OI\              & (-10,15) &  -   &  -   &  -        &  -   \\
\multicolumn{8}{c}{\rule[-3mm]{0mm}{8mm} {\bf outflow red}}\\ \hline
{\bf 3a} & 0.5/8        & southern cavity red \OI\  & (2,-20)  & 1.74 & 1.52 & 4.9 (7.0) & 4.9 (3.2) \\ \hline
\multicolumn{8}{c}{\rule[-3mm]{0mm}{8mm} {\bf high-velocity red}}\\ \hline 
{\bf 6}  & 8/25         & northern HV-red \OI\      & (10, 25) & 3.52  & 1.55 & 6.3 (6.3) & 6.0 (6.0) \\
{\bf 3a} & 8/25         & southern HV-red \OI\      & (2,-20)  & 2.11  & 1.61 & 5.3 (5.4) & 4.8 (4.5) \\
\hline\\
\end{tabular}
\end{center}
\end{table*}

\section{Analysis and discussion} \label{discuss} 

\subsection{Emission from different velocity ranges} \label{ranges} 

In this section we qualitatively and quantitatively discuss the
emission of the observed \CII, \OI, and CO lines in the different
velocity intervals. Each velocity range reflects different properties
of the gas in terms of radiation field, temperature, and density, and
the emission can have a different origin.  In particular the \OI\ and
high-J CO lines can originate from PDRs, i.e.  UV-heated clump
surfaces and cavity walls, and/or from shocks due to disk--envelope
interactions (accretion shock) and local shocks when the stellar winds
hit the cavity walls. One way to determine the origin of the emission
is by studying its velocity structure (see Sects.  \ref{hv-blue-red}
and \ref{outflow-blue-red}).  The excitation energies for \CII\ and
\OI\ emission are 91 and 228 K, respectively. These transitions are
easily excited by collisions with H and H$_2$ in a PDR with different
temperature layers. The critical densities for \OI\ and \CII\ are then
5 10$^5$ cm$^{-3}$ and 3 10$^3$ cm$^{-3}$, respectively. The high-J CO
lines (16$\to$15 and 11$\to$10) have higher excitation energies,
$\sim$750 K and $\sim$365 K, and when the emission is purely thermal
these lines probe hot gas associated with the PDR/molecular cloud
layer at critical densities of 2 10$^6$ cm$^{-3}$ and 4 10$^5$
cm$^{-3}$, respectively.  Though we discuss a possible origin from
shocks in the following subsections, we use the observed FIR line
ratios and the FIR flux only for PDR modelling.

The observed line intensities in the different velocity ranges and the
total line integrated intensity and the fraction of intensity are
listed in Table~\ref{tab1}.  All results from PDR modelling for the
individual velocity ranges are summarized in Fig.~\ref{Fig:ratio1} and
Fig.~\ref{Fig:ratio2} where we show the KOSMA-$\tau$ results for the
${\rm [CII]}_{158\mu m}/{\rm [OI]}_{63\mu m}$ and CO(16-15)/CO(11-10)
line ratios as iso-contours.  Dashed lines show contours of constant
FUV fields, while black contours represent lines of constant
density. In Table~\ref{tab:ratio1} and \ref{tab:ratio2} we summarize
the observed ratios and give the model parameters corresponding to
their position in the diagnostic Figs.~\ref{Fig:ratio1} and
\ref{Fig:ratio2}.

Using the observed line ratios as diagnostics for PDR modelling
involves several caveats which need to be kept in mind. Firstly, we
use a non-clumpy PDR model because a more sophisticated set-up, as was
used when studying the Orion Bar \citep{Andree2017}, is beyond the
scope of this paper. Secondly, the \OI\ 63 $\mu$m line and to a lesser
extent the \CII\ 158 $\mu$m line are affected by self-absorption. This
becomes obvious in the line profiles (Fig.~\ref{spectra-single}), and
is indicated by the low \OI\ 145 $\mu$m/63 $\mu$m ratio of $<0.2$
\citep{Schneider2003, Stock2015}.  If the \OI\ 63 $\mu$m line is only
moderately optically thick ($\tau \sim$1--2), \citet{Liseau2006}
showed that self-absorption is responsible for the low ratio. We thus
additionally constrained the PDR models using an \OI\ intensity for
all velocity ranges that is a factor 2, 4, and 10 higher. This is only
a first approximation, but it allows us to estimate how the density
and the radiation field change with varying \OI\ line intensity. Using
a much higher \OI\ intensity (4 and 10 times the observed value)
results in higher densities (typically 10$^{5-6}$ cm$^{-3}$) for all
positions but in a much lower radiation field (typically $\chi
\sim$10$^{2-3}$). Such a low radiation field is not supported by
observations; for example, we determine from {\sl Herschel} flux
measurements (see Sec.~\ref{compare}) a value of $\chi \sim$(2-4)
10$^4$ as a lower limit.  We thus only use the results of a PDR model
with the nominal \OI\ intensity, and one that is a factor of 2 higher
(given in Cols. 7 and 8 in Tables~\ref{tab:ratio1} and
\ref{tab:ratio2} in parentheses.)

\begin{figure*}
\centering
\includegraphics[width=7cm, angle=0]{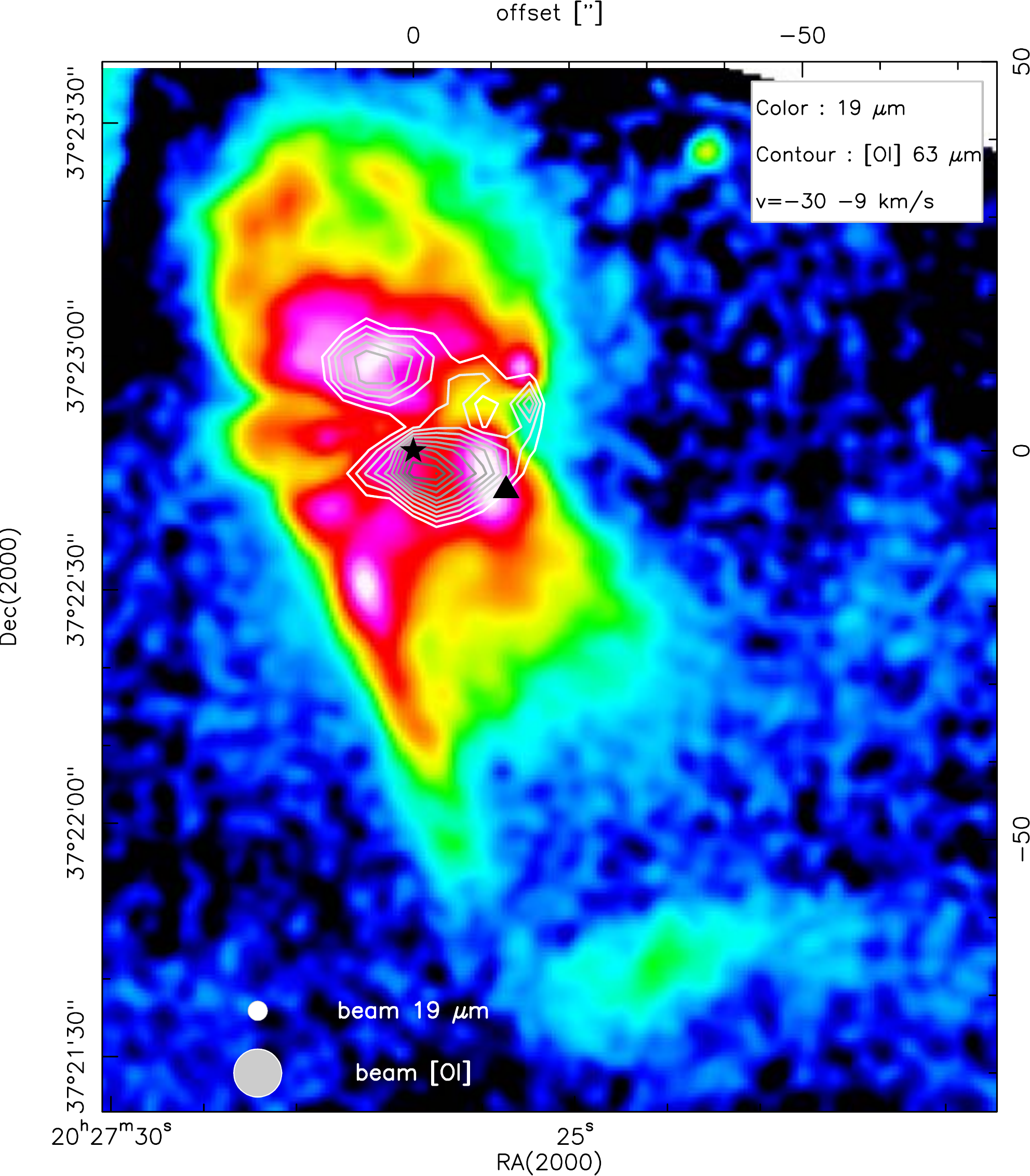}
\includegraphics[width=7cm, angle=0]{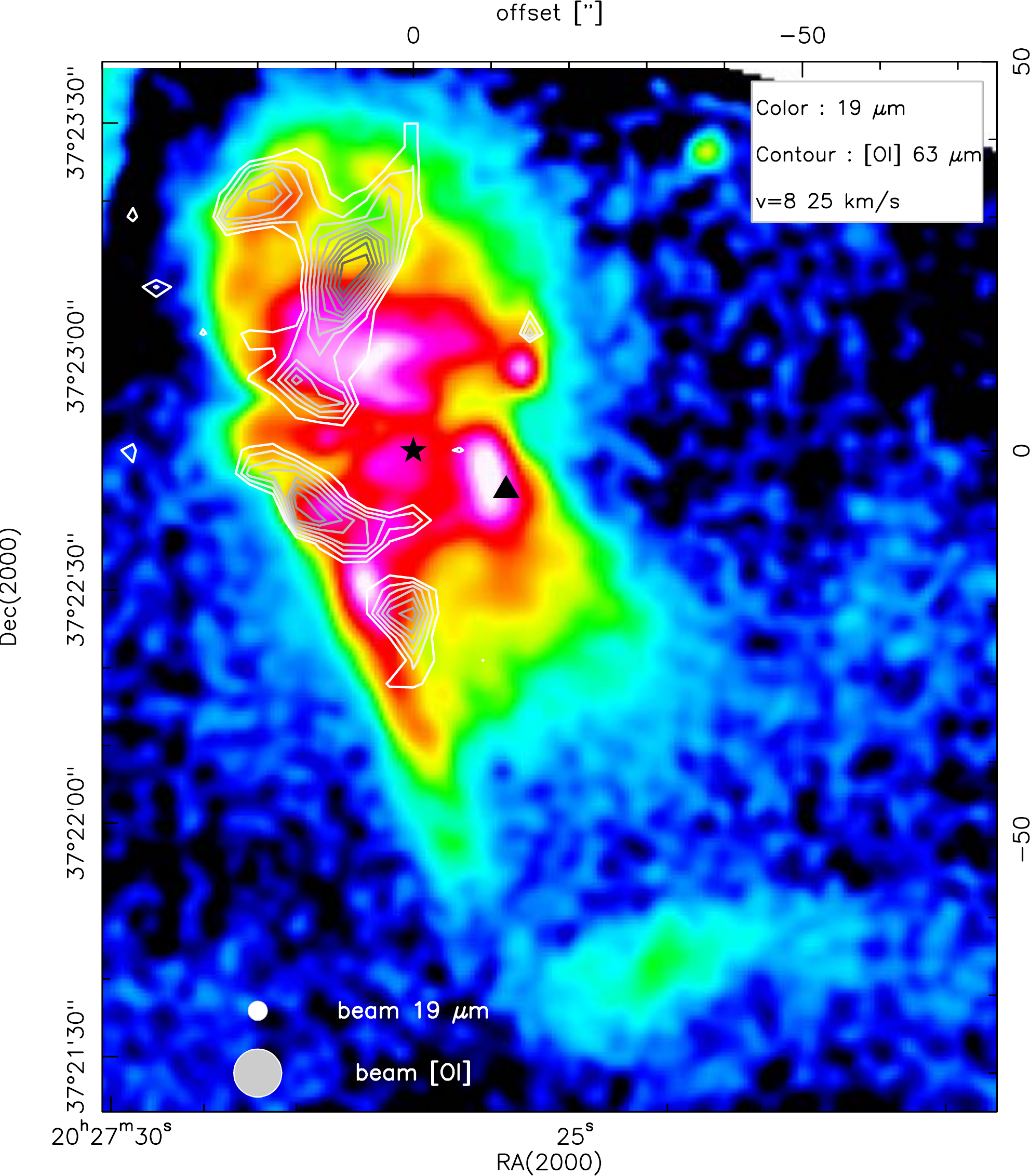}
\caption{\OI\ line integrated emission overlayed as contours on 19
  $\mu$m emission from dust \citep{Adams2015}. The left panel shows
  the high-velocity blue emission and the right panel the
  high-velocity red emission. The dust emission ranges between 0 and
  2.6 Jy/pix, the \OI\ contour lines go from 30\% to 100\% of maximum
  intensity (397 K km s$^{-1}$ for HV-blue and 121 K km s$^{-1}$ for
  HV-red).}
\label{OI-HV}
\end{figure*}

\begin{figure*}
\centering
\includegraphics[width=7cm, angle=0]{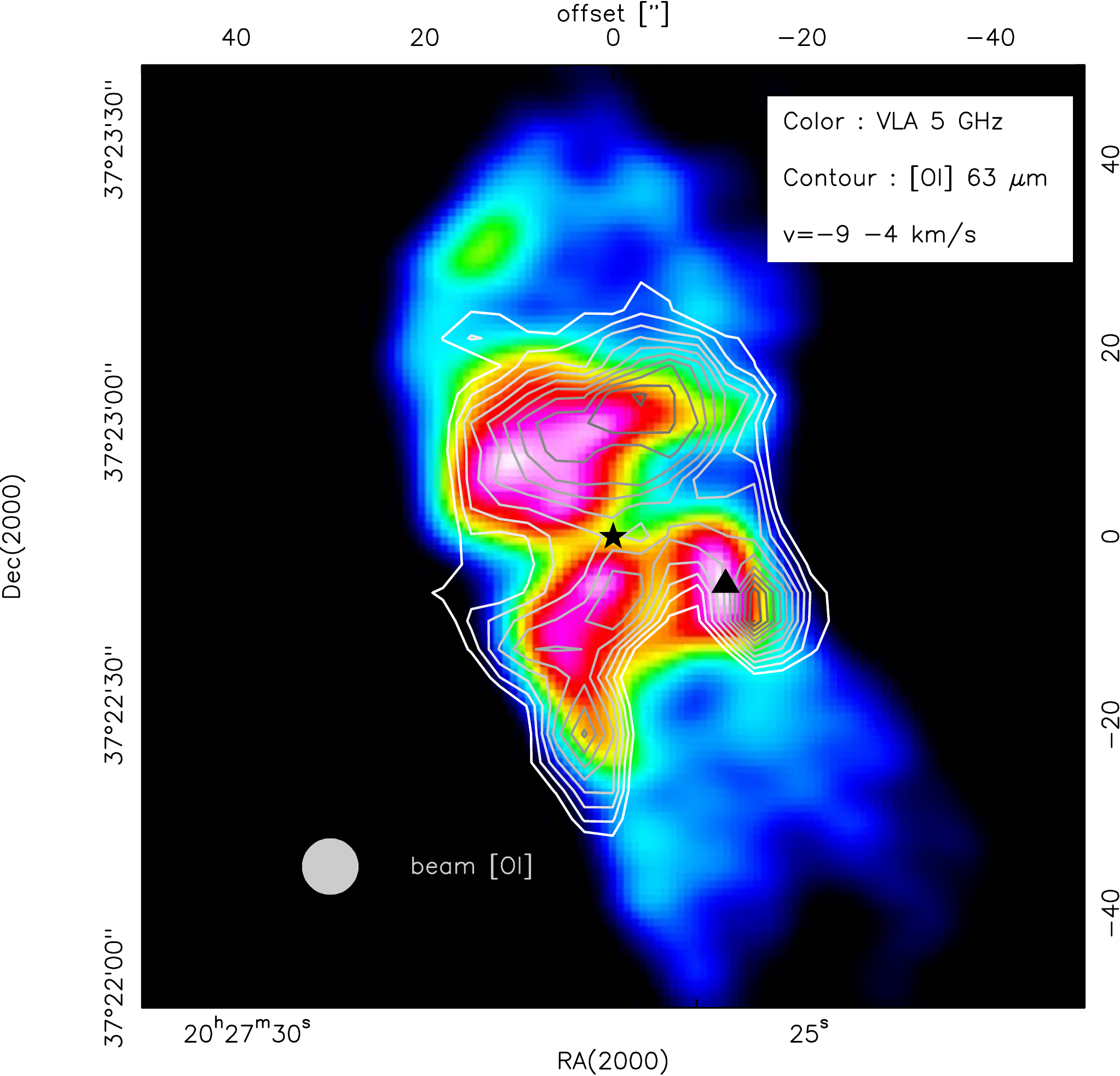}
\includegraphics[width=7cm, angle=0]{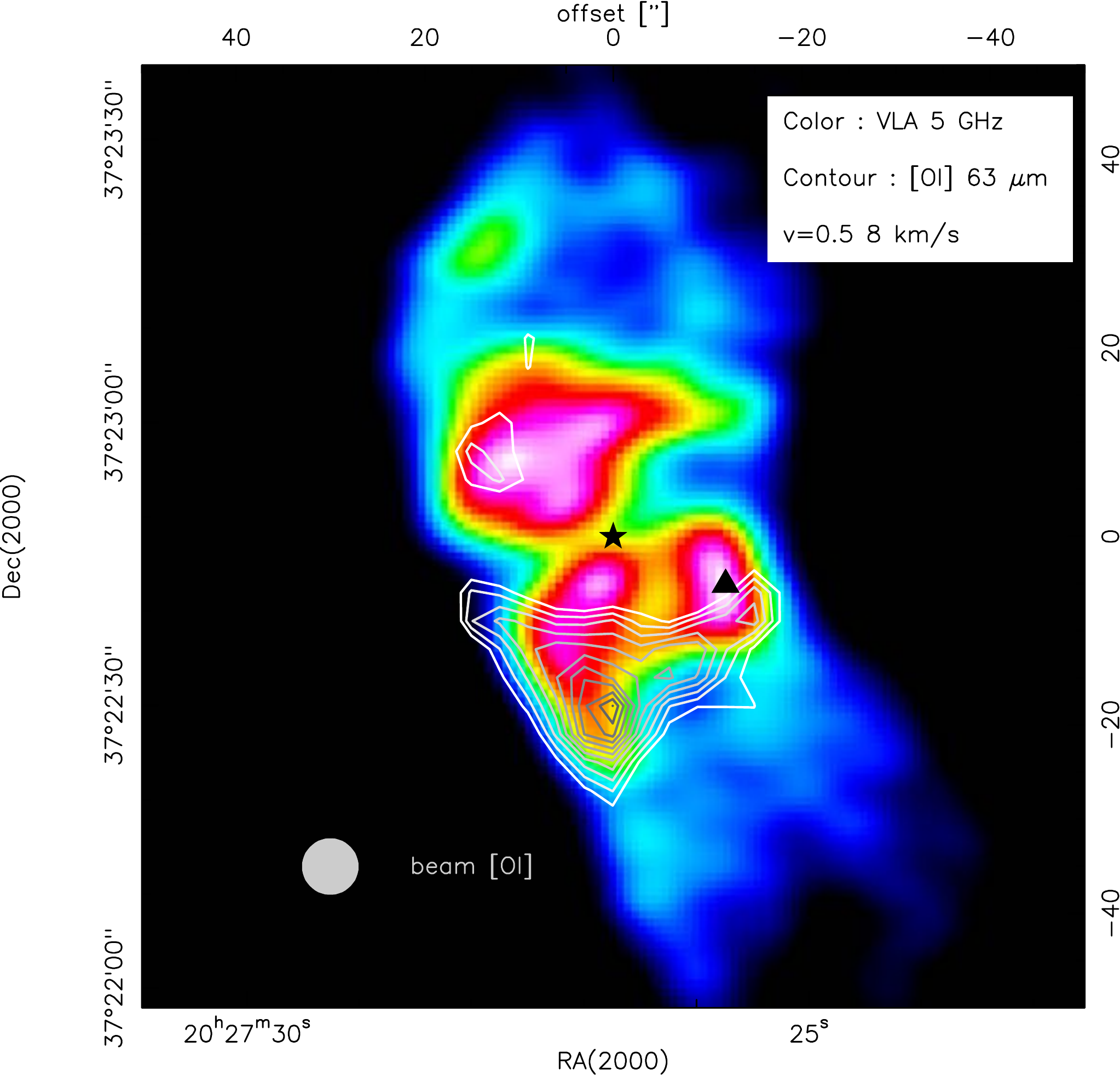}
\caption{\OI\ line integrated emission overlayed as contours on cm
  emission from the VLA \citep{Bally1983}. The left panel shows the
  blue outflow range and the right panel the red emission. The
  \OI\ contour lines go from 36 to 156 in steps of 10 K km s$^{-1}$ km s$^{-1}$
  for the blue outflow and and 40 to 180 in steps of 10 K km s$^{-1}$ for the
  red outflow. The star indicates the binary system S106
  IR and the triangle S106 FIR.}
\label{OI-outflow}
\end{figure*}

\subsubsection{High-velocity blue and red range} \label{hv-blue-red} 

The HV-blue (v$<$--9 km s$^{-1}$) \OI\ emission is spatially
concentrated in two regions north-east (Pos 1) and south-west (Pos 2)
of S106 IR (Fig.~\ref{OI-HV}).  Position 1 is correlated with a 19 $\mu$m
dust peak, and Pos 2 is located very close to S106 IR.  It is
remarkable that at Pos 1, the \OI\ emission at these high blue
velocities contributes $\sim$50\% of the whole line integrated
emission at this position (see Table~\ref{tab1}).  The \OI\ HV-blue
emission corresponds well to the \CII\ emission distribution
(Fig.~\ref{channel-hv-blue} in Appendix B), though the latter is more
extended and  outlines the cavity better. The $^{12}$CO 2$\to$1
emission in this velocity range (Fig.~\ref{channel-hv-blue} in
Appendix B) traces the molecular outflow and the cold gas of the dark
lane, but is probably more affected by beam dilution (11$''$ resolution
with respect to 6$''$ for \OI). The $^{12}$CO 16$\to$15 and 11$\to$10
lines show only very weak emission at some positions
(Fig.~\ref{spectra-single}).
  
The HV-red (v$>$8 km s$^{-1}$) \OI\ emission is more dispersed and has
two peaks along the eastern limb-brightened cavity wall, and a
northern peak that is not correlated with the dust. While the overall
\CII\ emission in this velocity range corresponds well with \OI, the
$^{12}$CO 2$\to$1 line is only visible at the position of the northern
clump (Fig. B.5).
                                                                                                                                                          
There are three possible explanations for the origin of the
high-velocity blue and red \OI\ emission: 1) disk wind in the context
of an accretion shock, 2) shocks caused by the stellar wind hitting
the inside of the dark lane, or 3) classical PDR emission from the
inner surface of the dark lane. Further details of each possibility
are given below: \\

\noindent {\bf 1.}  An {\bf {\sl accretion shock}} can occur because
S106 IR is surrounded by a disk-like structure. However, considering
the  distance of less than 0.2 AU between the  two stars, the
existence of a circumbinary disk is very unlikely;  it is probably
only one disk or its remains that were seen in
cm-interferometry. Adopting what is known for low-mass stars (see
\citealt{Hartmann2016} for a review), accretion of material from the
disk takes place in small-scale (sub-pc) flows along magnetic field
lines (see e.g. Fig.~1 in Hartmann et al.). The material has high
free-fall velocities (of the order of several hundred km s$^{-1}$),
that then causes shocks at the stellar surface. However, the resulting
shock heats the gas to very high temperatures  ($\sim$10$^6$ K), which
is far above the excitation of the \OI\ 63 $\mu$m line.  Moreover, the
HV \OI\ emission could originate from the disk that produces a bipolar
flow or jet driven by accretion energy. The observed HV-blue
\OI\ emission is indeed very collimated and would fit into this
scenario of an atomic jet, emanating from the disk--envelope--star
system. The projected velocity (towards the observer) is ${\rm
  v}_{lsr}\sim$25 km s$^{-1}$, so that the real velocity is ${\rm v} =
{\rm v}_{lsr}/\sin(\alpha),$ where $\alpha$ is the inclination angle
of the ionized lobes. \citet{Hippelein1981} and \citet{Solf1982}
determined the angle to be around 15$^\circ$ so that the velocity is
approximately 100 km s$^{-1}$ and thus consistent with high-velocity
dissociative J-type shocks (see \citealt{Hollenbach1985,
  Hollenbach1989} for theory, and \citet{Nisini2015,Leurini2015} for
observations).  However, there are two arguments against this shock
scenario on a sub-pc scale. Firstly, the HV-red \OI\  emission  is not
collimated in the same way as the HV-blue emission (Fig. ~\ref{OI-HV})
which would be expected in the case of a symmetric atomic jet. Moreover,
the emission is much more extended (up to several parsecs) and
correlates well with the cavity walls.  Secondly, the CO 16$\to$15
line as a typical J-type shock tracer has a very low line intensity
(Table~\ref{tab1}) in this velocity range and is only visible as an
extended wing (Fig.~\ref{spectra-single}).\\

\noindent {\bf 2.}  S106 IR consists of an OB star system that has a
strong {\bf {\sl stellar wind}} of around 200 km s$^{-1}$
\citep{Hippelein1981}. The wind is a source of high kinetic energy and
develops a shock when it expands into the ambient medium, in this case
hitting the inner part of the dark lane and the molecular cloud
fragments close to S106 IR. The shocked gas can then cool via the
\OI\ 63 $\mu$m line.  \\

\noindent {\bf 3.}  Given the proximity of the exciting S106 IR binary
system, the HV \OI\ emission can be explained by {\bf {\sl PDRs}} on
the surface of surrounding clumps and from the backside of the dark
lane. From the PDR model (Sec.~\ref{fit}), we derive a density of 5--6
10$^4$ cm$^{-3}$ and an UV field of $\chi \sim$7 10$^4$ from the ratio
\CII/\OI\ for S106 IR and the northern HV-blue and HV-red peak
emission positions. The southern HV-red \OI\ peak has a lower density
of 10$^4$ cm$^{-3}$ and a UV field of $\chi$=1.6 10$^4$ from
\CII/\OI. These density values increase to 2 10$^6$ cm$^{-3}$ (while
the radiation field becomes slightly lower, typically $\chi$ around a
few times 10$^4$) when the \OI\ intensity is increased by a factor of
two.  The density and radiation field obtained from the
CO(16$\to$15)/CO(11$\to$10) line ratio for the HV-blue emission are
lower (n=1.5 and 2.8 10$^4$ cm$^{-3}$ and $\chi$=2--4 10$^4$) and stay
below 10$^5$ for the density even if the \OI\ intensity
increases. Modelling the HV-red CO emission is tricky because of the
low line intensities (see Table~\ref{tab1}).  For Pos 6, we derive a
density of n$\sim$2 10$^6$ cm$^{-3}$ for both original and increased
\OI\ intensity, but the radiation field with a value of $\chi$=10$^6$
is uncharacteristically high.  \\
On first sight, it is puzzling that in the high-velocity blue range we
observe strong \OI\ emission but nearly no CO 16$\to$15 emission, even
though the two lines have similar critical densities. Because CO
16$\to$15 has a higher excitation temperature (750 K) than \OI\ (228
K), and emits much less prominently in this velocity range, we assume
that this high-velocity gas is not strongly heated and consists only
of gas with high densities ($>$10$^6$ cm$^{-3}$) at T$\sim$300 K. The
CO 16$\to$15 line is then mostly subthermally excited.

\subsubsection{Outflow blue and red \OI\ emission} \label{outflow-blue-red} 

The \OI\ emission in the blue and red outflow velocity range (-9 to -4
km s$^{-1}$ and 0.5 to 8 km s$^{-1}$, respectively) outlines precisely
the cavity of the bipolar nebula, illustrated in
Fig.~\ref{OI-outflow}, which shows \OI\ contours overlaid on a cm VLA
image \citep{Bally1983}, and in Fig.~B.2.  The cm emission delineates
the \HII\ region but avoids the dark lane. \citet{Bally1983} argue
that the density in the lane is so high that all ionizing radiation is
absorbed (see Sec.~\ref{lane}).  Similar to the \CII\ emission
distribution \citep{Simon2012}, the red \OI\ outflow emission is
absent in the northern lobe. Both \OI\ and \CII\ red outflow emission
thus stem mostly from swept-up gas of the cavity walls from the {\sl
  backside} of the southern lobe (since the front side does no longer
exists). In addition, we observe limb-brightened emission at the
cavity walls and, closer to S106 IR, the \OI\ emission in the blue
outflow range is also correlated with the clump containing S106
FIR. What is causing the gas dynamics is not clear. The stellar wind
of S106 IR can provoke shocks when it hits (mechanically) the cavity
walls. However, the radiation of S106 IR ionizes the interface layer
between the \HII\ region and the molecular cloud and creates a
classical PDR. A mixed scenario involving `irradiated shocks'
\citep{Lesaffre2013} and/or `moving PDRs' \citep{Stoerzer1998} is also
possible. From our modelling, we obtain average densities at the three
cavity positions (3a,b and 4) of 1.3--5.6 10$^4$ cm$^{-3}$ at a rather
constant radiation field of $\chi \sim$ a few 10$^4$, depending on
position and ratio (\CII\ and \OI\ or CO). Interestingly, the values
for density and radiation field do not change much if we increase the
\OI\ intensity.  Though the \CII\ line and to a lesser extent the
\OI\ line can also be excited by collisions with electrons, we assume
that only a very small fraction of \CII\ emission arises from the
\HII\ region because we detected only very weak \NII\ 205 $\mu$m
emission. Generally, this line arises only from the \HII\ region, so
that its detection only together with a sufficient \NII\ to \CII\ line
ratio leads us to conclude that a significant amount of \CII\ also
emerges from the ionized gas phase \citep{Heiles1994}. This is not the
case for S106.

\begin{figure}
\centering
\includegraphics[width=7cm, angle=0]{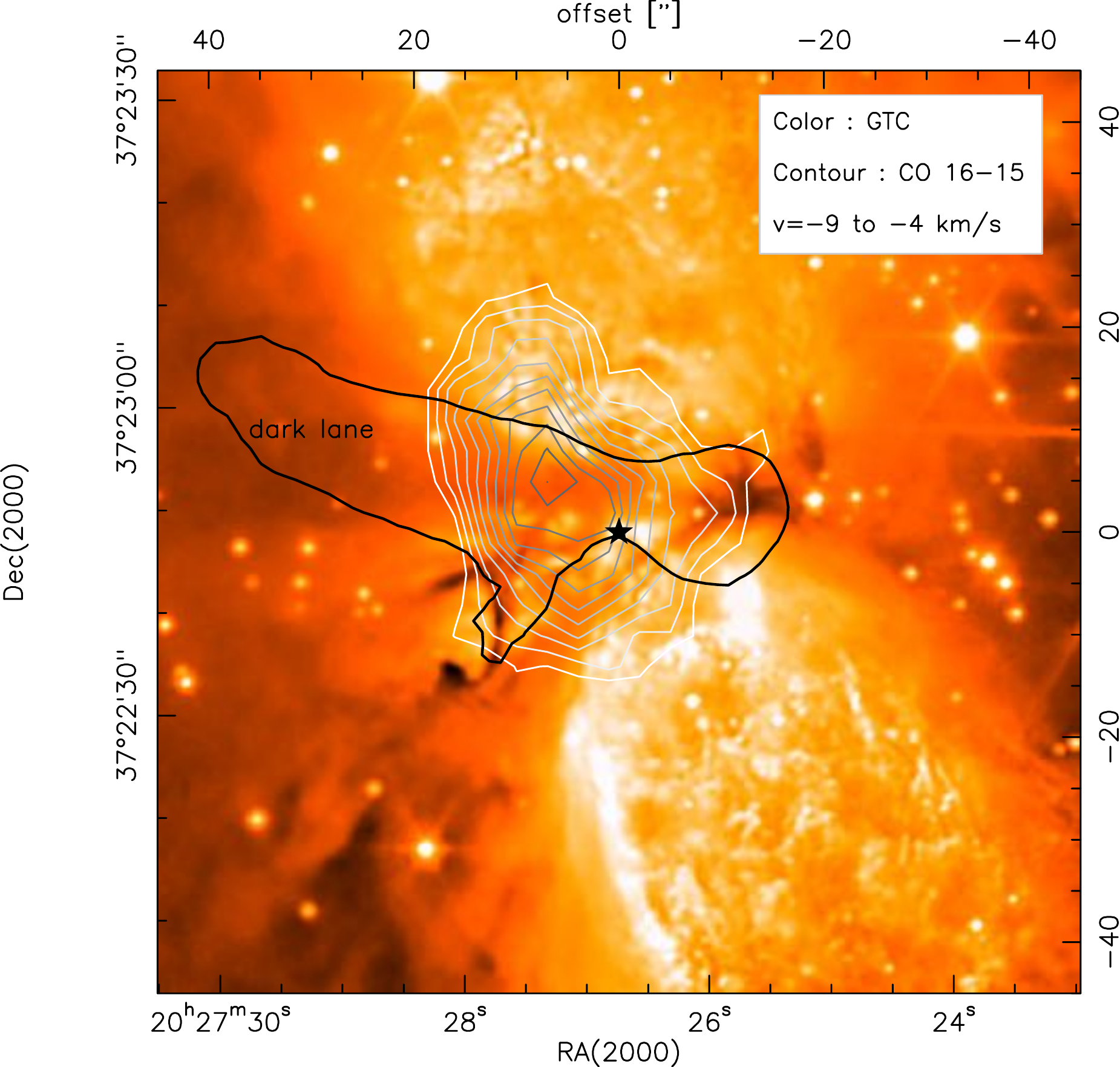}
\caption{Optical image from the Gran Telescopio Canaries (colour) with
  an overlay of CO 16$\to$15 line integrated emission (contour levels 18.6 to
  83.8 in steps of 6.5 K km s$^{-1}$) in the blue outflow velocity
  range (--9 to --4 km s$^{-1}$).  The dark lane is indicated by one
  black contour (3 Jy) of SHARC 350 $\mu$m emission (see Fig.~13).}
\label{blue-outflow-co1615}
\end{figure}

\begin{figure}
\centering
\includegraphics[width=7cm, angle=0]{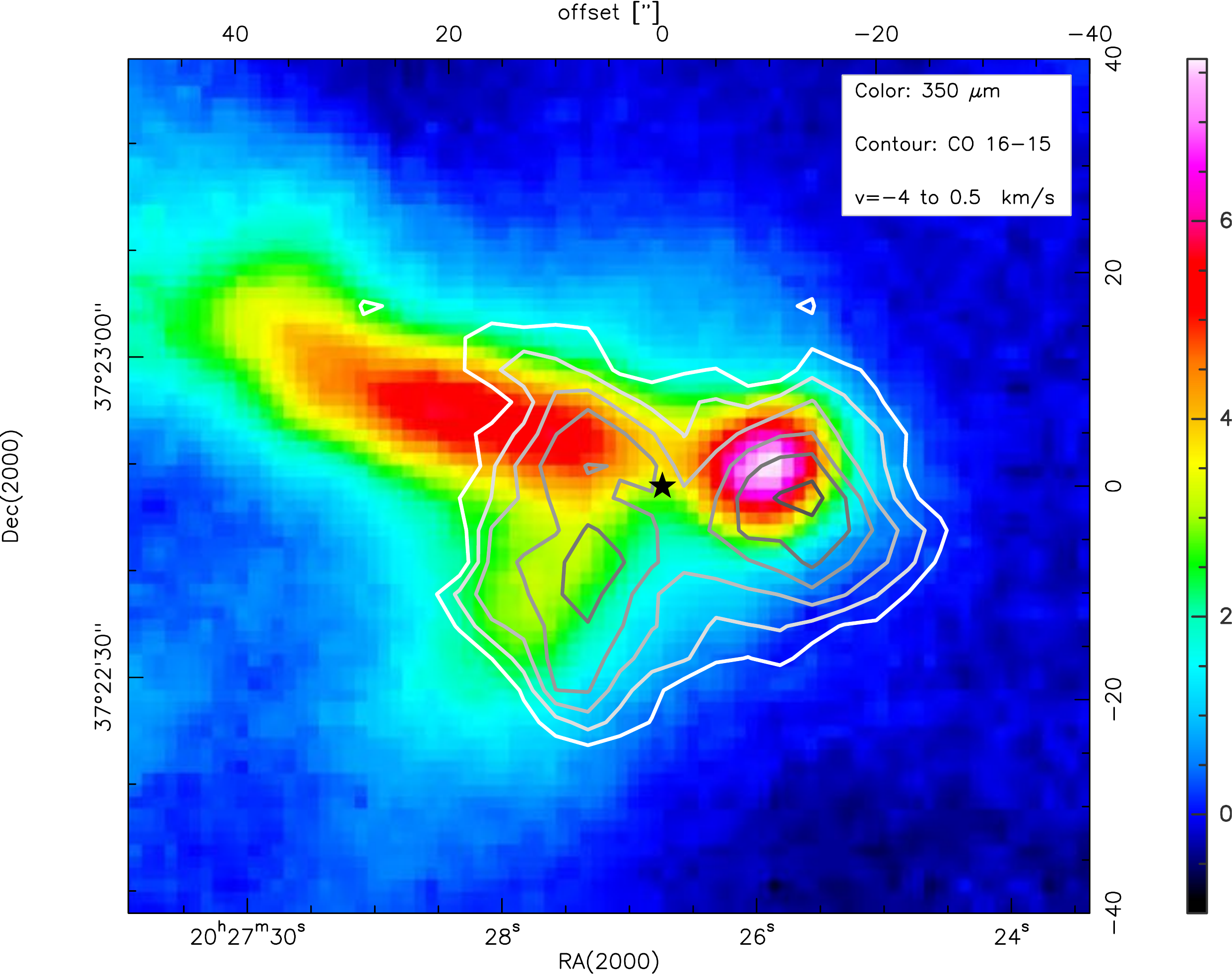}
\caption{SHARC 350 $\mu$m emission (colour wedge in Jy) with overlays
  of CO 16$\to$15 line integrated emission (levels 14 to 47 by 3 K km
  s$^{-1}$, grey contours) in the bulk velocity range (--4 to 0.5 km
  s$^{-1}$).}
\label{bulk-co1615}
\end{figure}

In contrast to \OI, the emission distribution of the CO 16$\to$15 line
(Fig.~\ref{blue-outflow-co1615}) in this velocity range shows a strong
spatial correlation with the `dark lane', peaking at position 3c, which
shows the highest values of CO 16$\to$15 emission
(Table~\ref{tab1}). Figure B.2 shows that the peak of emission in the
low-J CO lines ($^{12}$CO and $^{13}$CO 2$\to$1) is slightly shifted
towards S106 IR.  Because the lane appears prominently in the optical
image as a dark feature and consists of cold gas (see SHARC image,
Fig.~\ref{bulk-co1615}, and overlays with the low-J CO lines,
Fig.~B.2), the CO 16$\to$15 emission can only arise from a PDR layer
of warm and dense gas on the backside of the lane towards S106 IR. The
rather low line velocities and small line width argue against a
high-velocity shock origin of the high-J CO emission, but low-velocity
shocks cannot be excluded. From PDR modelling, using the CO
16$\to$15/11$\to$10 ratio at position 3c, we derive a density of 5.6
10$^4$ cm$^{-3}$ and a radiation field of $\chi$=7.4 10$^4$.  For the
red outflow velocity range (0.5 to 8 km s$^{-1}$) all lines with high
excitation temperature and high critical density (\OI, \CII, CO
16$\to$15) show a similar emission distribution (Fig.~B.4) with an
extended clump south of S106 IR. This velocity range is affected by
absorption so that the derived line intensities and ratios are not
reliable and we do not use those for PDR modelling.

Summarizing, we interpret the emission in the outflow velocity ranges
for all lines as arising from PDR surfaces of heated and UV
illuminated clumps around S106 IR, of ablated gas from the cavity
walls, and from a PDR surface on the backside of the dark lane.

\subsubsection{Bulk emission} \label{bulk} 

The velocity range from --4 to 0.5 km s$^{-1}$ is dominated by widespread
but clumpy \OI\ emission (see channel map, Fig.~\ref{oi-chan}) in
which the area around S106 IR is mostly devoid of emission (best
visible in panel v=--0.6 km s$^{-1}$).  The most prominent clump is
the western one associated with S106 FIR. Two other clumps north-west
and south of S106 IR (Fig.~B.3) are also clearly defined, and the
overall emission distribution corresponds very well with that of
\CII\ in this velocity range (Fig.~B.3). The peaks of emission in the
low-J CO lines and CO 16$\to$15 emission are close to the dark lane
and at the position of S106 FIR (Fig.~\ref{bulk-co1615} and B.3).  The
CO 16$\to$15 peak corresponds to Pos 5a, the NW peak of \OI\ emission
to Pos 5b, and the western \OI\ peak to Pos 3b, i.e. the continuum
source S106 FIR \citep{Richer1993,Little1995}, possibly a low- or
high-mass Class 0 YSO. The absolute line intensities of \OI\ and
\CII\ emission are high and the line fractional emission is 20--40\%,
indicating that these lines are the most important cooling lines for
this portion of the gas. The CO 16$\to$15 line is less strong at these
positions (in contrast to the CO 11$\to$10 line) and excited only in
the outermost surface PDR layer of the dark lane facing S106 IR.
Because the density and radiation field are not very high (see below),
it is  a fair assumption that this line is subthermally
excited. \\ From PDR modelling, we derive a density of 1.1 (1.6)
10$^4$ cm$^{-3}$ (from the \CII/\OI\ and CO ratios, respectively), and
a radiation field of $\chi$=1.1 (2.1) 10$^4$ for Pos 5a, i.e. the CO
16$\to$15 peak at the inner working surface of the dark lane. The
density and radiation field (n=3.0 10$^4$ cm$^{-3}$ and $\chi$=2.7
10$^4$) are similar at the position of the NW \OI\ peak. S106 FIR
shows peak emission in hot dust, but the gas density and radiation
field is not elevated, i.e. n$\sim$2 10$^4$ cm$^{-3}$ and $\chi
\sim$2.5 10$^4$. As for the blue and red outflow velocity range,
the density and radiation field do not change significantly when the
\OI\ intensity is increased in the PDR model.

In summary, the \OI\ emission in this velocity range seems to arise
from the inner cavity walls in combination with PDR surfaces of
individual clumps located within the cavity, close to S106 IR.

\subsection{Comparison between SOFIA results and earlier studies} \label{compare} 

This is the first time that a spectrally resolved map of the \OI\ 63
$\mu$m FIR cooling line is available for S106. A \CII\ 158 $\mu$m map
obtained with GREAT on SOFIA was presented by \citet{Simon2012}. Other
observations (not spectrally resolved) of the \CII\ and \OI\ lines
with the Infrared Space Observatory (ISO), the Kuiper Airborne
Observatory (KAO), and {\sl Herschel} were reported by
\citet{vandenAncker2000,Schneider2003,Stock2015}, respectively.  These
studies, however, have a much lower angular resolution (40$''$ to
80$''$) and thus contain the different physical regimes in S106
(\HII\ region, molecular cloud bulk emission, cavity, etc.) in one
beam\footnote{As a consistency check, we determined the total \OI\ 63
  $\mu$m and \CII\ 158 $\mu$m line intensity in a 40$''$ beam around
  the position of S106 IR and compared it to the value obtained with
  {\sl Herschel} \citep{Stock2015}. Our line intensities are slightly
  lower but agree well, considering the calibration uncertainties for
  both instruments (15--20\%). Our values of \OI\ and \CII\ emission
  are 2.2 10$^{-2}$ erg cm$^{-2}$ sr$^{-1}$ s$^{-1}$ and 3.6 10$^{-3}$
  erg cm$^{-2}$ sr$^{-1}$ s$^{-1}$, respectively, compared to 2.9
  10$^{-2}$ erg cm$^{-2}$ sr$^{-1}$ s$^{-1}$ and 3.9 10$^{-3}$ erg
  cm$^{-2}$ sr$^{-1}$ s$^{-1}$ from \citet{Stock2015}.}.
Nevertheless, all studies agree that there are different gas phases
with at least two temperature regimes. \citet{Schneider2003} proposed
a two-phase gas model with small ($<$0.2 pc) high-density (n=3 10$^5$
cm$^{-3}$ to 5 10$^6$ cm$^{-3}$) clumps with high surface temperature
(T$\sim$200-500 K) embedded in lower density (n$\sim$10$^4$ cm$^{-3}$)
gas. The radiation field in these two regimes is $\chi \sim$10$^5$ and
$\chi \sim$10$^3$, respectively. \citet{Stock2015}, based on their
PACS and SPIRE {\sl Herschel} study, support this scenario. They find
a two-phase regime with a hot (T$\sim$400 K) region characterized by a
high radiation field ($\chi >$10$^5$) and high density (n$>$10$^5$
cm$^{-3}$), based on CO rotational temperatures, and a less dense
region (n$\sim$10$^4$ cm$^{-3}$) with moderate UV field ($\chi
\sim$10$^4$) and a temperature of around 300 K, dominating the
emission of the atomic cooling lines. They argue that a third PDR gas
phase may be present with very high temperatures/radiation field and
densities to explain their excess emission in the CO rotational ladder
for the highest-J CO lines (J$>$20). From fitting the high-J CO lines
within a PDR model, they obtain a radiation field of $\chi \sim$10$^5$
and densities of 10$^8$ cm$^{-3}$.  The assumption of such high
densities is somewhat problematic as they are not observed in any
other way, thus excitation by shocks can also be the reason for the
strong CO emission. However, the contribution of shocks is not clear.
\citet{vandenAncker2000} and \citet{Stock2015} estimate that their
contribution is small, while \citet{Noel2005} attribute their H$_2$
observations to shocks.  In a subsequent paper on S106, we will apply
irradiated shock models and investigate whether they reproduce our
observed line intensities and ratios more closely.

In this study, we explain the observed \OI, \CII, and CO line ratios
with a PDR model of a single gas phase with densities of a few 10$^4$
cm$^{-3}$, exposed to a radiation field $\chi$ of a few 10$^{4}$. This
would be consistent with the lower density PDR component that was
found by \citet{Schneider2003} and \citet{Stock2015}. If we consider
self-absorption of the \OI\ line and approximate the missing emission
by doubling the \OI\ intensity, we find that the HV-blue and HV-red
emission has its origin in much denser gas of n$\sim$10$^{6}$
cm$^{-3}$, but with a similar radiation field\footnote{An independent
  measure of the radiation field is to use the {\sl Herschel} fluxes
  at 70 and 160 $\mu$m (see e.g. \citealt{Schneider2016}). With this
  method, we obtained a field of typically $\chi$=2-4 10$^{4}$ or
  S106.} of $\chi$=2-3 10$^{4}$. This can correspond to the
high-density gas component found by \citet{vandenAncker2000,
  Schneider2003,Stock2015}. Because they are velocity unresolved
observations, it was not possible to attribute one gas component to a
certain velocity range. Our data now point toward a possible scenario
in which the FIR line emission in the outflow and bulk velocity ranges
arises from PDRs at lower density (a few 10$^4$ cm$^{-3}$) exposed to
a radiation field of a few $\chi$=10$^{4}$, while the high-density gas
component is only found at the high blue and red velocities. This
finding emphasizes the importance of {\sl velocity resolved}
observations of \OI, \CII, and high-J CO lines.

Our findings also show that we do not  need to invoke a third, very
high-density gas phase (Stock et al.) nor a high-temperature two-gas
phase regime (at T$\sim$300 K and T$\sim$700 K) as was needed to model
outflow emission from low-mass protostars \citep{Kristensen2017}.

\begin{figure*}
\includegraphics[width=16cm, angle=0]{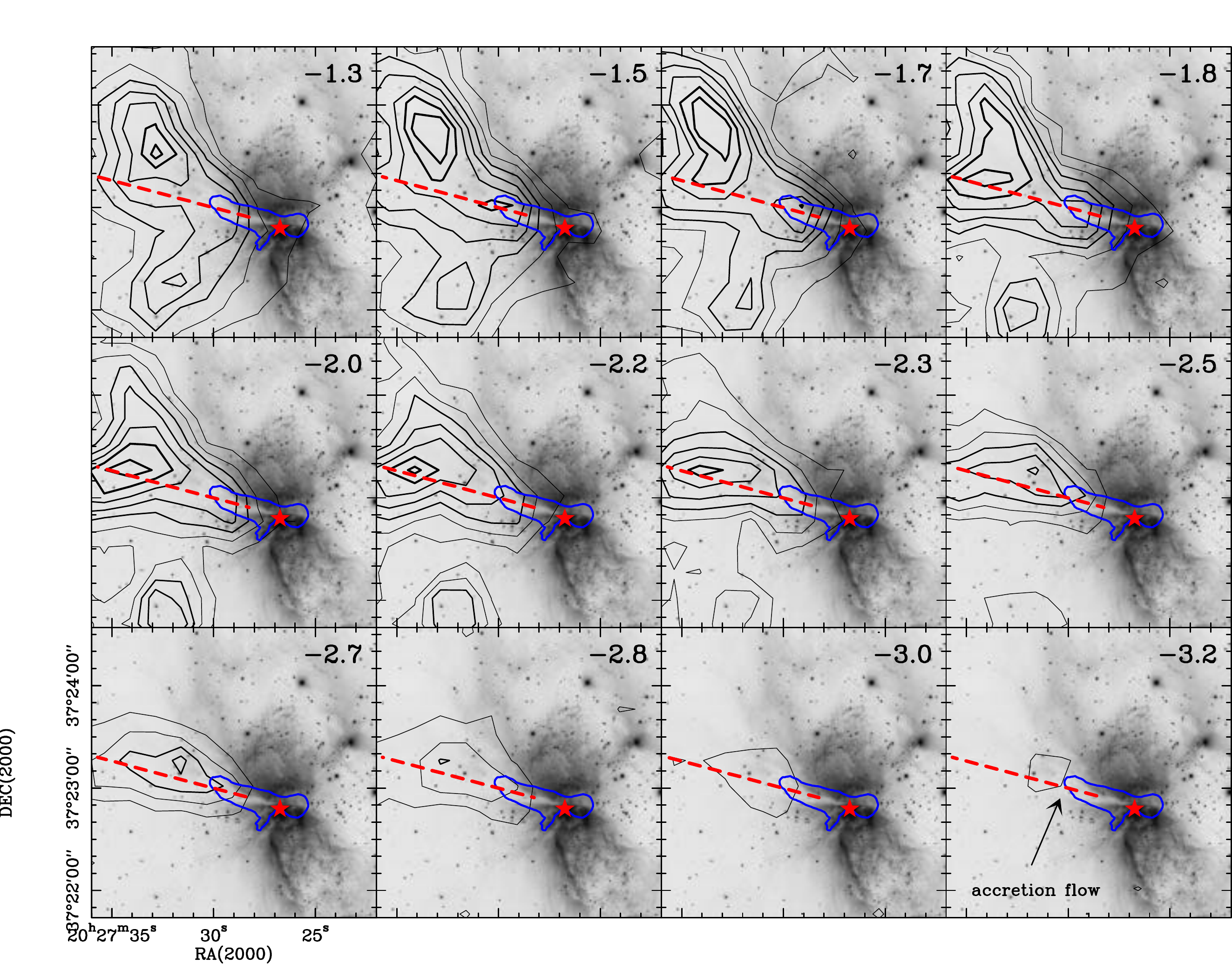}
\caption{Channel map of H$^{13}$CO$^+$ 1$\to$0 emission (contours in K
  km s$^{-1}$ from 0.25 to 1.75 in steps of 0.25) obtained with the
  IRAM 30m telescope. The background image is the Subaru IR image from
  Fig. 1.  The blue contour is the 3 Jy level of the SHARC 350 $\mu$m
  emission.  The star marks the position of the binary system S106 IR;
  the dashed red line indicates the run of the dark lane, i.e. the
  possible accretion flow.}
\label{flow}
\end{figure*}

\subsection{The dark lane: an accretion flow or an evaporating filamentary structure? } \label{lane}

It was shown by dust observations in the mm-wavelength range
\citep{Vallee2005,Motte2007}, in the mid-IR with FORCAST
\citep{Adams2015}, and in the FIR with SHARC \citep{Simon2012} at 350
$\mu$m, that the dark lane east of S106 IR is a high column-density
feature (see Fig.~\ref{bulk-co1615} for the SHARC 350 $\mu$m
map). From a column density map derived from a SED fit to the 160 to
500 $\mu$m flux maps of {\sl Herschel} \citep{Schneider2016}, we
obtain an average column density\footnote{The visual extinction in the
  lane has been estimated between A$_v$=12 mag and 21 mag
  \citep{Eiroa1979,Hodapp1991,vandenAncker2000}.}  of 4.5 10$^{22}$
cm$^{-2}$, a total mass of 275 M$_\odot$, and an average density of 3
10$^4$ cm$^{-3}$ for the lane.

\begin{figure*}
\centering
  \includegraphics[width=17cm, angle=0]{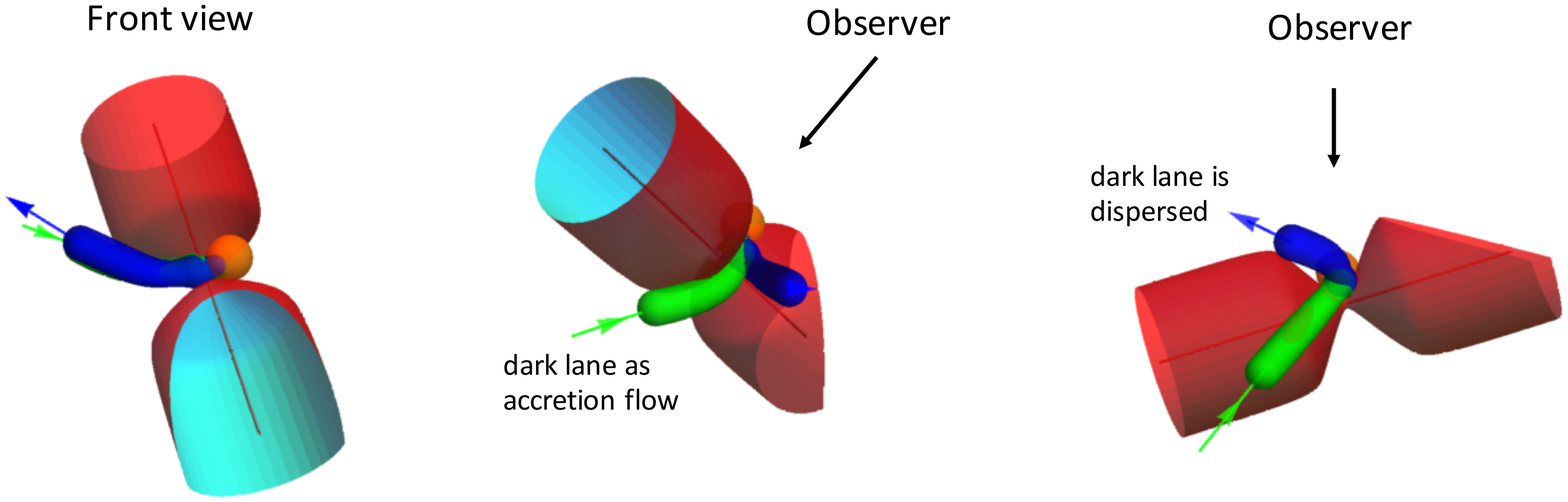}
  \vspace{-4cm}
  \caption{Schematic view of two possible scenarios (shown at the same
    time) explaining the nature of the dark lane.  The front view
    (left), as S106 is seen on the sky, does not allow us to distinguish
    between infall scenario (dark lane in green) or expansion scenario
    (dark lane in blue) of the gas in the dark lane. The middle view
    shows best the `accretion flow' scenario, i.e. how the (green)
    dark lane wraps around the equatorial waist of the
    hourglass-nebula. The right view shows how the (blue) dark lane,
    which is more detached from the nebula, is tilted towards the
    observer and dispersed.}
\label{cartoon-lane}
\end{figure*}

Our IRAM 30m observations of H$^{13}$CO$^+$ 1$\to$0 reveal the
dynamics of the dark lane more precisely than earlier $^{12}$CO and
$^{13}$CO data of the region \citep{Schneider2002}. Figure~\ref{flow}
shows a channel map of H$^{13}$CO$^+$ 1$\to$0 emission in the velocity
range between --1.3 km s$^{-1}$ and --3.2 km s$^{-1}$ overlaid on the
Subaru IR image (Fig~1).  The emission distribution has a close link
to S106 IR. It starts as a more widespread emission around --1.3 km
s$^{-1}$ with two lobes of emission and develops into a single feature
between --2.3 and --3.2 km s$^{-1}$. The peak of emission moves
gradually from a position further away north-east of S106 IR at --1.8
km s$^{-1}$ to a peak close to S106 IR at --3.2 km s$^{-1}$. This
velocity distribution is thus consistent with two possible geometries
that we illustrate in Fig.~\ref{cartoon-lane}.  Either the far end of
the lane is slightly tilted away from the observer and the gas flows
towards S106 IR (the accretion flow scenario, in green in
Fig.~\ref{cartoon-lane}) or the far end of lane is oriented towards
the observer and the gas is streaming off the lane (the dispersal
scenario, in blue in Fig.~\ref{cartoon-lane})\footnote{See also
  https://hera.ph1.uni-koeln.de/$\sim$nschneid/s106.html for an
  animated version of this cartoon.}. In both scenarios, the lane is
located in front of the \HII\ region because it is clearly visible as
a dark feature in front of a bright \HII\ region
(e.g. Fig.~\ref{flow}). While this is straightforward to understand in
the dispersal scenario because the lane is inclined towards the
observer, the flow scenario requires that the lane should be only
slightly tilted away from the observer so that it can wrap around the
equatorial plane of the hourglass-shaped \HII\ region, but still be
located in front of the ionized gas phase.

The {\sl {\bf dispersal scenario}} was originally proposed by
\citet{Hodapp1991}. They explain the bipolar \HII\ region as arising
from an anisotropic circumstellar wind, absorbing the ionizing
radiation in the equatorial plane, and postulate that the single
massive star S106 IR must be in a mass-loss evolutionary state and
not in an accretion phase. However, such a scenario requires an energy
transfer of the stellar wind onto the dense and cold gas deep in the
dark lane. We observe in \OI\ and \CII\ high-velocity gas around --30
km s$^{-1}$ that stems from PDRs and possibly shocks, caused by the
stellar wind and radiation hitting the inner working surface of the
dark lane. However, the cold gas within the dark lane occurs around
the bulk emission of the cloud, from --1 to --3 km s$^{-1}$, and is
thus decoupled from the high-velocity gas phase.

In contrast, an {\sl {\bf accretion flow scenario}}\footnote{We talk
  here of a parsec-scale flow that connects the molecular cloud with
  the disk(s) or the remains thereof, and not of possible sub-pc flows
  that channel material from the disk(s) onto the star (Sec.~5.1.1).}
was proposed by \citet{Bally1983} based on their VLA 5 GHz data (see
Fig.~\ref{OI-outflow}).  The area just around S106 IR shows only very
weak 5 GHz radio emission, indicating the absence of ionized gas
(because the 5 GHz emission is not attenuated by dust).  The gas at
very high densities is then able to absorb the emitted ionizing
radiation.  Our data show that the lane can be a flow that is
illuminated by S106 IR.

Both scenarios were also proposed in \citet{Balsara2001}, based upon
interferometric $^{13}$CO observations. They emphasize the importance
of the magnetic field that can channel material in filamentary
structures onto the highest gravitational potential well.  The
magnetic field was found to run parallel to the dark lane
\citep{Vallee2005}, thus consistent with an inflow of gas. A number of
recent observational studies support this dynamic scenario (e.g.
\citealt{Schneider2010,Kirk2013,Peretto2013,Rayner2017}). Despite the
higher angular resolution of their $^{13}$CO data compared to our
H$^{13}$CO$^+$ data at 30$''$ resolution, it is not possible to
discriminate between the two views, mostly because a low-J $^{13}$CO
line is not the best tracer for dense gas.  It is also not possible to
tell whether accretion is still occurring. Interferometric
observations are required with a high-velocity resolution of
high-density tracers such as H$^{13}$CO$^+$, H$^{13}$CN, and in
particular N$_2$H$^+$ (a probe for cold dense gas). This sort of
observation would enable us to study the velocity structure of the
dark lane and to explore the immediate environment around S106 IR.

Assuming that we indeed observe a flow towards S106 IR, the projected
length $l$ of the flow is roughly 2.2$'$, corresponding to $\sim$1 pc
at a distance of 1.3 kpc. The velocity difference $\Delta$v along the
flow, determined from the H$^{13}$CO$^+$ map, is $\sim$1.4 km
s$^{-1}$. We assume a random distribution of orientation angles and
thus an average angle of the flow to the line of sight of 57.3$^\circ$
so that the lifetime $t$ of the flow calculates as $t$ =
$l$/($\Delta$v $\tan(57.3))$.  The lifetime is then approximately 1.1
10$^6$ yr, leading to a mass input rate of 2.5 10$^{-4}$ M$_\odot$/yr
(using the {\sl Herschel} determined mass of the lane). This rate is
lower than that found for the massive subfilaments of the DR21 ridge
(a few 10$^{-3}$ M$_\odot$/yr, \citealt{Schneider2010}), but similar
to that observed for Mon R2 (a few 10$^{-4}$ M$_\odot$/yr,
\citealt{Rayner2017}).  We note, however, that this is a very crude
approximation, and does not consider the true inclination and bending
of the flow.

The flow scenario is consistent with simulations of
\citet{Peters2010a,Peters2010b} who modelled the collapse of rotating,
massive cloud cores including radiative heating by both non-ionizing
and ionizing radiation.  The simulations show fragmentation from
gravitational instability in the dense accretion flows in which either
a single massive star is formed or several massive stars with many
low-mass stars within the regions of the accretion flows. Because
there is competition of mass, but not in the classical meaning of
`competitive accretion' \citep{Bate2012,Bonnell2007}, they named this
process `fragmentation-induced starvation'. The simulations
\citep{Peters2010b} also show the non-uniform expansion of
\HII\ regions with the formation of bipolar lobes, very similar to
what is observed for S106. The bipolar outflow structure of the
\HII\ region visible at early time steps of their runs A (single sink)
and B (multiple sinks) reflects qualitatively what is seen for
S106. From what is known so far, S106 IR is a binary system
\citep{Comeron2018} and is associated with a large cluster of low-mass
stars, detected in the IR \citep{Hodapp1991}. In addition, several
dense pre-stellar cores have formed in the dark lane (S20 to S24 in
\citealt{Motte2007}), indicating ongoing fragmentation.
 
In the scenario of the dark lane as an ionized accretion flow in the
fragmentation-induced starvation model, the observed emission
distribution of \OI, \CII, and other tracers fit very well. In
particular the high-velocity blue emission is consistent with gas that
moves downward perpendicularly to the accretion flow, down the
steepest density gradient. This causes the very focused emission seen
in various tracers (Fig.~\ref{OI-HV} and B.1). Interestingly, the very
complex morphology and velocity distribution of jets and outflowing
gas -- as we observe it for S106 -- is well reproduced in the models
of \citet{Kuruwita2017}.  They produced simulations of the outflow
pattern of close (separation $<$10 AU) and wide ($>$10 AU) binaries
and showed that the geometry and velocity of the jets and outflows are
strongly modified with respect to a single star.

\section{Summary: a scenario for S106 IR} \label{summary} 

Summarizing our observational results and what is already known about
S106, we develop the following scenario:

S106 IR is a binary system with two stars that are sources of a strong
UV field and stellar wind.  The very small disk-like structure that
was detected in cm-interferometry can be an intact accretion disk that
is connected to a large-scale accretion flow, known as the dark lane
or the remains of a disk without a link to the dark lane. The lane is
illuminated by the more massive star of the system, presumably an O9
star with 20 M$_\sun$ (the companion has a preliminary classification
as a B8 star with $\sim$3 M$_\sun$; \citealt{Comeron2018}), and forms
a dense hot PDR that cools mostly via \OI\ 63 $\mu$m and CO 16$\to$15
emission. The PDR gas is highly dynamic; it flows fast and follows the
steepest density gradients of the dark lane, and `escapes' the lane
close to S106 IR, giving rise to the very collimated emission
distribution in the \OI\ lines at velocities from --30 to --9 km
s$^{-1}$ and 8 to 25 km s$^{-1}$. We obtain a radiation field of
$\chi$=2--7 10$^4$ at a density of 1.5--6 10$^4$ cm$^{-3}$ from PDR
modelling. Generally, modelling the \OI\ and \CII\ emission is
difficult because both lines show self-absorption features. By
increasing the \OI\ emission by a factor of 2, we obtain that the
densities in the HV emission ranges increase up to 10$^6$ cm$^{-3}$,
which is more consistent with earlier findings
\citep{Schneider2003,Stock2015}. If the emission in this velocity
range could also arise from a shock (disk-envelope interaction and/or
radiation and stellar wind hitting locally the dark lane) cannot yet
be answered and will be addressed in an upcoming study.

The gas in the blue and red outflow velocity ranges (from --9 to --4
km s$^{-1}$ and 8 to 25 km s$^{-1}$) has a very similar emission
distribution compared to the optical visible lobes. The \OI\ (and
\CII) emission distributions at the blue and red velocities indicate
that the emission mostly arises from the back side of the southern
lobe and the front side of the northern lobe. The outflow gas is
entrained in the wind of S106 IR and ablated by radiation (and
possibly shocks) from the cavity walls.  PDR modelling gives densities
typically of a few 10$^4$ cm$^{-3}$ at a radiation field of $\chi$ a
few 10$^4$. The low value for the density indicates that the CO
16$\to$15 line is most likely subthermally excited.

Molecular cloud clumps and possibly fragments of what once was the
larger-scale circumbinary disk around S106 IR are seen in the close
environment of the star. These clumps form PDRs on their surfaces (the
most prominent one is the western clump) that emit in all lines at the
cloud bulk velocity around --3 km s$^{-1}$. The clumps are exposed to
a radiation field $\chi$ of a $\sim$2 10$^4$ and the density is
$\sim$2 10$^4$ cm$^{-3}$.

S106 IR and its bipolar \HII\ region are embedded in a larger
molecular cloud that provides the gas reservoir for a possible
accretion flow onto S106 IR. Mapping of the high-density tracer
H$^{13}$CO$^+$ 1$\to$0 revealed a velocity gradient across the dark
lane that is consistent with either a flow onto S106 IR or gas
streaming away, depending on the geometry of the region. Only
interferometric observations can elucidate the nature of the dark
lane. The flow scenario is more consistent with the
fragmentation-induced starvation scenario of
\citet{Peters2010a,Peters2010b} than with the monolithic collapse
model of \citet{McKee2002}. As yet unclear is whether shocks driven by
the ionizing stellar wind that hits the accretion flow and the cavity
walls can also cause the observed emission of the FIR lines. It is
also not clear to what extent the \OI\ 63 $\mu$m line is
self-absorbed.  Higher line intensities will change the \CII/\OI\ line
ratios and thus modify the outcome of the PDR models. Observations of
the \OI\ 145 $\mu$m line, if this line is optically thin, may help to
tackle this problem.

Summarizing, this study shows that the new detection of high-velocity
emission in the \OI\ line, and the identification of various velocity
components in other FIR lines (\CII, high-J CO), which are only now
possible with the (up)GREAT receiver on SOFIA, help to diagnose more
precisely the physical properties of different gas phases in complex
star-forming regions.

\begin{acknowledgements}
This work was supported by the Agence National de Recherche
(ANR/France) and the Deutsche Forschungsgemeinschaft (DFG/Germany)
through the project `GENESIS' (ANR-16-CE92-0035-01/DFG1591/2-1).
N.S. acknowledges support from the BMBF, Projekt Number 50OR1714 (MOBS
- MOdellierung von Beobachtungsdaten SOFIA). This work is based on
observations made with the NASA/DLR Stratospheric Observatory for
Infrared Astronomy (SOFIA). SOFIA is jointly operated by the
Universities Space Research Association, Inc. (USRA), under NASA
contract NAS2-97001, and the Deutsches SOFIA Institut (DSI) under DLR
contract 50 OK 0901 to the University of Stuttgart.  This work is
based on observations carried out under project number 140-15 with the
IRAM 30m telescope.  IRAM is supported by INSU/CNRS (France), MPG
(Germany), and IGN (Spain).  N.S. acknowledges support from the
Deutsche Forschungsgemeinschaft, DFG, through project number Os
177/2-1 and 177/2-2, and central funds of the DFG-priority program
1573 (ISM-SPP). This work was supported by the German \emph{Deut\-sche
  For\-schungs\-ge\-mein\-schaft, DFG\/} project number SFB 956. We
thank B. Rumph for carefully reading the manuscript.
\end{acknowledgements}

%
%

\begin{appendix} 
\section{Chemical network}

\begin{table*}[htb]
\begin{center}
\caption{Chemical species included in the model} \label{species}
\resizebox{\textwidth}{!}{%
\begin{tabular}{ccccccccccc}
\hline\hline
\ce{e-}&\ce{H}&\ce{H2}&\ce{H+}&\ce{H2+}&\ce{H3+}&\ce{He+}&\ce{He}&\ce{O+}&\\
\ce{O}&\ce{C+}&\ce{C}&\ce{^{13}C+}&\ce{^{13}C}&\ce{OH+}&\ce{OH}&\ce{O2}&\ce{CO+}&\\
\ce{CO}&\ce{CH+}&\ce{CH}&\ce{^{13}CO+}&\ce{^{13}CO}&\ce{^{13}CH+}&\ce{^{13}CH}&\ce{HCO+}&\ce{H2O+}&\\
\ce{H2O}&\ce{H^{13}CO+}&\ce{CH2+}&\ce{^{13}CH2+}&\ce{H3O+}&\ce{C2+}&\ce{C2}&\ce{C^{13}C+}&\ce{C^{13}C}&\\
\ce{C3+}&\ce{C3}&\ce{C2^{13}C+}&\ce{C2^{13}C}&\ce{O2+}&\ce{HF+}&\ce{HF}&\ce{H2F+}&\ce{F+}&\\
\ce{F}&\ce{CF+}&\ce{^{13}CF+}&\ce{CH2}&\ce{C2H+}&\ce{C2H}&\ce{C^{13}CH+}&\ce{C^{13}CH}&\ce{^{13}CH2}&\\
\ce{CH3+}&\ce{CH3}&\ce{C3H+}&\ce{C3H}&\ce{C2H2+}&\ce{C2H2}&\ce{C2^{13}CH+}&\ce{C2^{13}CH}&\ce{C^{13}CH2+}&\\
\ce{C^{13}CH2}&\ce{^{13}CH3+}&\ce{^{13}CH3}&\ce{CH4+}&\ce{CH4}&\ce{^{13}CH4+}&\ce{^{13}CH4}&\ce{CH5+}&\ce{CH2CCH}&\\
\ce{CH2C^{13}CH}&\ce{CH2^{13}CCH}&\ce{C3H3+}&\ce{C2H4+}&\ce{C2H4}&\ce{C2^{13}CH3+}&\ce{C^{13}CH4+}&\ce{C^{13}CH4}&\ce{^{13}CH5+}&\\
\ce{^{13}CH2CCH}&\ce{C2H5+}&\ce{C2H5}&\ce{C^{13}CH5+}&\ce{C^{13}CH5}&\ce{C2^{13}CH5+}&\ce{HOC+}&\ce{HO^{13}C+}&\ce{HCO2+}&\\
\ce{HCO}&\ce{H3CO+}&\ce{H3^{13}CO+}&\ce{H2CO+}&\ce{H2CO}&\ce{H2^{13}CO+}&\ce{H2^{13}CO}&\ce{H^{13}CO}&\ce{CO2+}&\\
\ce{CO2}&\ce{CH3OH2+}&\ce{CH3OH+}&\ce{CH3OH}&\ce{^{13}CO2+}&\ce{^{13}CO2}&\ce{^{13}CH3OH+}&\ce{^{13}CH3OH}&\ce{HCOOCH3}&\\
\ce{HCOO^{13}CH3}&\ce{HC2O+}&\ce{HC^{13}CO+}&\ce{H5C2O2+}&\ce{H5C^{13}CO2+}&\ce{H^{13}COOCH3}&\ce{COOCH4+}&\ce{CH3CHO+}&\ce{CH3CHO}&\\
\ce{CH3^{13}CHO+}&\ce{CH3^{13}CHO}&\ce{^{13}CH3CHO+}&\ce{^{13}CH3CHO}&\ce{OCN+}&\ce{OCN}&\ce{O^{13}CN+}&\ce{O^{13}CN}&\ce{NO2+}&\\
\ce{NO2}&\ce{NO+}&\ce{NO}&\ce{NH4+}&\ce{NH3+}&\ce{NH3}&\ce{NH2+}&\ce{NH2}&\ce{NH+}&\\
\ce{NH}&\ce{N+}&\ce{N}&\ce{HNO+}&\ce{HNO}&\ce{HNCO+}&\ce{HNC}&\ce{HN^{13}CO+}&\ce{HN^{13}C}&\\
\ce{HCN+}&\ce{HCN}&\ce{H2NC+}&\ce{H2N^{13}C+}&\ce{H2CN}&\ce{H2^{13}CN}&\ce{H^{13}CN+}&\ce{H^{13}CN}&\ce{CN+}&\\
\ce{CN}&\ce{^{13}CN+}&\ce{^{13}CN}&\ce{N2O+}&\ce{N2O}&\ce{N2H+}&\ce{N2+}&\ce{N2}&\ce{SO2+}&\\
\ce{SO2}&\ce{SO+}&\ce{SO}&\ce{S+}&\ce{S}&\ce{OCS+}&\ce{OCS}&\ce{O^{13}CS+}&\ce{O^{13}CS}&\\
\ce{NS+}&\ce{NS}&\ce{HS+}&\ce{HS}&\ce{HCS+}&\ce{HCS}&\ce{H2S+}&\ce{H2S}&\ce{H2CS+}&\\
\ce{H2CS}&\ce{H2^{13}CS+}&\ce{H2^{13}CS}&\ce{H^{13}CS+}&\ce{H^{13}CS}&\ce{CS+}&\ce{CS}&\ce{C^{13}CS+}&\ce{C^{13}CS}&\\
\ce{^{13}CS+}&\ce{^{13}CS}&\ce{S2+}&\ce{S2}&\ce{Si}&\ce{Si+}&\ce{SiOH+}&\ce{SiH5+}&\ce{SiH4+}&\\
\ce{SiH4}&\ce{SiH3+}&\ce{SiH3}&\ce{SiH2+}&\ce{SiH2}&\ce{SiH+}&\ce{SiH}&\ce{SiC+}&\ce{SiC}&\\
\ce{Si^{13}C+}&\ce{Si^{13}C}&\ce{HCSi+}&\ce{HCSi}&\ce{H^{13}CSi+}&\ce{H^{13}CSi}&\ce{^{18}O+}&\ce{^{18}O}&\ce{O^{18}O}&\\
\ce{H3^{18}O+}&\ce{H2^{18}O+}&\ce{H2^{18}O}&\ce{^{18}OH+}&\ce{^{18}OH}&\ce{HC^{18}O+}&\ce{H^{13}C^{18}O+}&\ce{C^{18}O+}&\ce{C^{18}O}&\\
\ce{^{13}C^{18}O+}&\ce{^{13}C^{18}O}&&&&&&&&\\
\hline\\
\end{tabular}}
\end{center}
\end{table*}

\section{Overlays of different tracers in individual velocity ranges}

\begin{figure*}
\centering
\includegraphics[width=8.5cm, angle=0]{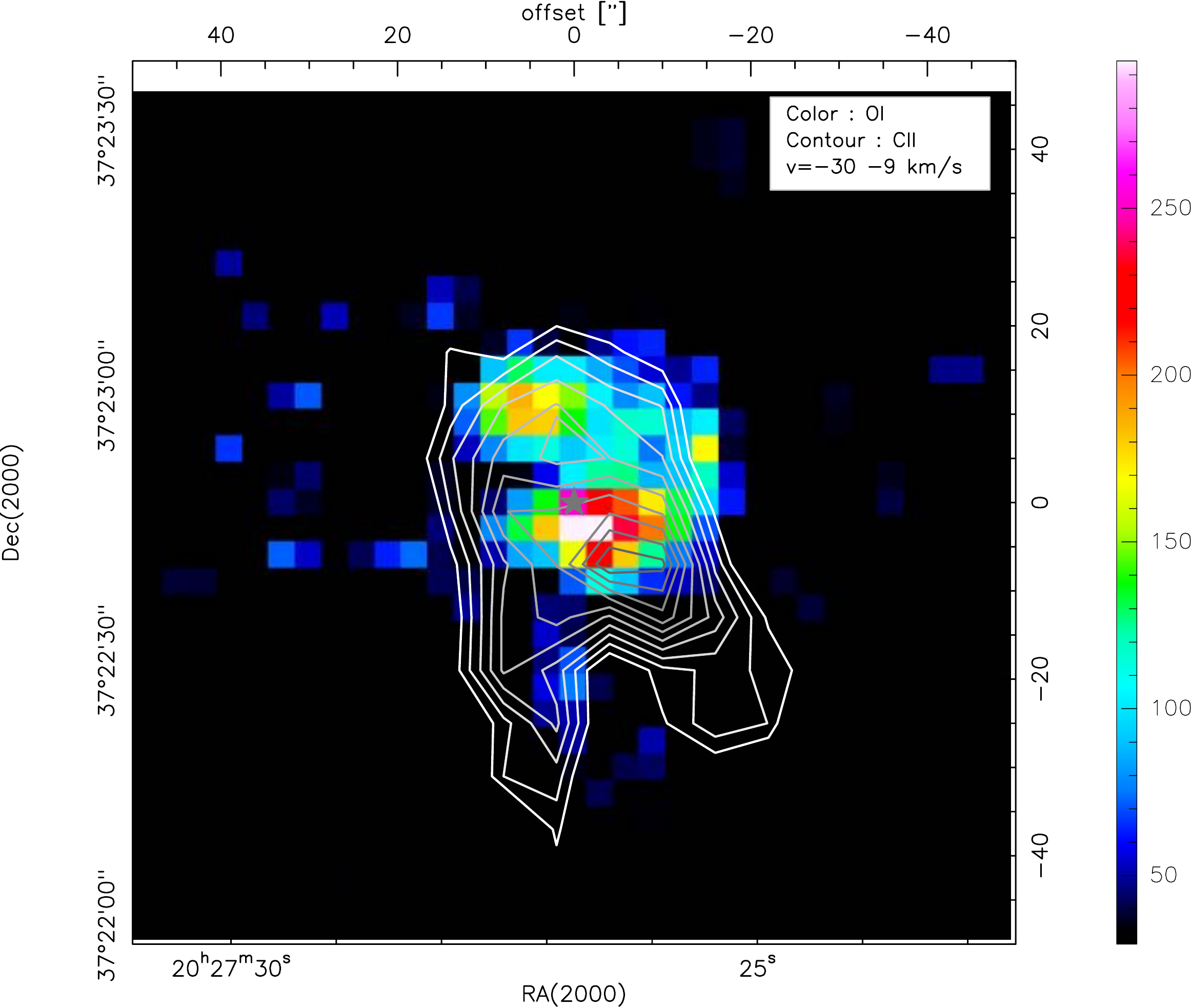}
\includegraphics[width=8.5cm, angle=0]{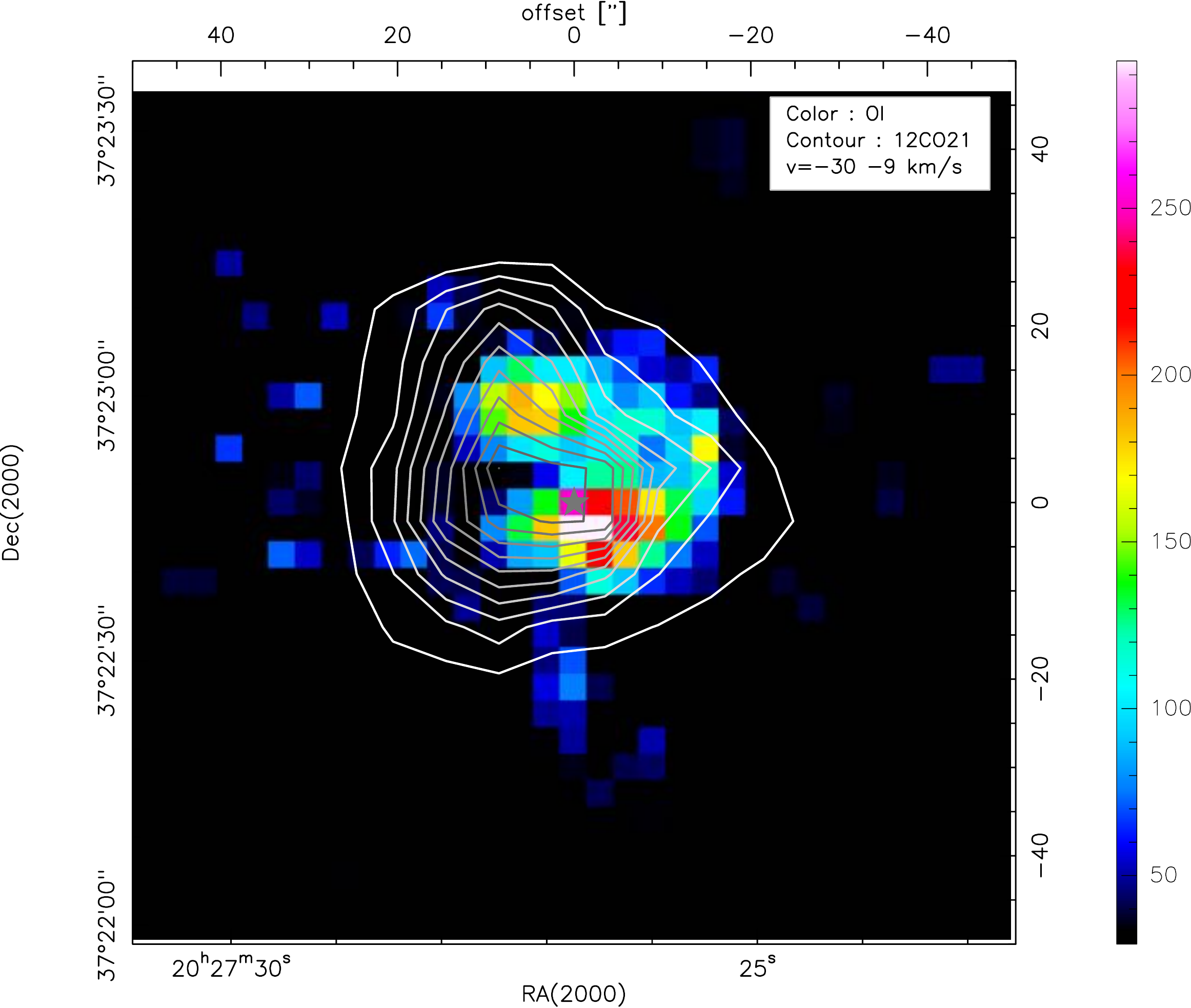}
\caption{High-velocity blue emission (--30 to --9 km s$^{-1}$):
  Contours of \CII\ (30 to 100 K km s$^{-1}$), and $^{12}$CO 2$\to$1
  (3.1 to 44.4 K km s$^{-1}$) emission overlaid on \OI\ emission in
  colour scale (wedge in K km s$^{-1}$).}
\label{channel-hv-blue}
\end{figure*}

\begin{figure*}
\centering
\includegraphics[width=8.5cm, angle=0]{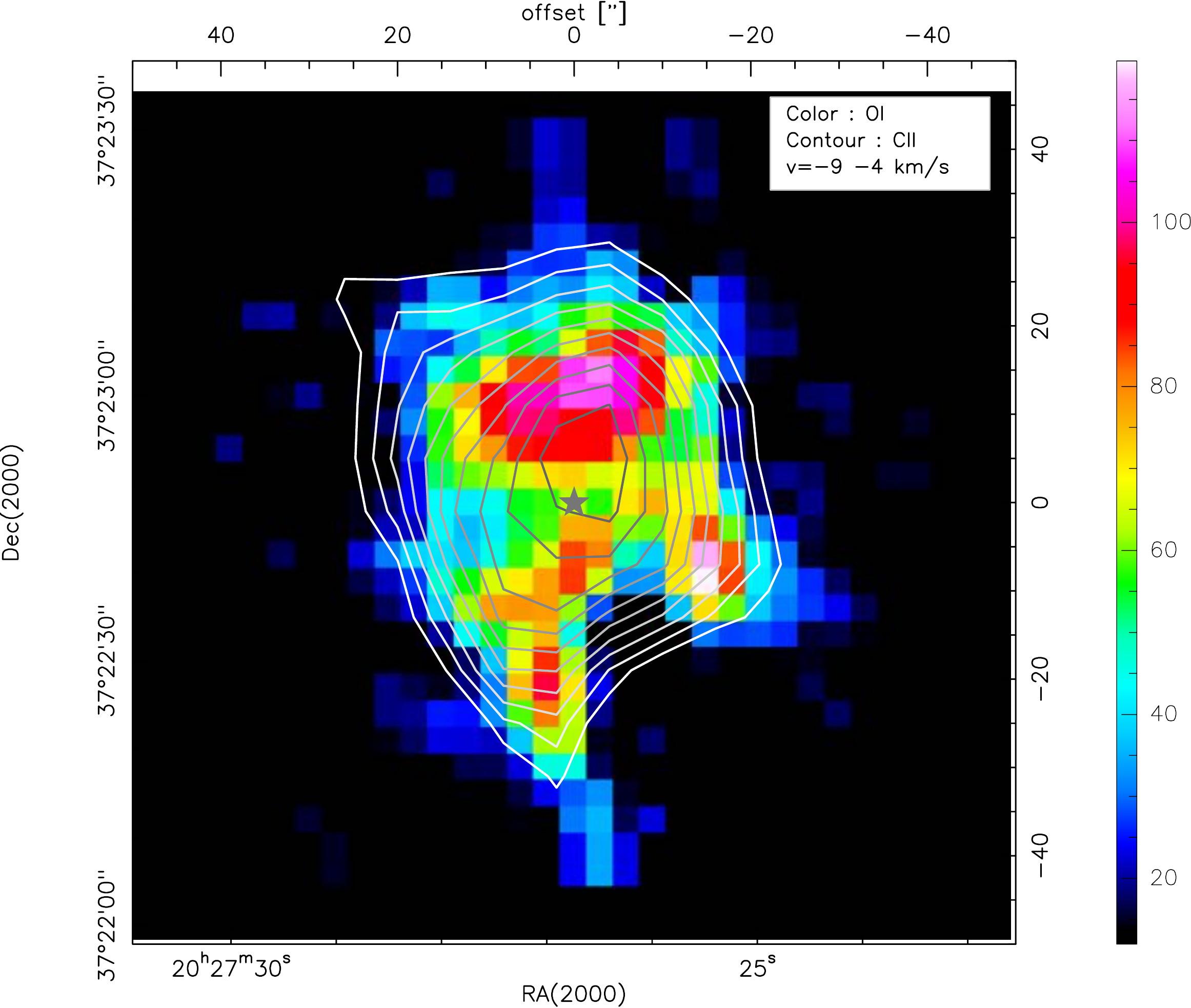}
\includegraphics[width=8.5cm, angle=0]{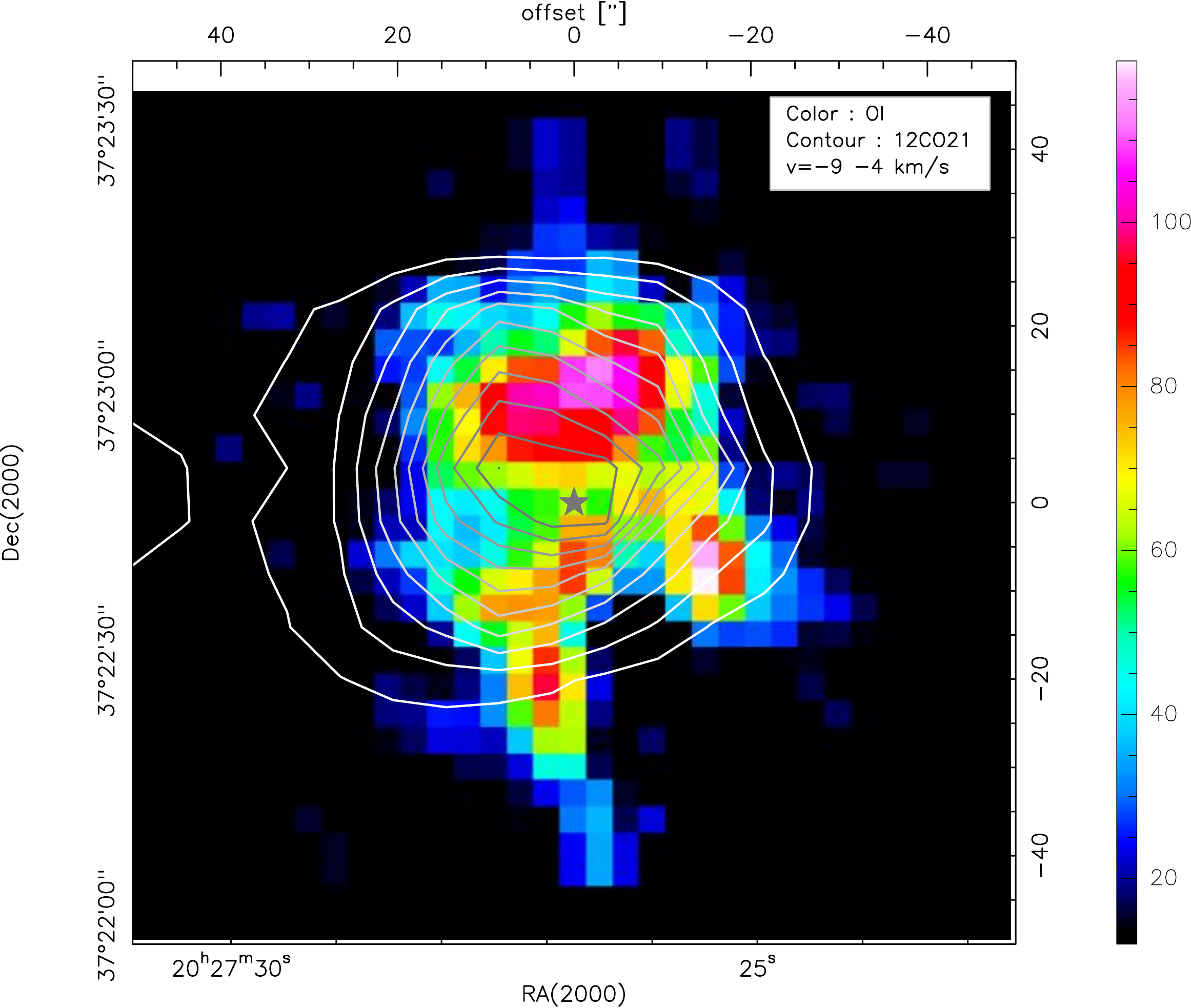}
\includegraphics[width=8.5cm, angle=0]{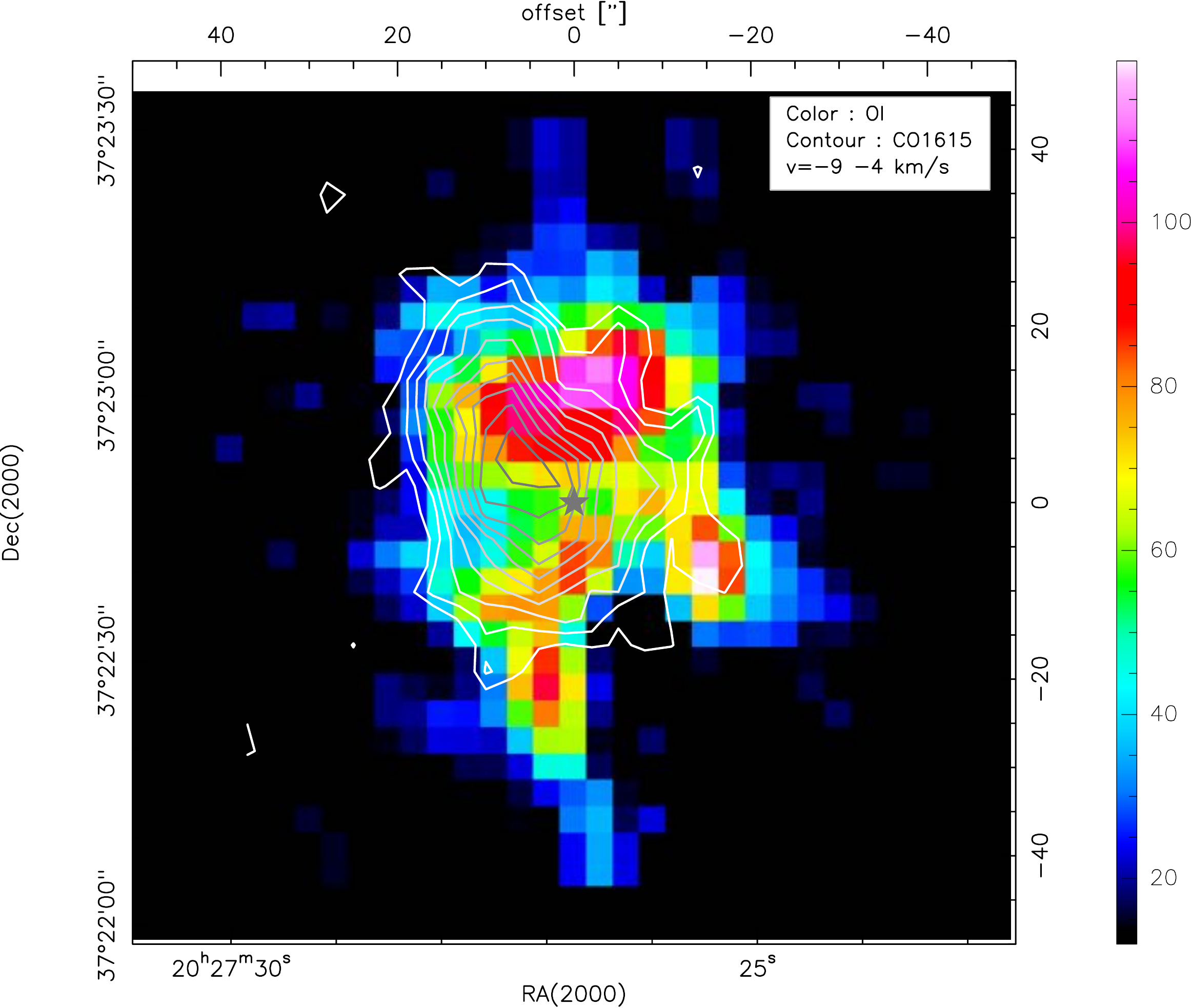}
\includegraphics[width=8.5cm, angle=0]{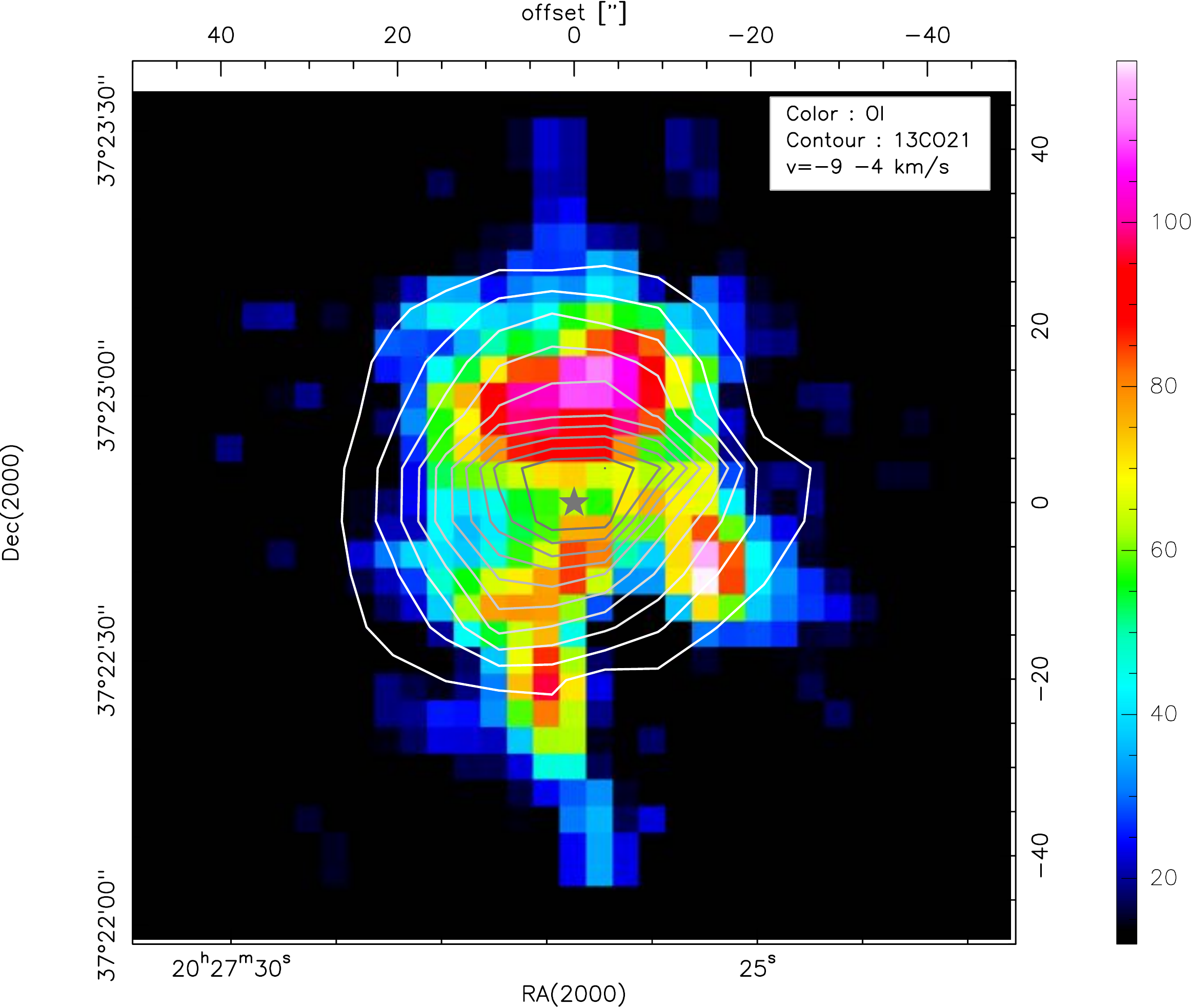}
\caption{Outflow blue emission (--9 to --4 km s$^{-1}$): Contours of
  \CII\ (59 to 192 K km s$^{-1}$), $^{12}$CO 2$\to$1 (3.1 to 44.4 K km
  s$^{-1}$), $^{12}$CO 16$\to$15 (13.1 to 191.1 K km s$^{-1}$), and
  $^{13}$CO 2$\to$1 (2.0 to 28.9 K km s$^{-1}$) emission overlaid on
  \OI\ emission in colour scale (wedge in K km s$^{-1}$).}
\label{channel-outflow-blue}
\end{figure*}

\begin{figure*}
\centering
\includegraphics[width=8.5cm, angle=0]{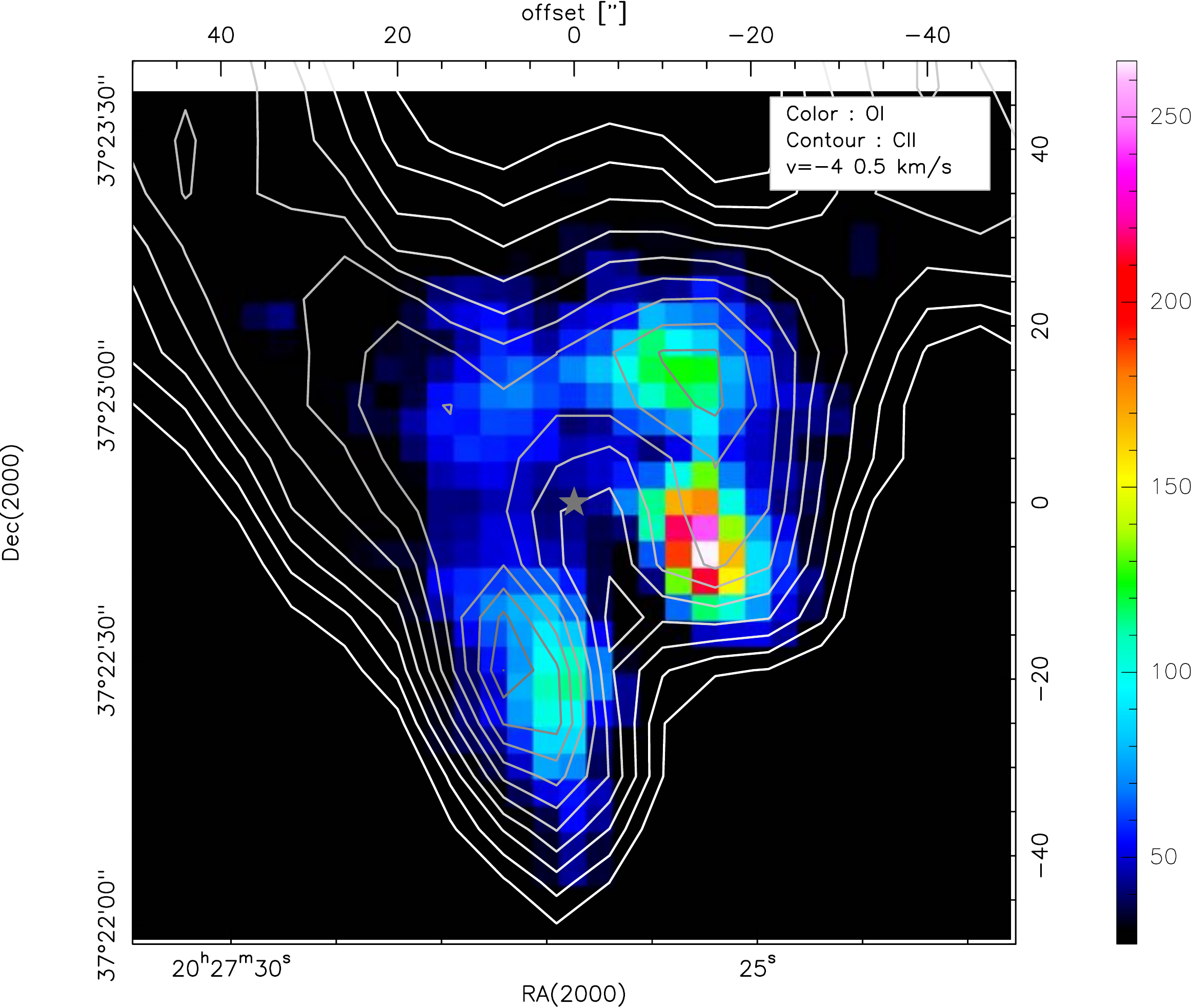}
\includegraphics[width=8.5cm, angle=0]{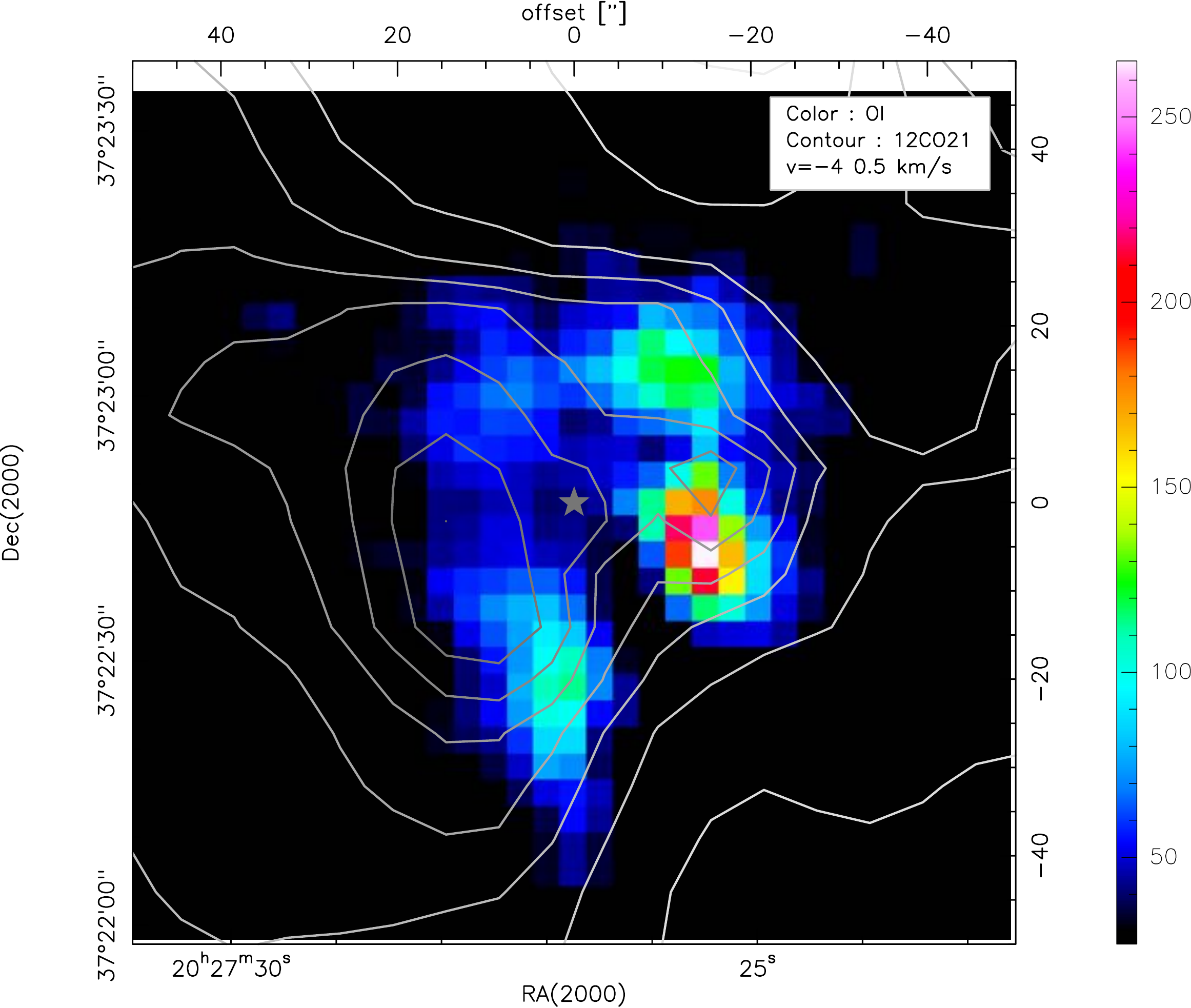}
\includegraphics[width=8.5cm, angle=0]{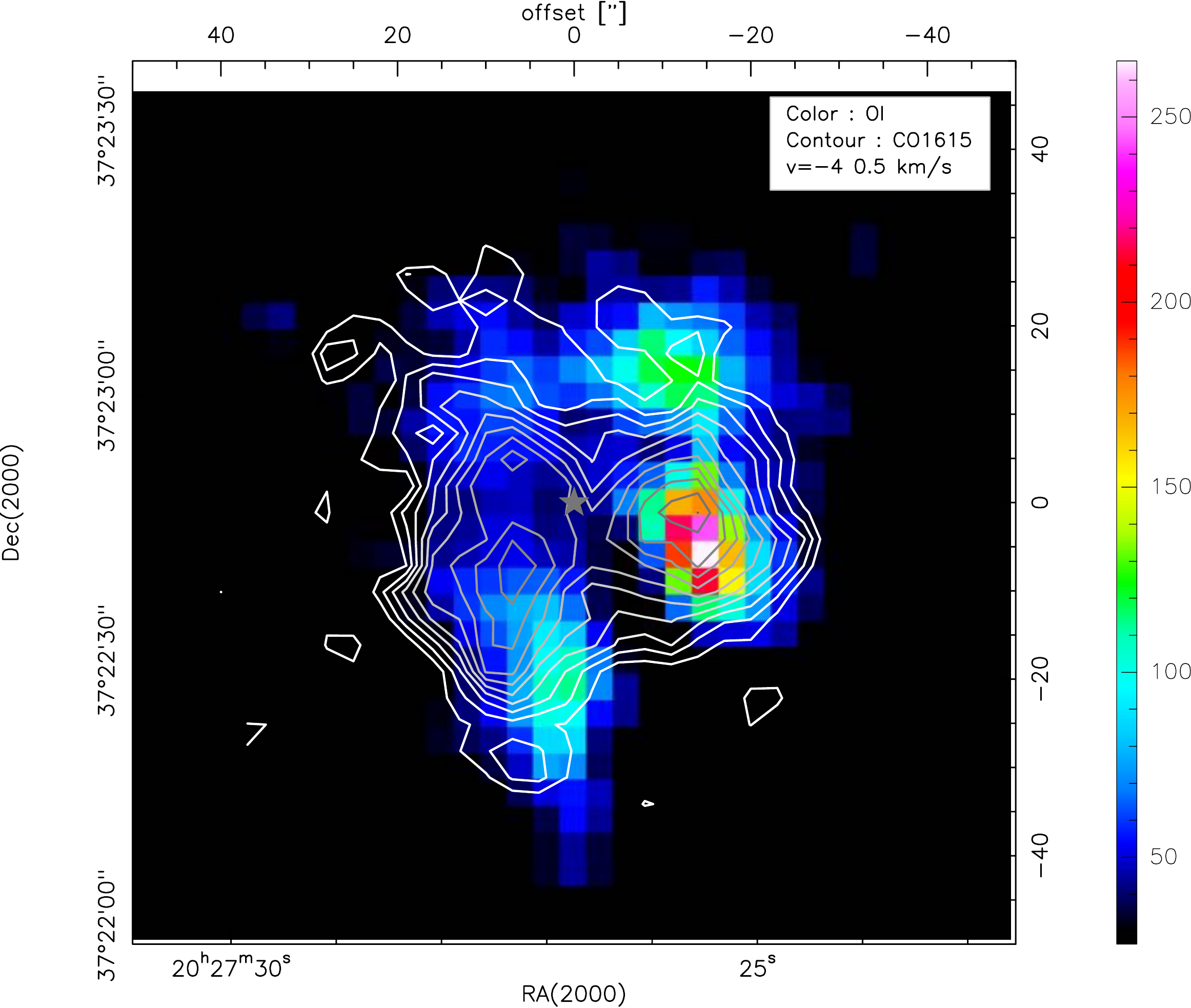}
\includegraphics[width=8.5cm, angle=0]{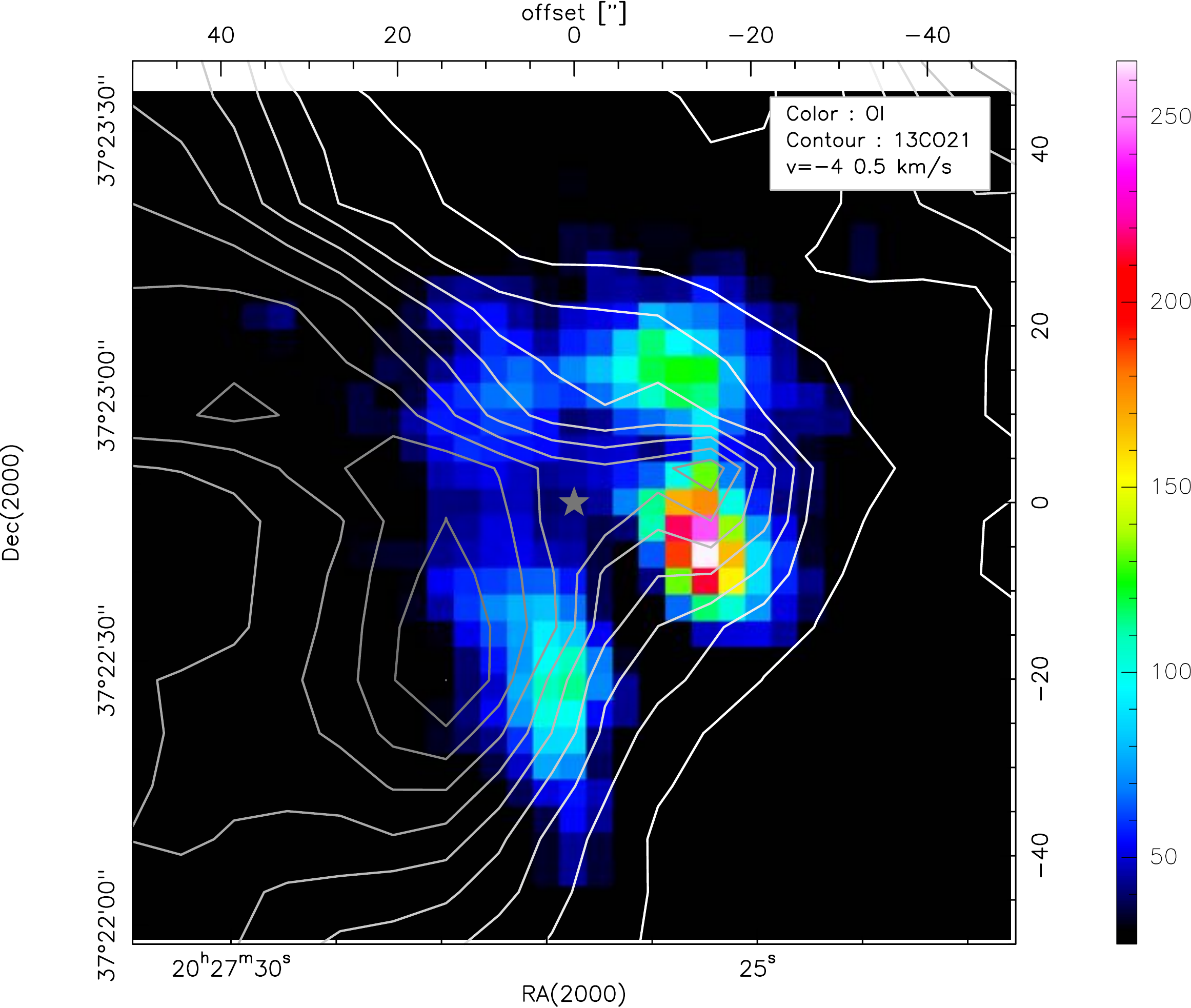}
\caption{Molecular cloud bulk emission (--4 to 0.5 km s$^{-1}$):
  Contours of \CII\ (72.5 to 235.5 K km s$^{-1}$), $^{12}$CO 2$\to$1
  (24.7 to 357.8 K km s$^{-1}$), $^{12}$CO 16$\to$15 (9.5 to 66.9 K km
  s$^{-1}$), and $^{13}$CO 2$\to$1 (5.8 to 84.5 K km s$^{-1}$) emission
  overlaid on \OI\ emission in colour scale (wedge in K km s$^{-1}$).}
\label{channel-bulk}
\end{figure*}

\begin{figure*}
\centering
\includegraphics[width=8.5cm, angle=0]{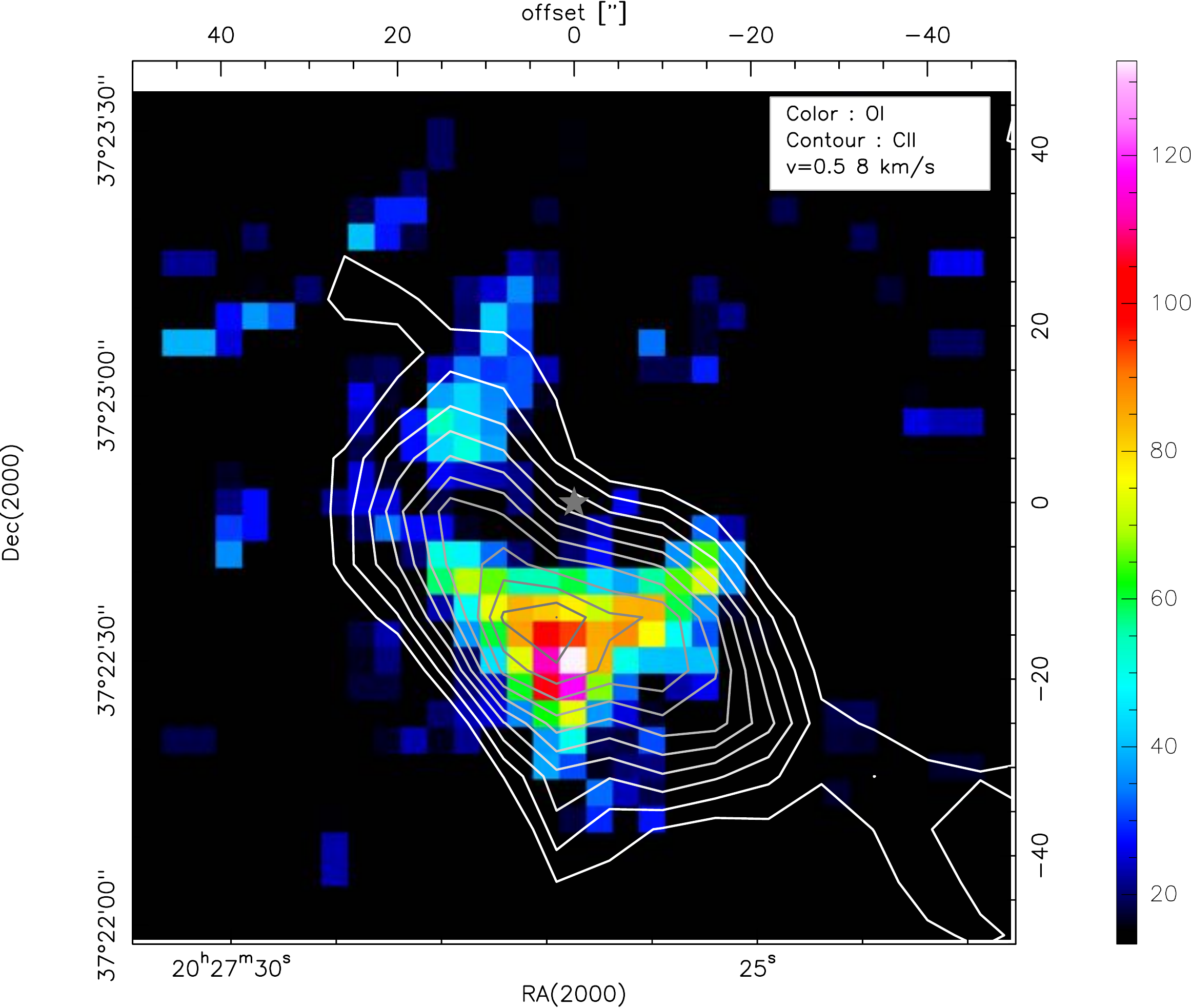}
\includegraphics[width=8.5cm, angle=0]{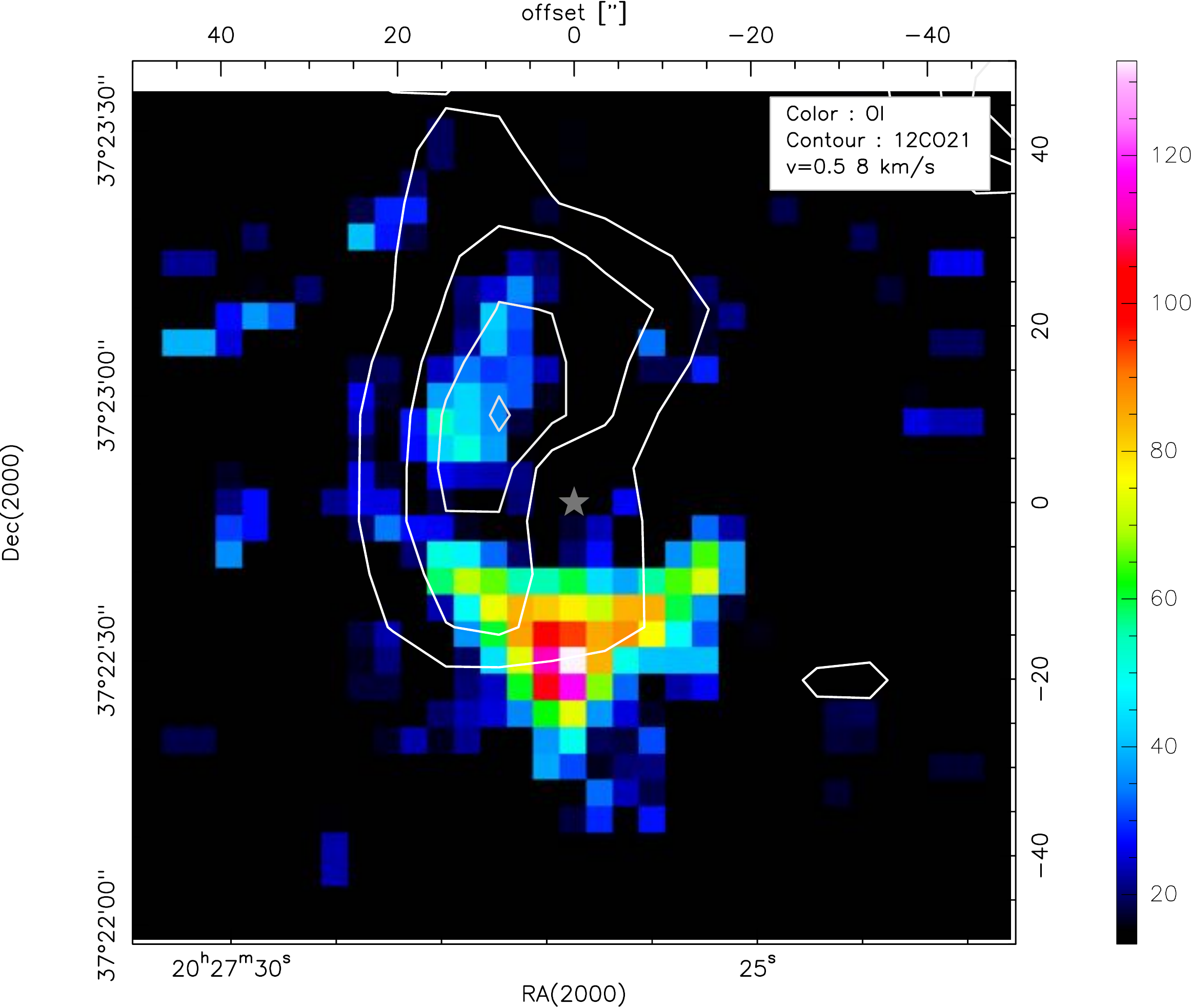}
\includegraphics[width=8.5cm, angle=0]{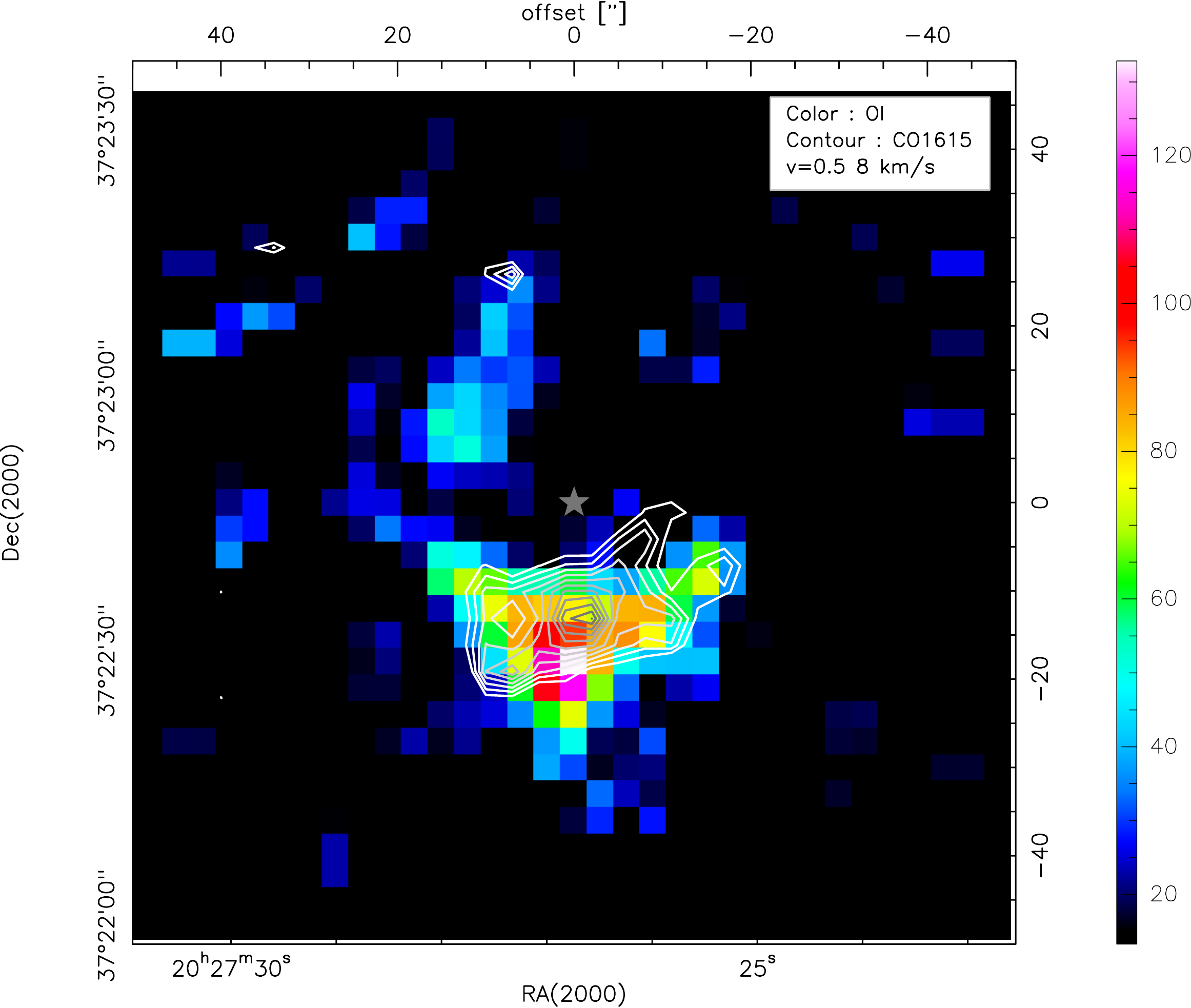}
\caption{Outflow red emission (0.5 to 8 km s$^{-1}$): Contours of
  \CII\ (58.5 to 190.2 K km s$^{-1}$), $^{12}$CO 2$\to$1 (3.2 to 46.6
  K km s$^{-1}$), and $^{12}$CO 16$\to$15 (15.0 to 48.9 K km s$^{-1}$)  emission overlaid on \OI\ emission in colour scale (wedge in K km
  s$^{-1}$).}
\label{channel-outflow-red}
\end{figure*}

\begin{figure*}
\centering
\includegraphics[width=8.5cm, angle=0]{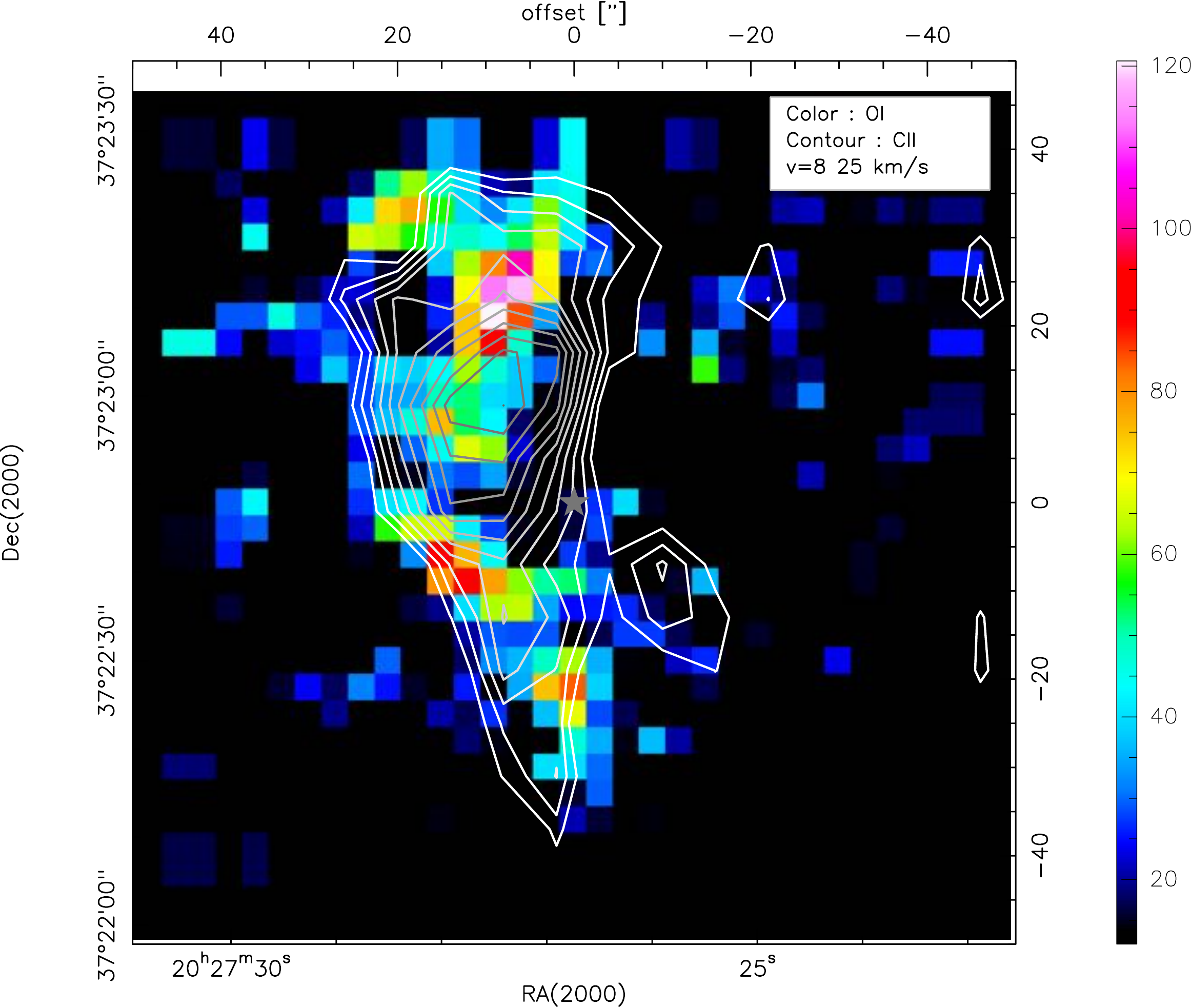}
\includegraphics[width=8.5cm, angle=0]{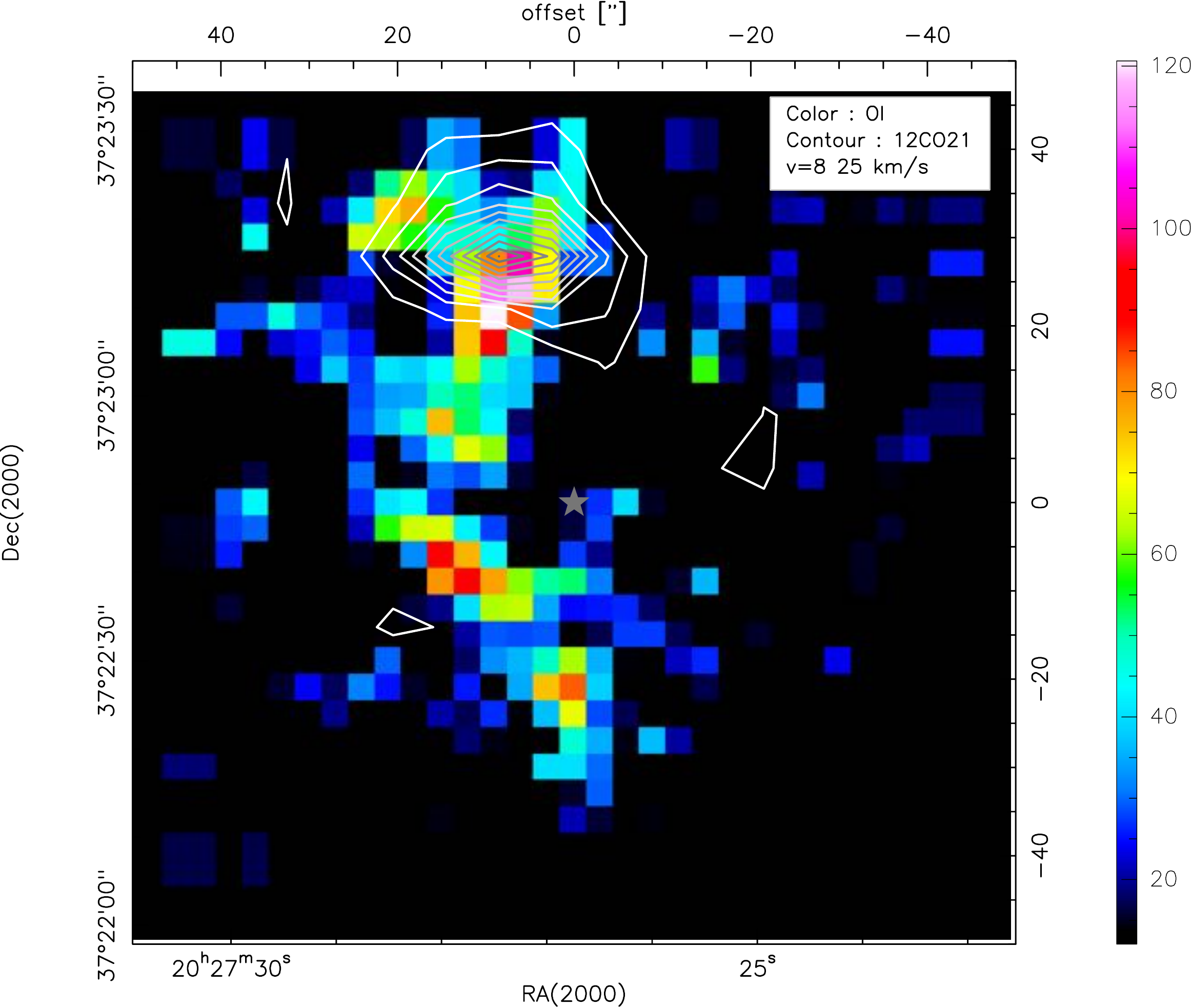}
\caption{High-velocity red emission (8 to 25 km s$^{-1}$): Contours of
  \CII\ (34.4 to 111.7 K km s$^{-1}$) and $^{12}$CO 2$\to$1 (3.1 to
  44.4 K km s$^{-1}$) emission overlaid on \OI\ emission in colour
  scale (wedge in K km s$^{-1}$).}
\label{channel-high-velocity-red}
\end{figure*}

Figures~\ref{channel-hv-blue} to \ref{channel-high-velocity-red} show overlays between \OI\ emission in colour scale and 
other tracers (\CII, CO 16$\to$15, $^{12}$CO 2$\to$1, and $^{13}$CO 2$\to$1) in the five main velocity ranges. 

\end{appendix} 

\end{document}